\documentclass[manuscript,nonacm]{acmart}

\AtBeginDocument{%
  }

\setcopyright{none}

\acmConference[]{Woodstock, NY}
\acmISBN{978-1-4503-XXXX-X/18/06}

\usepackage{booktabs}
\usepackage{caption}
\usepackage{hyperref}
\usepackage{makecell}
\usepackage{soul}
\usepackage{multirow}
\usepackage{graphicx}
\usepackage{accsupp}
\usepackage[font=small,labelfont=bf]{caption}
\usepackage{enumitem}
\usepackage[compact,small,noindentafter]{titlesec}
\usepackage[]{microtype}
\usepackage[defaultlines=2,all]{nowidow}
\newcommand{\rev}[1]{\textcolor{black}{#1}}
\newcommand{\revision}[1]{\textcolor{black}{#1}}
\newcommand{\blue}[1]{\textcolor{black}{#1}}
\usepackage{tikz}
\usepackage[noend]{algpseudocode}
\usepackage{algorithm}
\usepackage{arydshln}

\newtoggle{DEBUG}
\toggletrue{DEBUG}

\newcommand{\tabref}[1]{Table~\ref{#1}}
\newcommand{\figref}[1]{Fig.~\ref{#1}}
\newcommand{\alg}{\rev{\$M}}

\algnewcommand\algorithmicforeach{\textbf{for each}}
\algdef{S}[FOR]{ForEach}[1]{\algorithmicforeach\ #1\ \algorithmicdo}

\begin{document}


\title[\rev{\alg{} Recognizer for Mid-Air Gestures}]{\rev{\alg{} Recognizer for User-Defined Mid-Air Gestures}}

\author{Momona Yamagami}
\email{momona@rice.edu}
\affiliation{%
  \institution{Electrical and Computer Engineering, Rice University, Houston}
  \streetaddress{6100 Main St}
  \city{Houston}
  \state{Texas}
  \country{USA}
  \postcode{77005-1892}
}

\author{Claire L. Mitchell}
\affiliation{%
  \institution{The Information School | DUB Group, University of Washington, Seattle}
  \city{Seattle}
  \state{Washington}
  \country{USA}
  \postcode{98195-2350}
}

\author{Alexandra A. Portnova-Fahreeva}
\affiliation{%
  \institution{Paul G. Allen School of Computer Science \& Engineering, University of Washington, Seattle}
  \city{Seattle}
  \state{Washington}
  \country{USA}
  \postcode{98195-2350}
}

\author{Junhan Kong}
\affiliation{%
  \institution{The Information School | DUB Group, University of Washington, Seattle}
  \city{Seattle}
  \state{Washington}
  \country{USA}
  \postcode{98195-2350}
}

\author{H. Philo Katzman}
\affiliation{%
  \institution{Electrical and Computer Engineering, Rice University, Houston}
  \streetaddress{6100 Main St}
  \city{Houston}
  \state{Texas}
  \country{USA}
  \postcode{77005-1892}
}

\author{Jennifer Mankoff}
\affiliation{%
  \institution{Paul G. Allen School of Computer Science \& Engineering, University of Washington, Seattle}
  \city{Seattle}
  \state{Washington}
  \country{USA}
  \postcode{98195-2350}
}

\author{Jacob O. Wobbrock}
\affiliation{%
  \institution{The Information School | DUB Group, University of Washington, Seattle}
  \city{Seattle}
  \state{Washington}
  \country{USA}
  \postcode{98195-2350}
}

\renewcommand{\shortauthors}{Yamagami et al.}

\begin{abstract}
\rev{User-defined mid-air gestures hold promise to customize device interactions to a user's preferences, cultural background, and physical abilities.
This is especially true for individuals with upper-body motor disabilities, who may find standardized mid-air gestures inaccessible.}
    However, \rev{the biosignals that enable such mid-air gesture recognition} are multi-modal, multi-dimensional, and noisy, requiring domain expertise to design input features for gesture recognizers. 
    The \alg{}-recognizer enables \rev{mid-air} gesture recognition without needing expertise in biosignals or algorithms. 
    \alg{} resamples, normalizes, and performs dimensionality reduction to reduce noise and enhance signals relevant to the recognition.
    We evaluated \alg{} on 26 participants with and 8 participants without upper-body motor disabilities performing \rev{user-defined}
    ability-based gestures.
    For user-dependent and gesture articulation variability conditions, \alg{} outperformed our comparison algorithms (traditional machine learning with expert features and deep learning), with $>95$\% recognition rate.
    We tested \alg{}’s ease-of-implementation with six coders with minimal machine learning experience. Participants implemented \alg{} 53\% faster and with 23\% lower workload compared to deep learning. 
    \alg{} highlights the potential and feasibility of user-defined and accessible mid-air gesture interfaces.
\end{abstract}



\received{20 February 2007}
\received[revised]{12 March 2009}
\received[accepted]{5 June 2009}

\maketitle

\section{Introduction}

\begin{figure}[t]
    \centering
  \includegraphics[width=.8\linewidth,trim={0 .2cm 0cm .2cm},clip]{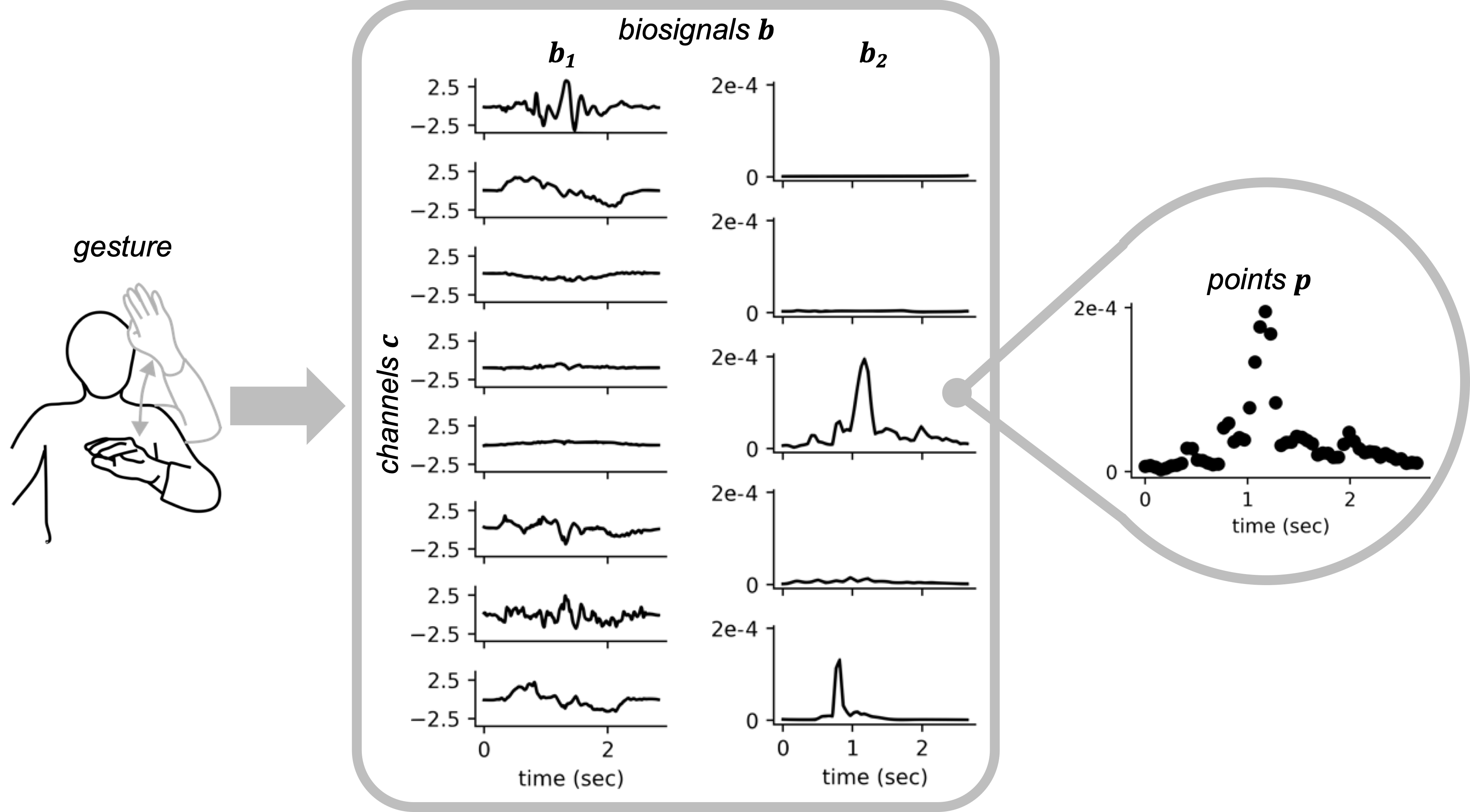}
  \caption{
  In biosignal interfaces, a user's gesture, whether a candidate gesture $G$ or a stored template gesture $T_i$, results in one or more biosignals $b$.
  Measured biosignal amplitudes can vary heavily depending on the sensor and measurement method. 
  In this example, the left biosignal (biosignal $b_1$) amplitudes range from $-2.5$ to $2.5$, whereas the right biosignal (biosignal $b_2$) amplitudes range from $0$ to $2\times 10^{-4}$.
  Each biosignal can have one or more biosignal channels $c$.
  Each channel contains one-dimensional signals composed of points $p$. 
  }
  \Description{
  A graphic overview of biosignal gestures. On the far left is an icon of a person moving their hand up and down. There is an arrow pointing away from this graphic to a graphic containing graphs of two biosignals b, b1 and b2. Each biosignal is associated with multiple amplitude vs time graphs. The x axis represents time and all graphs range from 0 to 3 seconds. b1 has 8 biosignal channels c all with amplitudes ranging from -2.5 to 2.5. b2 has 5 biosignal channels c all with amplitudes ranging from 0 to 2e-4. For both biosignals, some channels have amplitudes that are 0 for all times, while others have amplitudes that vary between the minimum and maximum amplitudes. Last, there is a magnifying glass pointing to one of the amplitude vs time graphs that is labeled "points p" and shows a magnified version of the same graph as a scatter plot.
  }
  \label{fig:teaser}
\end{figure}

Mid-air gesture interfaces 
are becoming more ubiquitous for device interactions, including with wearables~\cite{Xu2022-dw}, virtual and augmented reality~\cite{Albertini2017-fz}, and public displays~\cite{Schwarz2014-pi}.
\rev{
User-defined mid-air gestures that are personalized to each user's preferences, cultural background~\cite{Wu2019-fd}, and abilities~\cite{Yamagami2023-xj} have the potential to improve gesture memorability for the general population~\cite{Nacenta2013-ao} and device accessibility for people with disabilities~\cite{Zhao2022-gc,Fan2020-bu,Yamagami2023-xj,Wobbrock2018-cz, Wobbrock2011-wl,Khanna2024-gk}.
In particular, mid-air}
gesture interfaces can improve device accessibility for people with upper-body motor disabilities, as they enable people to interact with various body parts,
and---in inclusive interfaces---fully take advantage of their diverse abilities~\cite{Zhao2022-gc,Fan2020-bu,Yamagami2023-xj,Wobbrock2018-cz, Wobbrock2011-wl}. 
\rev{
Such gestures can be sensed using biosignals measured from sensors in, on, or around the body that record the user's actual (e.g., kinematic data captured using video cameras or accelerometers) or intended (e.g., electromyography or electroencephalography data that capture muscle and brain electrical activity) body movements.
}
However, translating biosignals into gesture classifications without domain expertise is challenging because biosignals are multi-modal, multi-dimensional, and noisy.
To ensure that emerging \rev{user-defined mid-air} gesture interfaces 
can be personalized to a specific user's \rev{preferences, cultural background, and} abilities, 
user interface developers need \rev{mid-air} gesture recognizers that can be integrated easily into prototypes without the need for extensive biosignal or algorithm development knowledge.
Almost 20 years ago, Wobbrock \textit{et al.} presented a \$1 unistroke recognizer~\cite{Wobbrock2007-ul} motivated by making pen gestures easier to recognize in user interface prototypes. Inspired similarly, here we offer \alg{}, a recognizer for \rev{user-defined} mid-air biosignal gestures (Fig.~\ref{fig:teaser}) collected from 34 participants, 26 of whom have upper-body motor disabilities.

Traditionally, biosignal gesture recognition has been applied in assistive technologies (e.g., an active prosthesis to replace a limb~\cite{Li2021-mr}) and rehabilitation (e.g., an exoskeleton for functional recovery after stroke~\cite{Ho2011-rv, Su2023-sy}). 
In these fields, feature selection by biosignal experts~\cite{Abbaspour2020-ei,Iqbal2018-vz,Dehzangi2018-ic} (e.g., number of zero-crossings, maximum signal amplitude) is an important component of gesture recognition and requires domain knowledge of important features of the specific biosignal. 
However, the need for deep domain expertise for feature selection makes biosignal gesture recognition challenging for those unfamiliar with biosignals.
More recently, deep learning has been used to train generalized models trained on large datasets and fine-tuned to individual users~\cite{Xu2022-dw,Shahi2024-bb,Ctrl-Labs_at_Reality_Labs2024-zb} or automatically derive important features from biosignal data~\cite{Atzori2016-ka,Triwiyanto2023-ub}. 
However, the large amounts of biosignal data needed for model learning~\cite{Bhasin2024-jg,Berenzweig2024-df, Eddy2025-ir} can be prohibitive to collect for accessibility applications and models trained on gesture data collected from individuals without motor disabilities~\cite{Ctrl-Labs_at_Reality_Labs2024-zb, Berenzweig2024-df,Shahi2024-bb,Xu2022-dw} may not extend well to individuals with motor disabilities or easily translated to the many coding languages that may be used in HCI prototypes.

For 2-D stroke-gesture interfaces, such as with pens, fingers, or mouse cursors, the \$-family gesture recognizers~\cite{Wobbrock2007-ul,Vatavu2018-lb, Vatavu2017-lx, Vatavu2012-da, Anthony2010-en, Anthony2012-tk} have had a tremendous impact on supporting lightweight gesture recognition prototyping, being successfully adopted in both research and commercial applications.
However, biosignal gesture recognizers, especially those that leverage data from multiple channels and \rev{potentially} support users with diverse abilities, have unique constraints that make the \$-family gesture recognizers unsuitable:
1) noise inherent to biosignals (e.g., human movement, poor contact between human and biosensor) can make it challenging to scale and rotate the dataset\rev{~\cite{De_Luca1997-jw,De_Luca2010-ul}};
2) not all channels are equally important in biosignal interfaces (e.g., information collected from left hand does not contain relevant information if the individual only moves their right hand); 
3) biosignals are often highly redundant;
4) different biosignals can be measured on vastly different scales (e.g., measurement amplitude, sampling frequency)\rev{~\cite{Rizzoglio2021-gj,Yamagami2018-cg}}.
All of the aforementioned issues make the \$-family recognizers unsuitable for biosignal gesture recognition.

In this work, we designed \alg{}, a new biosignal gesture recognizer for mid-air gestures.
\rev{
}
In contrast to recent state-of-the-art biosignal recognizers that utilize sophisticated deep learning techniques~\cite{Shahi2024-bb,Xu2022-dw,Ctrl-Labs_at_Reality_Labs2024-zb} or biosignal-specific features~\cite{Abbaspour2020-ei,Iqbal2018-vz,Dehzangi2018-ic}, \alg{} is simple and uses known techniques like resampling, normalization, and principal component analysis. 
Like the popular \$1 recognizer~\cite{Wobbrock2007-ul} before it, this simplicity makes \alg{} easily approachable to non-experts \rev{who may be unfamiliar with machine learning or biosignal-specific features}, and has the potential to enable \rev{user-defined} and accessible
gesture recognition for next-generation prototypes, including for virtual and augmented reality, gaming, and wearables applications.
Additionally, the simplicity of \alg{} makes it easy for generative AI tools to take in the pseudocode and generate the code in a language of the prototyper's choice.
For example, consider a college student who is developing an accessible virtual reality game in C++ using kinematic input obtained from video data. 
Without \alg{}, the student may have to investigate and implement kinematics features or replicate state-of-the-art algorithms like~\cite{Shahi2024-bb}, a significant endeavor. 
Or, the student can implement \alg{} with or without AI tools, run short tests to choose two parameters (number of time points, principal components), and continue development.
Prior work has addressed this challenge through the development of toolkits~\cite{Lee2019-jq, Kohlsdorf2013-qz, Ashbrook2010-fi} and libraries~\cite{Christ2018-tf}.
However, many prototyping languages and tools lack such toolkits and libraries; 
our work aims to fill this gap through the development and evaluation of \alg{}, a simple, language-agnostic algorithm that can be easily integrated into HCI prototypes. 

We tested \alg{} on mid-air gestures collected from 34 participants, 26 of whom had upper-body motor disabilities.
Overall, \alg{} outperformed our comparison algorithms---1) a support vector machine applied to  expert-derived biosignal features (\rev{SVM}) and 2) a multi-layer convolutional and long-short term neural network on normalized data (\rev{CNN+LSTM})---when the models \rev{were trained on user-defined gestures and} personalized to each user. 
Specifically, \alg{} had a recognition error rate of 9\% ($SD$=9), whereas \rev{SVM} had 18\% ($SD$=13) and \rev{CNN+LSTM} had 
17\% ($SD=14$). 
When the same gestures were articulated at a different time, faster, or larger than originally performed, \alg{} continued to outperform \rev{SVM and CNN+LSTM}, with recognition error rates of 24\% ($SD$=31), 62\% ($SD$=32), and 44\% ($SD$=29),
respectively. 
Lastly, although \alg{} did not perform as well as \rev{CNN+LSTM} for user-independent models overall 
(recognition error rate of 29\% ($SD$=19) vs. 22\% ($SD$=17),
respectively), the two algorithms performed similarly for participants with disabilities (recognition error rate of 
34\% ($SD$=23) vs. 30\% ($SD$=18),
respectively). 
We tested the ease-of-implementation of \alg{} with six participants with minimal machine learning experience. 
All participants found implementing \alg{} easier than implementing \rev{CNN+LSTM}.
Thus, as with prior \$-family recognizers, \alg{} is a simpler, more straightforward approach to recognizing gestures that yield better or similar performance as more complex, expert methods. 




In this work, we offer the following research contributions~\cite{Wobbrock2016-to}:
\begin{itemize}[leftmargin=*]
    \item \textit{Artifact.} The introduction of \alg{}, an algorithm inspired by the prior \$-family recognizers that takes into account the unique challenges associated with multi-dimensional \rev{mid-air} gestures. 
    The pseudocode can be found in Appendix~\ref{sec:appdx:M}.
    Our biosignal dataset and code are publicly available online~\cite{Yamagami2024-qg}.
    \item \textit{\rev{\alg{} Evaluation and Analysis.}} Evaluation of \alg{}, \rev{its components, and ablations} for 26 participants with and 8 participants without upper-body motor disabilities \rev{with inertial measurement unit (IMU) and electromyography (EMG) data}, demonstrating a low recognition error rate (9\%, $SD$=9) compared to 
    \rev{SVM with expert features}
    (18\%, $SD$=13)
    and 
    \rev{CNN+LSTM}
    (17\%, $SD$=14) 
    algorithms for \rev{user-defined gestures}.
    The \alg{} recognizer is robust against variations in how a gesture is articulated and performs comparably when tested across participants with disabilities.
    \item \textit{\rev{Ease-of-Implementation Study.}}
    Ease-of-implementation assessment of \alg{} compared to \rev{CNN+LSTM}, finding that all participants thought implementing \alg{} was easier than implementing \rev{CNN+LSTM}.
    \item \textit{Dataset.} A released dataset of 26 participants with and 8 participant without upper-body motor disabilities performing \rev{user-defined gestures}
    that are personalized to their abilities, when the same gestures are performed analogously, and when the same standardized gestures are performed across different participants ~\cite{Yamagami2024-qg}. 
    Our distributed electromyography and inertial measurement unit measurement system enabled us to collect data and recognize movements from individuals with significant motor disabilities and envision accessible futures.
\end{itemize}

Our work improves the ability to incorporate \rev{user-defined mid-air} gestures into user interfaces for new and emerging technologies without the need for domain expertise.


\section{Related Work}

Our work builds on prior research on gesture customization in HCI, both using \rev{mid-air} biosignal interfaces as well as more commonly used interfaces such as touchscreens~\cite{Wobbrock2007-ul}.
The prior work demonstrates that currently available biosignal interfaces in HCI are not accessible for people with upper-body motor disabilities. 
Further, we discuss prior research on body-machine interfaces for health and rehabilitation, which many of the biosignal HCI applications build on. 
While body-machine interfaces provide a starting point for biosignal interfaces for HCI applications, direct translation of body-machine interface approaches to HCI are challenging and require novel adaptations for translation to HCI applications~\cite{Eddy2023-iz}.

\subsection{Gesture Customization in HCI}

Gesture customization \rev{via supporting user-defined gestures} is important for improving device accessibility~\cite{Yamagami2023-xj}, accounting for cultural differences~\cite{Wu2019-fd}, and improving memorability of the gesture meaning compared to standardized gestures~\cite{Nacenta2013-ao}. 
In recent years, gesture recognition and customization for mid-air gesture input have received increasing academic attention~\cite{Xu2022-dw,Mezari2018-tn,Shahi2024-bb,Ctrl-Labs_at_Reality_Labs2024-zb,Saponas2008-dt,Huang2015-uh}.
Many focus on wrist-worn IMU sensors which record linear acceleration and rotational velocity in three-dimensions~\cite{Xu2022-dw,Mezari2018-tn, Shahi2024-bb,Yeo2020-zp, Khanna2024-gk}.
Other work focuses on non-invasive neural interfaces like electromyography (EMG)~\cite{Ctrl-Labs_at_Reality_Labs2024-zb,Saponas2008-dt,Huang2015-uh,Fan2018-ng,Zhang2021-vr,Rahimian2021-sq}, which measure muscle electrical activity.
Preliminary work with \rev{mid-air} biosignal interfaces takes inspiration from research in prosthetics to use expert-derived features~\cite{Saponas2008-dt} or distance metrics~\cite{Huang2015-uh} to classify gestures. 
These works highlight the potential of using biosignal interfaces for HCI applications.
\rev{However, expert-derived features requires deep knowledge and expertise in the specific biosignal of interest, making it challenging for someone with limited knowledge of biosignals to choose which features will result in the best performance.}

More recent works leverage large datasets (500+~\cite{Xu2022-dw} to 5000+~\cite{Ctrl-Labs_at_Reality_Labs2024-zb} participants) to develop deep learning models that are shared among the end-users. 
For work that supports gesture customization, one- or few-shot learning (an approach that leverages a pre-trained model to extend to unseen samples~\cite{Wang2020-ua}), has been used to enable gesture customization from few samples~\cite{Xu2022-dw,Shahi2024-bb,Tong2021-pr}.
While these algorithms can achieve up to 97\% accuracy even with only one template gesture~\cite{Shahi2024-bb}, there are several limitations with these approaches. 
First, the algorithms are complex to train and integrate into a project for developers not well-versed in algorithm development and deep learning.
Second, the algorithms require large datasets~\cite{Xu2022-dw,Ctrl-Labs_at_Reality_Labs2024-zb} or applying novel transformations to the dataset~\cite{Shahi2024-bb,Zhang2022-nw,Zhou2024-io,Deldari2022-nm}, which requires additional expertise.
Third, such gesture customization methods rely on pre-trained models that use demonstrations from people without motor disabilities, which may not extend well to people with upper-body motor disabilities.
Our work aims to fill this gap and make mid-air gesture customization accessible for user interface developers, who may not have access to expertise in algorithm development or large datasets.

In HCI, the \$-family recognizers for touch-based gestures has received wide acclaim in both academic and industry settings~\cite{Wobbrock2007-ul,Vatavu2018-lb, Vatavu2017-lx, Vatavu2012-da, Anthony2010-en, Anthony2012-tk}. 
The \$-family focuses on algorithms that are easy to implement, contain minimal code, and require only a small number of templates to customize gestures.
Towards this goal, the \$-family recognizers focus on transforming gestures by scaling and rotating and then comparing distances between candidate and template gestures for gesture recognition.
Crucially, the algorithms are simple to implement, requiring little expertise in gesture recognition and can identify gestures with only a few templates.
This work is the first in the \$-family to extend these principles to mid-air biosignal interfaces. 
As emphasized in our introduction and algorithm section, biosignal interfaces pose unique challenges compared to touchscreen gestures that require new innovations in transforming and comparing the candidate and template gestures. 

\subsection{(In)Accessible \rev{Mid-Air} Biosignal Input in HCI}

Biosignal usage has propagated extensively in HCI, enabling novel interaction techniques~\cite{kimura_2022_SilentSpeller}, advanced activity recognition~\cite{zhu_2018_EIS, li_2024_WheelPoser, guan_2017_ensembles}, social connection~\cite{liu_2017_supporting}, and continuous health monitoring~\cite{hao_2017_MindfulWatch,gashi_2020_detection}. 
During this transition, biosignal input has become a major focus due to its inherent ability to reflect the wearer's intentions and movements. 
Recent commercial advances have demonstrated the feasibility of using biosignals for input interaction including hand gestures for interaction with AssistiveTouch on Apple Watch~\cite{Apple_Support_undated-gs} and hand tracking on Meta Oculus~\cite{Meta_undated-hw}.  
%
%
However, despite the proliferation of devices that use biosignals and upper-body movements for interactions in recent years, these interactions remain inaccessible for people with upper-body motor disabilities due to the lack of gesture customization~\cite{Mott2020-io,Malu2018-uo} and assumptions underlying these gesture interfaces~\cite{Yamagami2023-xj}. 
People with motor disabilities experience challenges like difficulty with performing the motion precisely, fatigue, and not being able to do the desired motion at all due to their disability~\cite{Malu2018-uo,Yamagami2023-xj}.

Much of the research on improving accessibility to biosignal interfaces have been formative, focusing on understanding the accessibility limitations of existing devices~\cite{Malu2018-uo}, characterizing the effect of upper-body motor disability on biosignal input~\cite{Vatavu2022-qm}, or defining new gestures for device interaction~\cite{Fan2020-bu,Zhao2022-gc,Yamagami2023-xj}.
For example, \citet{Vatavu2022-qm} investigated the accessibility of stroke and motion gestures on wearable devices worn on the wrist, finger, and head, finding that participants with disabilities find motion-gestures performed in mid-air much easier than stroke-gestures performed on wearable touchscreens.
\citet{Carrington2014-wz, Carrington2016-by} explored and developed a set of input devices (``chairables'') designed specifically for individuals who use wheelchairs.
Other work investigated people's preferences on input modality~\cite{Li2022-he} or defined above-the-neck~\cite{Zhao2022-gc}, eyelid~\cite{Fan2020-bu}, and upper-body~\cite{Yamagami2023-xj} gesture sets that are inclusive of people with upper-body motor disabilities.
While these formative studies are critical to understanding the needs and challenges of individuals with motor disabilities, they fall short of addressing the challenge that there are no algorithms designed for non-expert developers that support accessible and customizable biosignal gesture recognition in HCI. 

\subsection{Body-Machine Interfaces for Health and Rehabilitation}

While \rev{mid-air} biosignal interfaces have only recently started to be investigated in HCI~\cite{Eddy2023-iz}, these signals have been extensively studied and implemented for body machine interfaces (BoMIs) in health and rehabilitation applications~\cite{Li2021-mr,Ho2011-rv, Su2023-sy,Abbaspour2020-ei,Iqbal2018-vz,Dehzangi2018-ic,Atzori2016-ka,Triwiyanto2023-ub}.
For example, inertial measurement units (IMU) and electromyography (EMG) sensors can be used to detect motion intention for lower-limb stroke rehabilitation~\cite{Su2023-sy}. 
Similarly, EMG has been used extensively for active prosthesis control to replace lost limbs~\cite{Li2021-mr}.

To perform gesture recognition for BoMI applications, windowed gesture data (e.g., 200 ms of data) are continuously classified, usually by extracting expert-derived features such as the number of zero-crossing or maximum signal amplitude~\cite{Abbaspour2020-ei,Iqbal2018-vz,Dehzangi2018-ic}.
Which expert features to extract depends on the biosignal type and application, and there is no one universal set of expert features.
More recently, deep learning has been successfully used in BoMI applications for gesture recognition~\cite{Atzori2016-ka,Li2021-mr} and continuous control~\cite{Berman2023-qu}.
Deep learning can be used to learn nonlinear relationships between the expert features and the gesture recognition~\cite{Li2021-mr} or to extract features directly from the biosignal data in a data-driven manner~\cite{Atzori2016-ka,Xu2022-dw}.

Despite these advances, direct translation of the techniques developed in this field to HCI applications is difficult for several reasons~\cite{Eddy2023-iz}.
First, research in BoMIs predominantly focuses on uninterrupted and continuous gesture recognition because the BoMI is being constantly used throughout the day.
Therefore, in health and rehabilitation applications, biosignal gestures are usually continuously recognized (e.g., classify biosignals every 200 ms), rather than the complete gesture being classified (e.g., detect when a gesture begins and ends, and then recognize the entire gesture trajectory).
In HCI applications, however, device interactions
\blue{can be}
intermittent,  which \blue{may require} new approaches for biosignal gesture recognition~\cite{Eddy2023-iz}.
\rev{For example, \citet{Khanna2024-gk} extracted a gesture ``nucleus'' using a change-point detection algorithm for recognizing mid-air gestures performed by individuals who are blind.}
Additionally, BoMI researchers and practitioners are experts in biosignal interfaces, with domain expertise of appropriate features and algorithms for their biosignal and task.
Our work aims to support user interface prototypes and developers who do not have this domain expertise so that they can still incorporate \rev {mid-air} biosignal gesture recognition into their HCI prototypes.

\section{The \alg{} Recognizer}


%
%
In this section, we present an overview of \alg{}, a ``\rev{mid-air} gesture recognizer'' \rev{for user-defined gestures}. 
We describe the four components of this recognizer \rev{(Step 1: Resample Each Time Channel; Step 2: Normalize; Step 3: Compute Template PCA; Step 4: Compute Template with Best Score)} and detail how the unique properties of biosignals were considered during development of the recognizer. 
A pseudocode of the algorithm can be found in Appendix~\ref{sec:appdx:M}.


%


\subsection{\rev{\alg{} Design Considerations}}

In mid-air biosignal recognition, gestures are multi-dimensional signals sampled over time, resulting in high-dimensional signals. 
These high-dimensional signals, if used in \$-family recognizers~\cite{Vatavu2018-lb, Vatavu2017-lx, Vatavu2012-da, Anthony2010-en, Anthony2012-tk,Wobbrock2007-ul} or other conventional HCI gesture recognition algorithms (e.g.,~\cite{Rubine1991-yo}) without further processing or an understanding of the biosignal type, could result in misrecognized gestures due to the data being manipulated without considerations for the unique challenges of \rev{mid-air} biosignal gestures. 

First, biosignals may be measured on different scales
and the inherent characteristics of the different sensors must be considered.
Some types of sensors may have a higher sampling frequency than others, leading to an unequal sample number for a given gesture.
This can be resolved with resampling over time.
Measurement amplitudes for different biosignals may be orders of magnitude apart from one another.
For example, EMG data is usually measured in volts, with signal amplitudes around $10^{-4}$ volts, whereas accelerometer data is usually measured in terms of the gravitational constant, $g$ (where $1g=9.81 m/s^2$), with amplitudes around 1 to $10g$.
Without proper resampling and scaling, it would be difficult to compare different biosignals (\figref{fig:teaser}, \figref{fig:normalize}). 
To normalize the amplitudes, one can employ a calibration scheme where users are asked to flex their muscles as hard as they can~\cite{Yamagami2018-cg} or move their arms in a random pattern~\cite{Rizzoglio2021-gj}.
However, this adds additional burden on the user to perform calibration and is not desirable in a user interface.

\begin{figure}[t]
  \centering
  \includegraphics[width=.4\linewidth,alt={A single magnitude vs time (sec) graph. The magnitude ranges from -2 to 2, and the time ranges from 0 to 3 seconds. There are three lines, a dark blue line, a light blue line, and a light grey line corresponding to "signal 1", "signal 2", and "noise" labels. The dark blue signal 1 line starts at 0, increases in amplitude from 0 seconds to 0.7 seconds to a magnitude of 2, and then decreases in amplitude back to 0 around 1.2 seconds. It then decreases in amplitude to -2 around 1.7 seconds before going back to 0 around 2.5 seconds. The light blue signal 2 line follows the same pattern has the dark blue line but with the magnitude flipped, such that the signal decreases in amplitude first to -2 before increasing in amplitude to 2. The light grey noise line stays at 0.}]{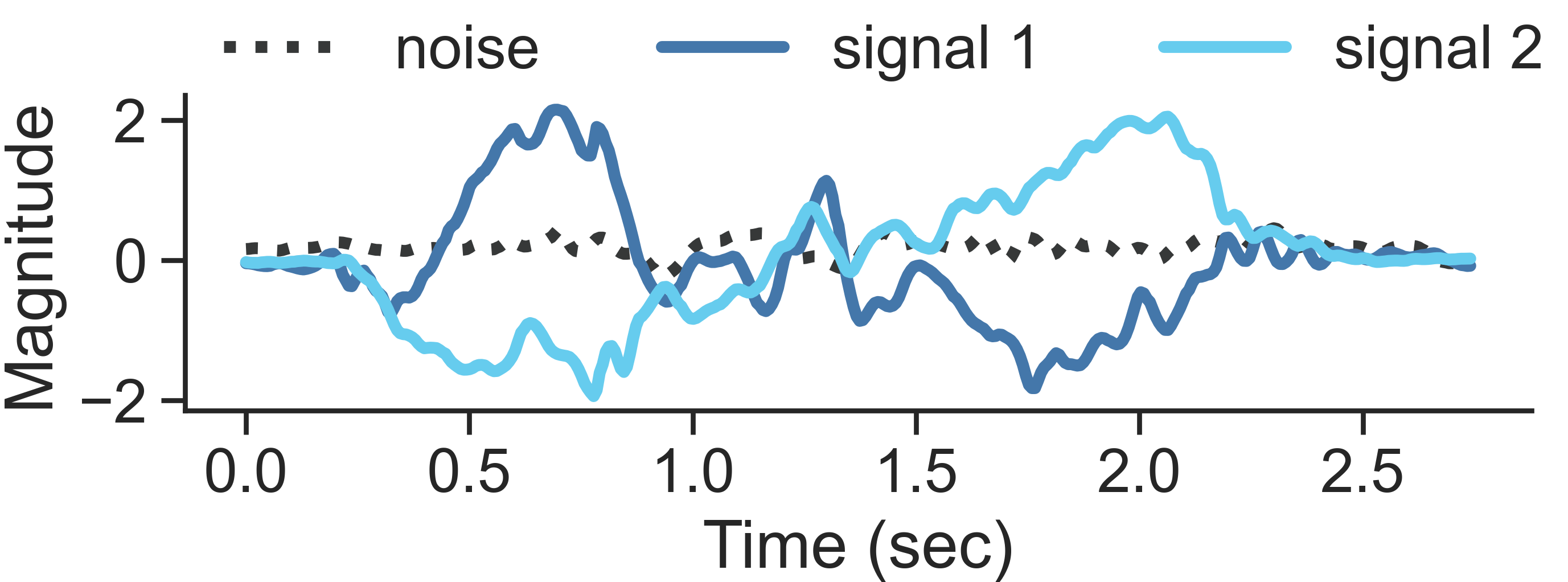}
  \caption{Biosignals can have relevant (signals 1 and 2 in dark and light solid blue) or irrelevant (noise in dotted black) channels for recognition.
  Both relevant and irrelevant biosignals contain noise that must be reduced to improve recognition accuracy.
  Additionally, biosignals are often correlated.
  For example, signals 1 and 2 contain similar information in opposite orientations.
  As signal 1 increases in magnitude, signal 2 decreases in magnitude and vice versa.
  Decreasing signal correlations through dimensionality reduction reduces noise and improves recognition accuracy.
  }\label{fig:considerations}
  \Description{
  A single magnitude vs time (sec) graph. The magnitude ranges from -2 to 2, and the time ranges from 0 to 3 seconds. There are three lines, a dark blue line, a light blue line, and a light grey line corresponding to "signal 1", "signal 2", and "noise" labels. The dark blue signal 1 line starts at 0, increases in amplitude from 0 seconds to 0.7 seconds to a magnitude of 2, and then decreases in amplitude back to 0 around 1.2 seconds. It then decreases in amplitude to -2 around 1.7 seconds before going back to 0 around 2.5 seconds. The light blue signal 2 line follows the same pattern has the dark blue line but with the magnitude flipped, such that the signal decreases in amplitude first to -2 before increasing in amplitude to 2. The light grey noise line stays at 0.
  }
\end{figure}

Second, biosignals can be extremely noisy, and even channels that contain important information may be corrupted (\figref{fig:considerations}, dark and light blue lines). 
Sensor failure, human movement, or poor connection between the sensor and the human are all potential sources of noise~\cite{De_Luca2010-ul, De_Luca1997-jw}.
Additionally, some biosignals such as electromyography (EMG) signals, which quantify muscle electrical activity, can be highly stochastic and motor disabilities could cause unintended large signal amplitudes causing outliers.
This can make it challenging to adopt simple scaling methods, as is done in previous \$-family recognizers~\cite{Wobbrock2007-ul,Vatavu2018-lb, Vatavu2017-lx, Vatavu2012-da, Anthony2010-en, Anthony2012-tk}. 

Third, in biosignal interfaces, not all channels are equally important (\figref{fig:considerations}, black dotted line).
In the simplest example, consider a biosensor that captures movement from both arms, but the user only moves their left arm to perform a gesture.
The biosignal data collected from the right arm contains no useful information to recognize the gesture.
Therefore, simply adapting the amplitude of the points to fit a bounding box may not be beneficial if the entire biosignal channel is solely noise.
Similarly, biosignal interfaces can be highly redundant and contain similar information within multiple channels (\figref{fig:considerations}, dark and light blue lines).
Consider the scenario where a biosensor that captures movement information is placed on the wrist and the forearm. 
If the individual is moving their entire arm to perform a gesture, the two sensors contain redundant information about the movement. 
While redundant information is not necessarily an issue for gesture recognition, noise in multiple redundant channels can make it challenging to accurately compare template gestures with the candidate gesture.

To alleviate the large number of channels and noise associated with biosignal data, dimensionality reduction methods that reduce the biosignal channels to a lower-dimensional \textit{latent space} must be employed.
However, when performing this transformation, it is important to ensure that gestures from the same gesture class are close in the latent space, so they can be correctly recognized as the same gesture. 
This is because, depending on how the dimensionality reduction method is applied, noise inherent in biosignal data may affect the transformation and make it challenging to recognize similar gestures in the latent space.

Therefore, in addition to the design criteria originally laid out for the \$1 recognizer~\cite{Wobbrock2007-ul}, the \alg{} recognizer must:
\begin{enumerate}[leftmargin=*]
    \item Scale biosignal data  without the need for user calibration while maintaining differences in signal amplitude between relevant and irrelevant channels and ignoring biosignal measurement outliers;
    \item Decrease biosignal noise and redundancy prior to gesture recognition; and
    \item Employ dimensionality reduction methods that transform high-dimensional biosignal gestures to a lower-dimensional \textit{latent space} such that gestures from the same gesture class are close to each other in the latent space.
\end{enumerate}
With these design needs in mind, \alg{} performs \rev{mid-air} biosignal gesture recognition in four steps corresponding to the pseudocode in Appendix~\ref{sec:appdx:M}.



\subsection{The \alg{} Algorithm}
\rev{
\alg{} is a supervised machine learning algorithm, meaning that training samples (i.e., templates $T_i$) are needed to train \alg{} to recognize testing samples (i.e., candidate gesture $G$). 
\alg{} assumes that both candidate gestures $G$ or templates $T_i$ have the same number of biosignal channels $c$ which span some variable time $t$ (\figref{fig:teaser}).}
Each channel, depending on its signal origin type, is of biosignal type $b_j$.
The \alg{} recognizer resamples, translates, normalizes, and decomposes the $c$ channels of the candidate gesture $G$ or template $T_i$. 
Candidate gestures $G$ and templates $T_i$ are treated the same for Steps 1 and 2, with both steps being applied when the template points are initially read or when the gesture is performed.  

\subsubsection{Step 1: Resample Each Time Channel.}
To make template gestures and candidate gestures directly comparable even with multiple biosignal inputs or different movement speeds, we first resample each biosignal channel such that all channels have the same number of points $n$. 
The number of points $n$ to resample each channel to is a parameter that must be explored and defined by the prototyper based on the biosignal of interest. 

\begin{figure}[t]
  \centering
  \includegraphics[width=.5\linewidth]{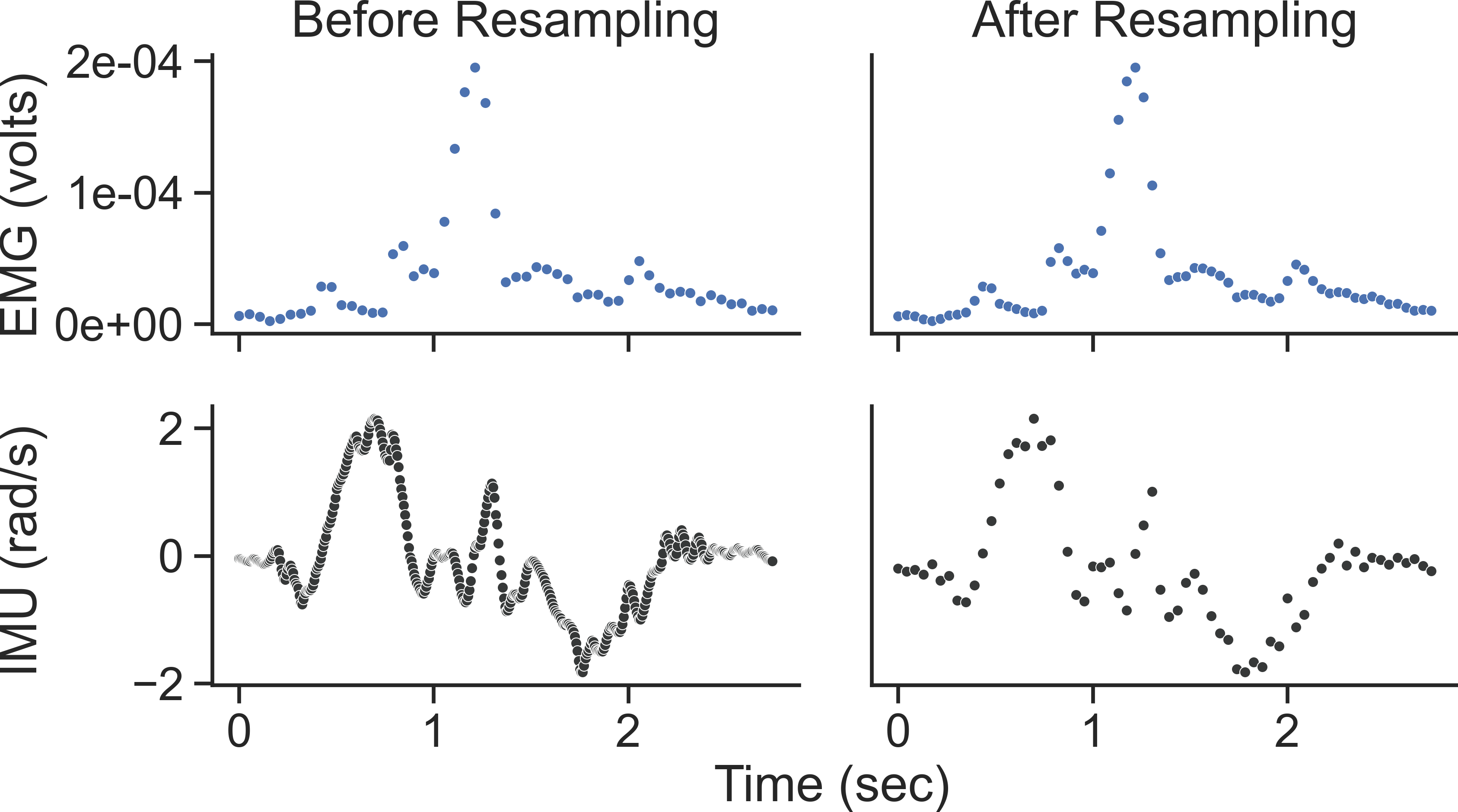}
  \caption{Two different biosignals---(top, EMG) electromyography, measures muscle electrical activity; (bottom, IMU) inertial measurement unit, measures linear acceleration and rotation velocity---may have different sampling rates (left), and must be resampled (right) before comparisons are made.
  Note the two different measurement scales for the EMG signal (around $10^{-4}$ volts) and the IMU signal (around $2$ g).
  }\label{fig:resample}
  \Description{
  Four biosignal plots over time arranged in 2 rows and 2 columns. The rows represent electromyography (EMG) (top) and inertial measurement unit (IMU) (bottom) biosignals, the columns represent before resampling (left) and after resampling (right). Both before and after EMG signals look similar, with a scatter plot indicating a rapid spike in amplitude around 1.5 seconds to 2e-4 V. Both IMU plots have the same shape and vary in amplitude between -2 and 2 g. However, the density of the points are different, with the "after" graph appearing as a downsampled version of the "before" graph. In the "after" graph, both EMG and IMU signals appear to have a similar number of points representing the biosignals.
  }
\end{figure}

\subsubsection{Step 2: Normalize.}
\begin{figure}[t]
  \centering
  \includegraphics[width=.5\linewidth]{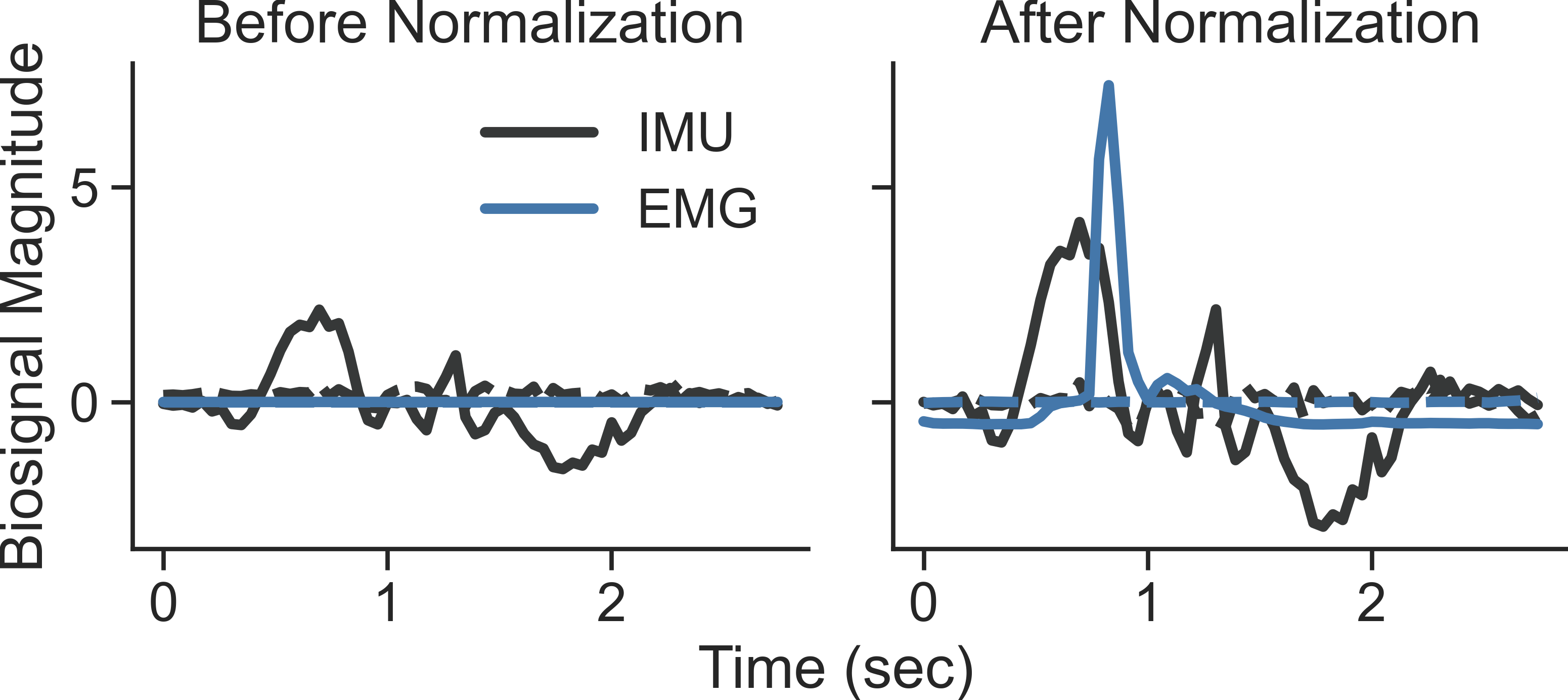}
  \caption{EMG (blue lines) and IMU (black lines) signals before (left) and after (right) normalization.
  The solid line represents relevant channels for recognition.
  The dotted line represents irrelevant noise channels. 
  Before normalization (left), the EMG signal and noise are both significantly smaller in amplitude than the IMU signal and noise.
  After normalization (right), the EMG signal is now on the same scale as the IMU signal. 
  Both EMG and IMU noise channels remain small in magnitude. 
  }\label{fig:normalize}
  \Description{
  Two biosignals plots, one on the left and another on the right. The y axis represents signal magnitude ranging from -2 to 5, and the x axis represents time ranging from 0 to 3. On the left, before normalization, the blue lines representing EMG activity can barely be seen, and the main line that can be seen varying in amplitude is the IMU signal in light grey. The dotted grey line is also hard to see and does not vary in amplitude. On the right, after normalization, both the blue and grey solid lines, representing EMG and IMU signals, can be seen varying in amplitude. The grey dotted line and blue dotted line does not change in amplitude, and is a flat horizontal line at 0.
  }
\end{figure}

After resampling, the gesture is normalized such that each channel has a mean of 0 and 
a given biosignal across all channels has a standard deviation of 1 (\figref{fig:normalize})~\cite{Hye2023-xa}.
\revision{For example, if both IMU and EMG data are collected, for a given gesture, concatenated IMU and EMG channels are normalized to 1 separately.}
This ensures that after normalization, channels relevant to the recognition are the same order of magnitude across different biosignals and channels irrelevant to the recognition remain small (\figref{fig:normalize}, right).
\revision{The normalization step occurs on a per-gesture basis, not on a per-user basis.
As each gesture is added to \alg{}, the biosignal channels are normalized based solely on the added gesture itself.
Therefore, there is no calibration step when new users are added. 
}

\subsubsection{Step 3: Compute Template PCA.}
In this step, only template $T_i$ principal components and transformed points are computed and saved after the templates' points are read in, resampled, and normalized.
The template principal components will be used to transform the candidate points before recognition in Step 4. 

Principal component analysis (PCA) is a dimensionality reduction method that transforms high-dimensional data into a low-dimensional \textit{latent space}~\cite{Greenacre2022-co}.
PCA reorients high-dimensional data such that the data's variance (i.e., how much a signal fluctuates from its mean) is maximally explained using only a few major components. 
It can be a useful tool to extract biosignals relevant to recognition, which usually has a high variance, and remove noise, which usually has a low variance (\figref{fig:considerations}).
It is also a useful tool to decrease correlations and redundant information within a dataset.
For this step, we compute the principal components of the template $T_i$ and apply the transformation to the templates' points. 
Note that the prior normalization step also served as an important preprocessing step for PCA.

To compute the principal components, all biosignals $b$ and channels $c$ must first be vertically concatenated.
This results in a data matrix $D$ with $c$ rows and $n$ columns, where $c$ represents the total number of channels across all measured biosignal types and $n$ represents the number of resampled time points.
At this point, a built-in PCA function (e.g., Python: \texttt{sklearn.decomposition.pca}; Java: \texttt{weka.attributeSelection.PrincipalComponents};  Matlab: \texttt{pca}; \\R: \texttt{stats.princomp}) can also be used to compute the principal components $U$ and the transformed $points$ in the latent space. 
\revision{
The $points$ can then be flattened into a one-dimensional vector of length $nPC\times n$.
The final vector $points$ can be used to compare the templates $T_i$ against the candidates gestures $G$ through a point-by-point comparison.
A more detailed explanation of the PCA steps can be found in Appendix~\ref{sec:appdx:pca}.}

\subsubsection{Step 4: Compute Template with Best Score.}

Lastly, a candidate gesture $G$ is compared against templates $T$ to find a template match $T'$ that minimizes the distance between the candidate and template $points$. 
First, the candidate gesture $G$ is read-in, resampled, and normalized with Steps 1 and 2.
Next, all biosignals $b$ and channels $c$ for the candidate gesture $G$ are vertically concatenated such that $G$ is a matrix with $c$ rows and $n$ columns, where $c$ represents the total number of channels across all measured biosignal types, and $n$ represents the number of resampled time points.
Last, $G$ is compared against each template $T_i$. 
For each template, the principal component of the template $U$ computed in Step 3 is read-in. 
Then, $G$ is transformed to the template latent space via matrix multiplication ($G^TU$)
and the resulting matrix is flattened into a one-dimensional vector $points$ of length $nPC\times n$. 
Then, the gesture $points$ can be compared against the template points $T_i$ by computing the sum of the absolute value of the difference between the two (i.e., $d=\sum_{k}|points[k] - T_i[k]|, k=1$ to $nPC\times n$)~\cite{Pasqual2014-xt}. 
The template with the smallest distance will be considered a match. 

\begin{figure}[t]
  \centering
  \includegraphics[trim={4em 12em, 6em 17em},clip,width=.6\linewidth]{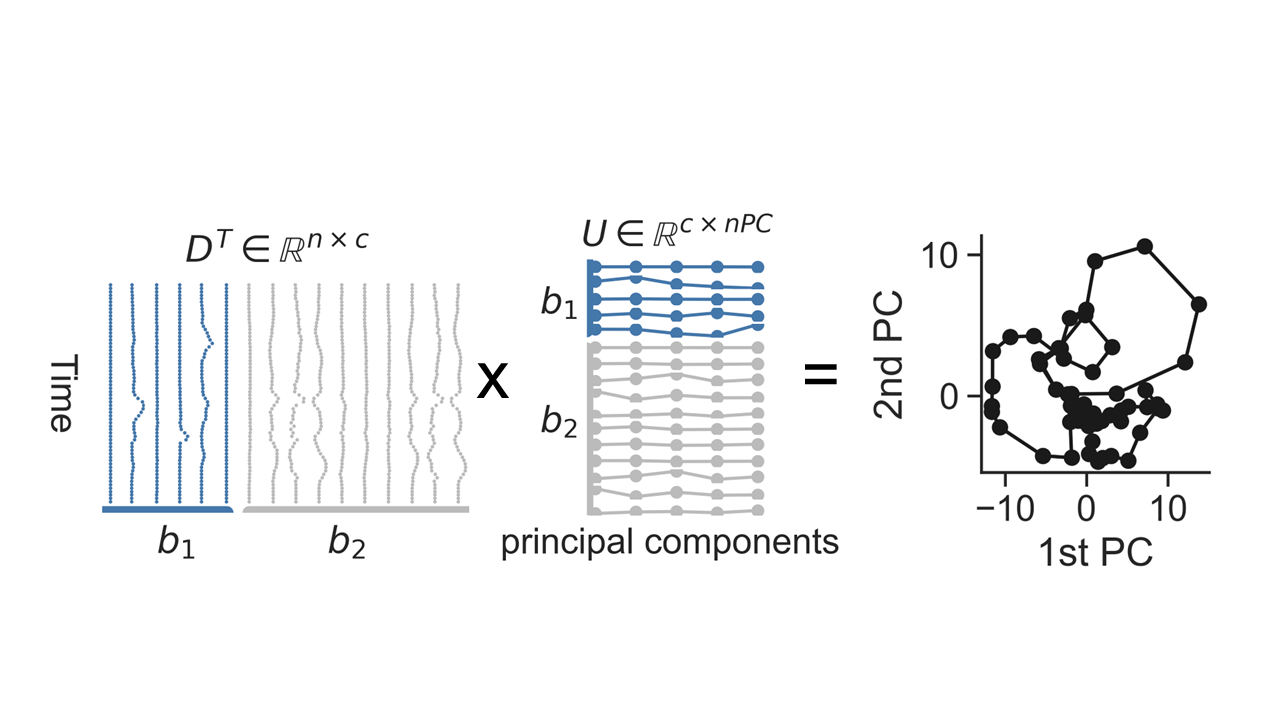}
  \caption{In principal component analysis (PCA), high-dimensional signals (left) are decomposed into its principal components (middle). 
  Multiplying them results in the signals being translated to a lower-dimensional \textit{latent space} (right), where signal redundancy and noise have been reduced and signal variability is maximized. 
  The original time-domain biosignals $D^T$ (left), or the transpose of matrix $D$, is a matrix with $n$ rows representing time and $c$ columns representing biosignal channels.
  The biosignal channels may represent multiple biosignals $b_1$ (blue), $b_2$ (grey).
  The principal components $U$ (middle), is a matrix with $c$ rows representing biosignal channels and $nPC$ columns representing the number of principal components to extract. 
  $nPC$ is a parameter for the \alg{} that must be tuned for a given dataset.
  In this figure, $nPC = 5$. 
  Multiplying $D^T$ with $U$ will transform the high-dimensional signals in a lower-dimensional latent space (right) with $n$ rows representing time and $nPC$ columns representing the number of principal components.
  The right plot represents the transformed data plotted over the first two principal axes, which represent directions of greatest variance.
  }\label{fig:PCA}
  \Description{
  Three graphs from left to right, there is an x sign indicating multiplication between the first two graphs, and an equal sign representing the result of matrix multiplication between the second two graphs. The first graph is a visual depiction of the transpose of matrix D. The transpose of matrix D has time on the y axis and biosignal channels b1 and b2 on the x axis. There are 6 lines representing b1 in blue and 10 lines representing b2 in gray. The next graph is a visual depiction of U, the principal components of D. The graph depicts the amplitudes of biosignal channels b1 and b2 on the y axis and 5 principal components on the x axis. Each biosignal changes in amplitude according to it's contribution to the first five principal components. THe last graph plots the first two principal components on the x and y axis. The time-domain signal appears to form two circles on this latent space representation of D.
  }
\end{figure}

\subsection{\revision{Example Use Cases of \alg{}}}\label{sec:B:usecase}

This work presents \alg{} and its evaluation on data collected from people with and without motor disabilities.
We anticipate that readers will have many possible use cases for \alg{} that use various coding languages, such as:
\revision{
\begin{itemize}[leftmargin=*]
    \item A college student who is developing an accessible virtual reality game using hand motion data collected from cameras (e.g., C\#).
    \item An app developer who wants to enable gesture input for their application using smartphone sensors (e.g., Swift iOS).
    \item A graduate student who is creating an exercise gaming system for older adults using smartwatch \rev{or neural data} (e.g., JavaScript).
\end{itemize}
}
\rev{\noindent A key feature of \alg{} is that only two parameters need to be tuned (number of resampled points and number of principal components), supporting rapid prototyping. 
When choosing these parameters, we encourage developers to systematically evaluate a range of parameter choices and choose the one with the highest performance based on preliminary data.
Due to the math behind PCA, the maximum number of principal components possible is either the number of resampled points or the number of biosignal channels, whichever is smaller. }

We deliberately tested \alg{} on high-quality biosignal data (i.e., IMU and EMG data collected from 16 sensors spread across the participants' upper-body (Fig.~\ref{fig:sensorPlacement});
although this setup is not desirable for everyday use, we wanted to initially develop and test \alg{} with the highest-quality data possible.
We chose combined IMU and EMG sensors rather than kinematic data extracted from cameras to be inclusive to participants who are not able to generate changes in pose that are needed to collect kinematic data~\cite{Yamagami2023-xj}.
Our \alg{} demo 
uses a laptop camera and MediaPipe~\cite{Lugaresi2019-zi}, supporting translation of \alg{} to more ecologically feasible sensors.
With the demo, \$M obtained a 76\% accuracy with a single template for each of the ten features on a single participant, comparable to our results in this work (Fig.~\ref{fig:main}).

\subsection{\rev{Design Considerations and Limitations of \alg{}}}

\subsubsection{Why PCA as a Dimensionality Reduction Method?}

For high-dimensional noisy signals like what is measured from biosignal interfaces, dimensionality reduction is a critical step for gesture recognition to reduce noise and redundancies between signals.
For the \alg{} recognizer, we chose to apply principal component analysis (PCA), but there are other ways of reducing the dimensionality of the data and extracting relevant information from noisy data.
Here, we briefly summarize our justification and compare PCA to other common dimensionality reduction methods.

First, what does it mean to reduce the dimensionality of a gesture? 
In touch-based gestures, the gesture results in a two-dimensional $x,y$ points array over time. 
Both directions usually contain unique information relevant to gesture recognition.
In biosignal gestures, sensors measure and collect multi-dimensional points over time. 
For example, even a ubiquitous sensor such as an accelerometer collects acceleration data over three channels: $x, y, $ and $z$. 
If multiple sensors are used (e.g., an accelerometer on each body segment), the number of channels can increase significantly. 
To decrease the number of dimensions (i.e., channels) and decrease noise associated with biosignals, dimensionality reduction is a critical step in \rev{mid-air} biosignal gesture recognition. 

There are many ways to decrease noise and enhance the signal from high-dimensional data, including channel reduction and linear and non-linear dimensionality reduction methods. 
Channel reduction is perhaps the most straightforward way to reduce high-dimensional data. 
In channel reduction, channels whose largest amplitude signal falls below a certain threshold (e.g., 5\% of the maximum amplitude measured for a given biosignal) are ignored when matching template gestures to candidate gestures~\cite{Sharma2023-gf}. 
For example, if an individual only uses their right hand for mid-air gestures, biosignals measured from the left side can be ignored. 
The method is simple to implement and eliminates channels that only contain noise. 
However, it is not possible to combine
redundant information with this method.
In contrast, linear (e.g., principal component analysis~\cite{Greenacre2022-co}) and non-linear (e.g., autoencoder~\cite[Ch~14]{Goodfellow2016-xz} ) dimensionality reduction methods compress high-dimensional data into a lower-dimensional \textit{latent} space that reduces noise and combines channels with redundant information. 
Linear dimensionality reduction methods are generally faster to train and easier to implement compared to non-linear dimensionality methods; therefore, we chose PCA for \alg{}.

\subsubsection{Limitations of the \alg{} Recognizer.}

Because we apply the same transformation to the latent space for both the template and candidate gestures, \alg{} cannot account for mirrored gestures.
For example, if an individual wants mirrored gestures to be recognized but only stores templates for the right hand, \alg{} will not consider the same gesture performed with the left hand a match. 
Similarly, if the sensors are moved or placed in a different orientation (e.g., an accelerometer is rotated 90 degrees on the person's wrist) without storing new templates, \alg{} will not recognize candidate gestures accurately. 
%
Conversely, this design choice also enables users to map mirrored gestures to two separate recognition labels.
For example, a tap with the right and left index fingers could represent ``\textit{select}'' and ``\textit{duplicate}'', respectively.
This could be useful in scenarios where users have limited movement and want to maximize the number of labels associated with a small gesture set.
\section{\alg{} Evaluation: Method}

The goal of this evaluation is to compare the performance of \alg{} to other commonly used \rev{mid-air} gesture recognizers, namely traditional machine learning with expert features (\rev{SVM}) and deep learning (\rev{CNN+LSTM}).
To ensure a diverse sample of gestures and to ensure that \alg{} is ability-inclusive~\cite{Wobbrock2018-cz, Wobbrock2011-wl}, we collected personalized gestures from participants with ($N=26$) and without ($N=8$) motor disabilities.
Due to the promise of \rev{user-defined mid-air} biosignal interfaces to enable accessible device input~\cite{Yamagami2023-xj}, we wanted to develop a recognizer that worked for both people with and without upper-body motor disabilities.  
\blue{A more detailed analysis of a subset of the data (\textit{Personalized}, see below), including the types of movements the participants came up with, was previously published in~\citet{Yamagami2023-xj}.}\footnote{\blue{The quantitative analysis in \citet{Yamagami2023-xj} solely demonstrated the muscles that participants used to perform the user-defined movements and the potential for data-driven methods to identify user-defined movements using data collected from the \textit{Personalized} condition.
This work focuses on applying the complete dataset, including \textit{Personalized, Variations,} and \textit{Standardized} conditions to validate \alg{} and its components.
The full dataset of all three conditions are published with this paper.
Therefore, relevant details of the experiment from~\citet{Yamagami2023-xj} are repeated here for completeness and clarity.
}}

\subsection{Participants}

\begin{table}[]
\caption{Participant Demographics}
  \label{tab:participants}
\begin{tabular}{ccc}
\multicolumn{1}{c|}{\textbf{Characteristic}} &
  \multicolumn{1}{c|}{\textbf{\begin{tabular}[c]{@{}c@{}}with disability\\ ($N=26$)\end{tabular}}} &
  \textbf{\begin{tabular}[c]{@{}c@{}}without disability\\ ($N=8$)\end{tabular}} \\ \hline\hline
\multicolumn{1}{c|}{\textbf{\begin{tabular}[c]{@{}c@{}}Age: yrs, \\ mean $\pm$ SD\end{tabular}}} &
  \multicolumn{1}{c|}{$50\pm18$} &
  $27\pm8$ \\ \hline\hline
\multicolumn{3}{c}{\textbf{Gender}}                           \\ \hline 
\multicolumn{1}{c|}{Woman}           & \multicolumn{1}{c|}{12} & 5 \\ \hdashline 
\multicolumn{1}{c|}{Non-Binary}      & \multicolumn{1}{c|}{2}  & 0 \\ \hdashline 
\multicolumn{1}{c|}{Man}             & \multicolumn{1}{c|}{12} & 3 \\ \hline\hline
\multicolumn{3}{c}{\textbf{Race/Ethnicity}}                   \\ \hline
\multicolumn{1}{c|}{White/Caucasian} & \multicolumn{1}{c|}{24} & 4 \\ \hdashline
\multicolumn{1}{c|}{\begin{tabular}[c]{@{}c@{}}Black/\\ African-American\end{tabular}} &
  \multicolumn{1}{c|}{2} &
  0 \\ \hdashline
\multicolumn{1}{c|}{Asian}           & \multicolumn{1}{c|}{1}  & 4 \\ \hline\hline
\multicolumn{3}{c}{\textbf{Handedness}}                       \\ \hline
\multicolumn{1}{c|}{Right-handed}    & \multicolumn{1}{c|}{20} & 6 \\ \hdashline
\multicolumn{1}{c|}{Left-handed}     & \multicolumn{1}{c|}{2}  & 2 \\ \hdashline
\multicolumn{1}{c|}{\begin{tabular}[c]{@{}c@{}}Right-handed \\ before disability\\ onset, but now \\ left-handed\end{tabular}} &
  \multicolumn{1}{c|}{4} &
  N/A
\end{tabular}
 \Description{
 A table representing participant demographics. The column labels are "characteristic", "with disability (N=26)" and "without disability (N=8)". Each of the 4 rows represents a different characteristic. For each characteristic, a value for "with disability" (first number reported) and "without disability" (second number reported) are listed. Row 1, "age: yrs, mean plus or minus SD": 50 plus or minus 18, 27 plus or minus 8. Row 2, "gender", sub-row 1, "woman": 12, 5; sub-row 2: non-binary: 2, 0; sub-row 3: man: 12, 3. Row 3, "race or ethnicity", sub-row 1, "white or Caucasian": 24, 4; sub-row 2, "Black or African-American": 2, 0; sub-row 3, "Asian": 1, 4. Row 4, "handedness", sub-row 1, "right-handed": 20, 6; sub-row 2, "left-handed": 2, 2; sub-row 3, "right-handed before disability onset, but now left-handed": 4, not applicable.
 }
\end{table}

We recruited 26 participants with and 8 without upper-body motor disabilities from the greater 
Seattle
community \rev{by distributing flyers to relevant community groups} (\tabref{tab:participants}). 
Our participants with disabilities ($N=26$) self-reported various motor disabilities that affected their ability to physically interact with technology including: spinal-cord injury (13), muscular dystrophy (3), peripheral neuropathy (3), essential tremor (3), and other disabilities (4). 
Participants' average age and standard deviation since disability onset was $17\pm15$ years, with the minimum and maximum time since disability onset being 2 and 52 years, respectively.
The average age and standard deviation of the participants, when their disability started, was $33\pm20$, with some participants having their disability since birth and the oldest disability onset at 68 years. 
We quantified the level of disability of the participants using the \textit{Quick Disabilities of the Arm, Shoulder and Hand} instrument (Quick DASH)~\cite{Gummesson2006-xx}. 
It provides a quantitative [0,~100] measure of upper-body function, and greater values indicate that a disability has a greater impact on daily life.\footnote{We modified the wording of the Quick DASH to mitigate ableist assumptions that the person's disability is a \textit{problem}.
For example, one question asks ``...to what extent has your arm, shoulder or hand \textit{problem} interfered with your normal social activities...?''
This was replaced with ``...to what extent has your arm, shoulder or hand \textit{affected by your disability and/or chronic condition} interfered with your normal social activities...?''} 
The average Quick DASH score and standard deviation was $55\pm18$, with the minimum measured score being 14 and the maximum being 82. 
A detailed analysis of the participant demographics, types of movements that
participants came up with, and detailed prompts and instructions \blue{for a subset of the conditions (\textit{Personalized}}, see below) can be found in~\citet{Yamagami2023-xj}.

\subsection{Apparatus}

\begin{figure}[t]
    \centering
    \includegraphics[trim={6em 6em 10em 5em},clip,width=.65\columnwidth]{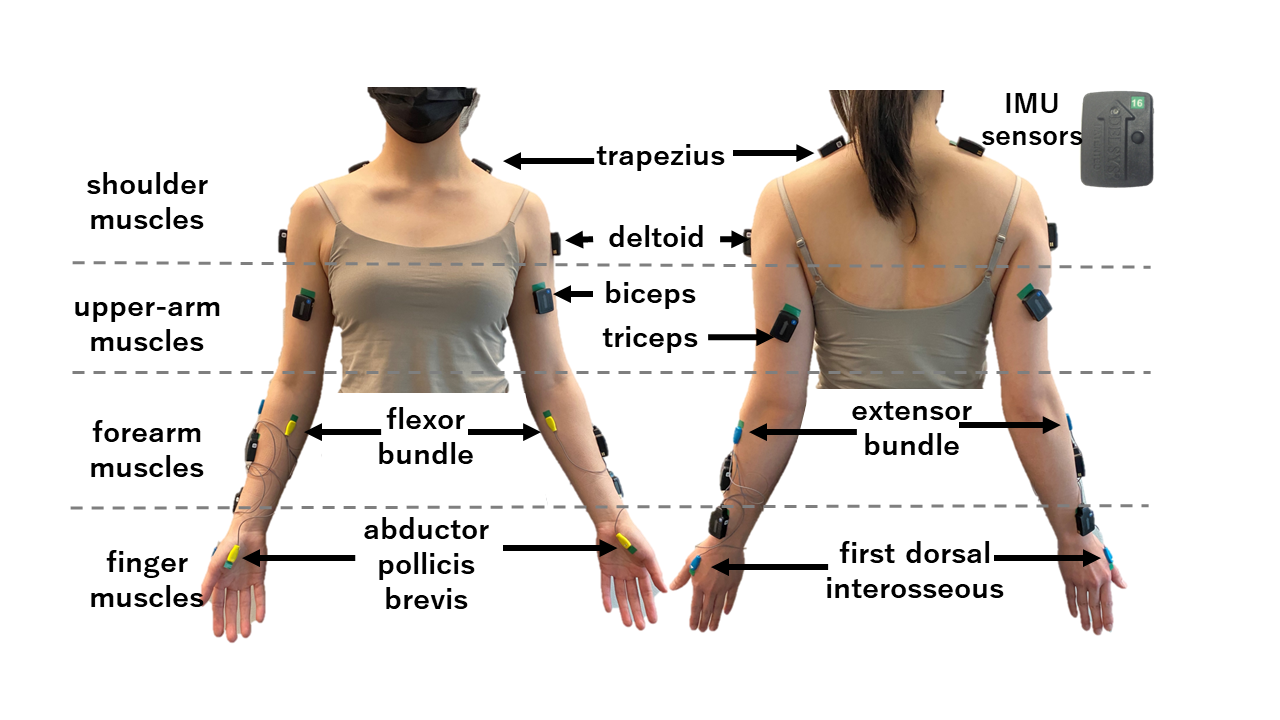}
    \caption{Combined IMU and EMG sensors were placed on the participant's shoulders (trapezius, deltoid) and upper arms (biceps, triceps).
    The large black sensors on the forearms and wrists only sense IMU signals. 
    The small yellow and blue sensors on the person's forearms (flexor, extensor) and hands (abductor pollicis brevis, first dorsal interosseous) only sense EMG signals.
    The placements were chosen to sense movement or muscle activation from the participant's upper body, including the fingers, arms, shoulders, neck, and head.
    From~\cite{Yamagami2023-xj}.}
    \Description{
    Photo of a person with IMU and EMG sensors placed on their shoulders, upper-arms, forearms, and fingers. A large black box is placed on the person's shoulder, upper-arm, forearms, and wrist, and smaller sensors are placed on the person's forearms and fingers. The sensors are distributed across the person's upper-body.
    }
    \label{fig:sensorPlacement}
\end{figure}

We placed combined Delsys\footnote{Delsys, Inc., Boston, MA, USA. \url{https://delsys.com/}} EMG and IMU sensors on the participants' upper-body (\figref{fig:sensorPlacement}).
12 inertial measurement unit (IMU) sensors were placed on participants' wrists, forearms, upper arms, shoulders, and neck to measure linear and rotational movement.
16 electromyography (EMG) sensors were placed on participants' finger muscles, forearm muscles, upper-arm muscles, and shoulder or neck muscles to measure muscle activations, which can occur without movement. 
Delsys Trigno Duo were placed on the smaller finger and forearm muscles and  Delsys Trigno Avanti sensors were placed on the rest of the muscles. 
We used the Delsys software development kit to stream and save the biosignal data. The Avanti and Duo sensors collected IMU and EMG data at slightly different frequencies (Avanti: \rev{148} Hz and 1259 Hz, respectively; Duo: 74 Hz and 1926 Hz, respectively). The software development kit synchronized all the data and upsampled the IMU signals to $148$ Hz and the EMG signals to 2000 Hz. 
The data was collected using a customized data collection interface in  Unity.\footnote{Unity Technologies, San Francisco, CA, USA. \url{https://unity.com/}}

\subsection{Procedure}

Participants provided consent according to our protocol approved by the 
University of Washington's 
Institutional Review Board.
All participants provided informed consent that their data will be publicly released.
\rev{Study completion times ranged from 2-4 hours.}
Participants filled out a short demographics questionnaire that asked about their age, gender, race, ethnicity, and handedness.
Participants with motor disabilities also reported on their disability, time since the onset of their disability or chronic condition that affects their ability to interact with technology, and answered questions from the Quick DASH instrument~\cite{Gummesson2006-xx} to quantify the extent to which their upper body affects their everyday lives.

\begin{figure*}
  \includegraphics[width=\textwidth, trim={.5cm 1.4cm .5cm 3.4cm},clip]{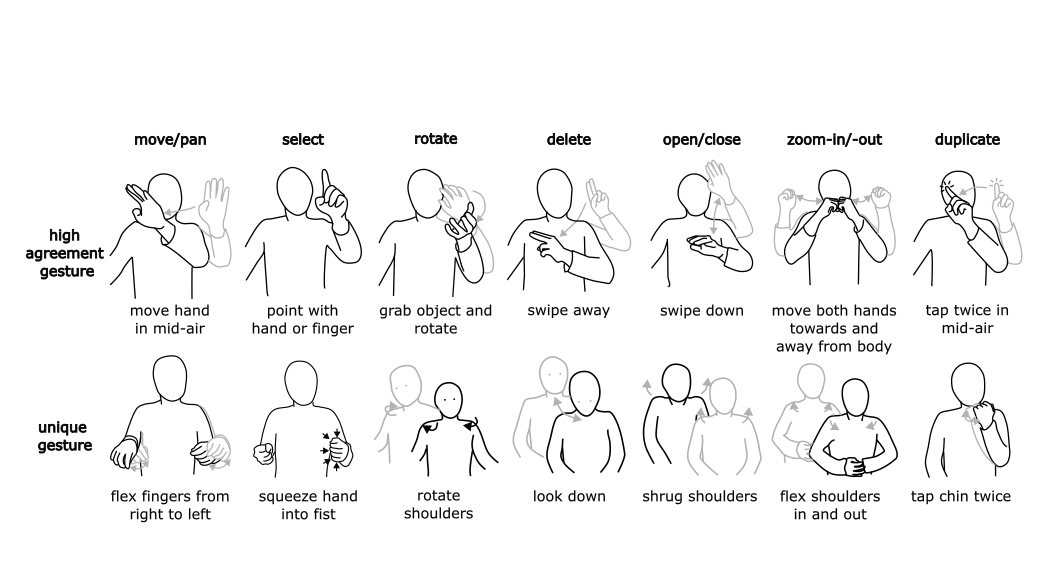}
  \caption{Upper-body gestures authored by 26 participants with upper-body motor impairments in \textit{Personalized} condition. 
  The top row represents gestures that multiple people designed and the bottom row represents a unique gesture that only one participant came up with. 
  At least four participants came up with unique gestures for each function.
  Participants performed high agreement gestures in different ways depending on their disability, such as pointing their whole hand at the screen instead of their index finger.
  From~\cite{Yamagami2023-xj}.}
  \Description{14 drawings of participant gestures organized as 2 rows and 7 columns. The top row represents high agreement gestures, and the bottom row represents unique gestures that participants designed. The columns are functions (from left to right): move/pan, select, rotate, delete, open/close, zoom-in/-out, duplicate. For move/pan, the high agreement gesture was moving hand from left to right in mid-air and the unique gesture was flexing fingers from left to right. For select, the high agreement gesture was pointing with hand or finger, and the unique gesture was squeezing hand into fist. For rotate, the high agreement gesture was grabbing object and rotating about the person's wrist and the unique gesture was rotating shoulders. For delete, the high agreement gesture was swiping away and the unique gesture was looking down. For open/close, the high agreement gesture was swiping down, and the unique gesture was shrugging shoulders. For zoom-in/-out, the high agreement gesture was moving both hands towards and away from body, and the unique gesture was flexing shoulders in and out. For duplicate, the high agreement gesture was tapping twice in mid-air in two separate locations with the hand and the unique gesture was tapping chin twice.}
  \label{fig:exgestures}
\end{figure*}

We then had our participants perform gestures for each of the three conditions below in sequence. 
We did not randomize conditions or repeated gestures due to concerns about participant dropout, time constraints, and participant cognitive ability concerns. 
To account for time fluctuations, participants repeated one gesture (rotate) in the \textit{Variations} condition, introducing natural variations in gesture data.
Participants did 10 gestures for each prompted function. 
A ``function'' was an action a user might take with an interactive system, for example, \textit{zoom-in}. 
Participants completed each of the three conditions:
\begin{enumerate}[leftmargin=*]
    \item \textit{Personalized}: Participants were asked to come up with 10 \rev{user-defined} gestures corresponding to 10 functions originally proposed in Wobbrock et al.~\cite{Wobbrock2009-md}: (1) \textit{move}, (2) \textit{select}, (3) \textit{rotate}, (4) \textit{delete}, (5) \textit{pan}, (6) \textit{close}, (7) \textit{zoom-in}, (8) \textit{zoom-out}, (9) \textit{open}, (10) \textit{duplicate}.\footnote{Two participants without disabilities 
    were prompted to perform gestures for \textit{shrink} and \textit{enlarge} instead of \textit{duplicate} and \textit{open}. 
    One participant performed the same gesture for \textit{shrink} and \textit{zoom-out}, as they felt that the two were identical functions.
    Both participants mentioned feeling as though \textit{shrink}/\textit{enlarge} and \textit{zoom-in}/\textit{zoom-out} were effectively the same functions.
    Therefore, we changed \textit{shrink} and \textit{enlarge} to \textit{duplicate} and \textit{open} for subsequent participants.} 
    Participants were told that they could come up with any gesture they wanted that involved their upper body, including their fingers, arms, shoulders, and head or neck.
    \item \textit{Variations}: Participants were asked to articulate the \textit{rotate} gesture in different ways
    to determine \alg{}'s performance when the same gestures are articulated differently. 
    We chose only one gesture to articulate in different ways to decrease fatigue and minimize our study duration. 
    At the beginning of the variations condition, each participant was shown a video of themselves performing their rotate gestures in the personalized phase to ensure they remembered the gesture they came up with.
    For the ``time'' variation, we asked participants to do the gesture in the same way they did in the personalized condition. 
    This was to test the effect of time on biosignals.
    For the ``speed'' variation, we asked participants to do the same rotate gesture, but as fast as possible.
    For the ``size'' variation, we asked participants to do the rotate gesture as large as possible.
    \item \textit{Standardized}: Participants were asked to perform five standardized gestures currently integrated into the Meta Oculus virtual reality platform~\cite{Meta_undated-hw} and Microsoft HoloLens augmented reality platform~\cite{Microsoft2021-zx}: (1) \textit{point and pinch}, (2) \textit{pinch and scroll}, (3) \textit{palm pinch}, (4) \textit{air tap}, and (5) \textit{wrist tap}. 
    Items 1-4 are one-handed gestures, while item 5 is a two-handed gesture.
    For each gesture, participants were shown videos of an avatar performing the gesture. 
    Participants without motor disabilities were asked to do the gesture exactly as shown in the video.
    Participants with motor disabilities could choose whichever hand they preferred to do one-handed gestures.
    Individuals whose disabilities only affected function on one side were asked to use their affected side to do the one-handed gestures.
\end{enumerate}
Participants without motor disabilities ($N=8$) completed all three conditions. 
All of our participants with motor disabilities ($N=26$) completed the \textit{personalized} condition, but only 25 and 10 participants completed the \textit{variations} and \textit{standardized} conditions, respectively,
due to fatigue and participants' physical abilities.
Fig.~\ref{fig:exgestures} provides examples of the types of upper-body gestures that were authored by the 26 participants with upper-body motor disabilities in the \textit{Personalized} condition. 


\subsection{Model Architectures and Hyperparameters}

We tested \alg{} against two commonly used recognition methods in \rev{mid-air} gesture recognition: (1) traditional machine learning with expert feature extraction (\rev{\textit{SVM}; support vector machine})~\cite{Abbaspour2020-ei,Iqbal2018-vz,Dehzangi2018-ic,Fan2018-ng}, and (2) deep learning \rev{(\textit{CNN+LSTM}; convolutional neural network with long short-term memory)}~\cite{Atzori2016-ka,Li2021-mr,Xie2018-gz,Karnam2022-wf}. 
%
As highlighted in previous sections, the two methods represent different expertise that is usually required for \rev{mid-air} gesture recognition using biosignals.
Expert feature extraction requires biosignal expertise, both the knowledge of which features to extract as well as how to implement the feature extraction in code. 
Some of the features can be complex and specific to the particular biosignal. 
Conversely, deep learning does not require biosignal expertise, as the deep learning algorithm will automatically learn the features that are most relevant to the recognition task. 
However, deep learning requires expertise in designing the model architecture, tuning hyperparameters, and writing and running the training code, which is a non-trivial task. 
Both methods require expertise in different ways that motivate the need for \alg{} to be implementable by non-experts. 

\subsubsection{Traditional ML \rev{(SVM)}.}

In traditional ML, features that are specific to specific biosignals and have previously been demonstrated to have discriminatory properties are extracted~\cite{Abbaspour2020-ei,Krasoulis2020-ix,Fan2018-ng,Li2023-ca}. 
Although there is no clear consensus on what expert features for EMG and IMU data are most suitable for HCI applications~\cite{Eddy2023-iz,Yamagami2023-xj,Fan2018-ng}), feature extraction for EMG and IMU data has been extensively studied for prosthetic control~\cite{Abbaspour2020-ei,Krasoulis2020-ix}.
Therefore, we chose to implement expert features that have been validated to have high recognition accuracy for prosthetic hand control. 
Based on results from~\citet{Abbaspour2020-ei}, we chose 10 time-domain EMG features.\footnote{ These features were mean absolute value, standard deviation, difference absolute mean value, integrated absolute value, variance, waveform length, correlation coefficient, Hjorth complexity parameter, and Hjorth mobility parameter~\cite{Abbaspour2020-ei}.} 
Before the features were extracted, the raw EMG signals were filtered with a fourth-order Butterworth bandpass filter with a cutoff frequency of 10 Hz and 500 Hz~\cite{Abbaspour2020-ei}. We additionally computed six IMU features that were previously demonstrated to improve recognition accuracy compared to using solely EMG expert features~\cite{Krasoulis2017-jb}.\footnote{These features were the mean values of three accelerometer and three gyroscope outputs~\cite{Krasoulis2017-jb}.} All features were computed over the entire time window in which the gesture occurred.
Therefore, each gesture was distilled into 232 features total (10 EMG features $\times$ 16 EMG sensors + 6 IMU features $\times$ 12 IMU sensors).
We used a multi-class support vector machine (SVM) with a linear kernel for recognition; 
\rev{the rest of the paper refers to this algorithm as \textit{SVM}}.
Hyperparameters were chosen based on performance during pilot testing.

\subsubsection{Deep Learning \rev{(CNN+LSTM)}.}

In \textit{deep learning}, nonlinear models are used for feature extraction and gesture recognition~\cite{Atzori2016-ka,Li2021-mr,Xu2022-dw}.
Deep learning has recently been successfully applied to both prosthetic~\cite{Atzori2016-ka,Li2021-mr} and HCI~\cite{Xu2022-dw} gesture recognition tasks. 
The input data can be either expert features~\cite{Li2021-mr} or the biosignal data itself~\cite{Atzori2016-ka,Xu2022-dw}.
Both simple (e.g., feedforward neural network~\cite{Hye2023-xa}) and complex model architectures have been used (e.g., convolutional neural network~\cite{Atzori2016-ka}, recurrent neural network~\cite{Xie2018-gz}, long-short term memory network~\cite{Xie2018-gz}).
Prior work has demonstrated that extracting relevant biosignal features with a convolutional neural network (CNN) and then passing those features to a recurrent neural network such as a long-short term memory network (LSTM) results in 
high accuracy classifiers \cite{Karnam2022-wf,Xie2018-gz}. 
Although we initially tested the three-layer one-dimensional CNN followed by a three-layer LSTM evaluated in~\citet{Xie2018-gz}, we found that shallower models resulted in better performance for the \textit{Variations} condition during pilot testing.
Therefore, we chose to implement a similar, but shallower deep learning model compared to~\citet{Xie2018-gz};
\rev{the rest of the paper refers to this algorithm as \textit{CNN+LSTM}}.

Our final neural network consisted of a single-layer one-dimensional CNN (kernel size: 2; stride size: 1) followed by a two-layer LSTM (hidden size: 64) and a feedforward neural network (hidden size: 64) for classification (10 for \textit{Personalized} and \textit{Variations}, 5 for \textit{Standardized}). 
Before inputting the biosignal data into the CNN, we preprocessed the data as described below and followed the first two steps of \alg{} (resampling and normalizing) for better comparison.
Therefore, the input to the CNN was the 88 normalized IMU and EMG channels resampled to 64 time points for a given gesture. 
The rectified linear unit (ReLU) was used as the nonlinear activation for all layers. 

We used the Adam optimizer~\cite{Kingma2014-wg} with a learning rate of 0.001 and a batch size of 10 to train our neural network for each of the 20 epochs. Hyperparameters were chosen based on performance during pilot testing.

\subsubsection{\alg{}.}
Our hyperparameters for \alg{} were time points per channel, $n=64$ points, and number of principal components, $nPC=50$, which were chosen based on initial hyperparameter exploration. 
\rev{Detailed analysis of the influence of the number of time points per channel and number of principal components on recognition error rate can be found in Section~\ref{sec:analysis}}.

\subsection{Data Preprocessing and Segmentation}

\subsubsection{Data Preprocessing.}

We preprocessed the biosignal data that was the inputted to the \alg{} recognizer and the \rev{CNN+LSTM} algorithm in the same way. For the IMU data, no pre-processing steps were applied as the data already had low noise and was collected at only 148 Hz. EMG data, on the other hand, was collected at 2000 Hz, and were stochastic and noisy; therefore, we computed the EMG linear envelope by filtering and downsampling. 
This pre-processing step provides a smooth trajectory that can be used as input to the algorithms. 
\revision{To aggressively filter the EMG signals and ensure less noise}~\cite{Yamagami2018-cg,Peters2018-sv}, we first filtered the EMG signal with a high-pass filter (fourth-order Butterworth filter at 40 Hz cutoff frequency), computed the absolute value of the EMG data, and then applied a low-pass filter (also a fourth-order Butterworth filter at 40 Hz cutoff frequency).\footnote{\revision{Note that because filters gradually roll off from the cutoff frequency, this filtering method still leaves the majority of the signal intact.}}
We then downsampled the EMG data from 2000 Hz to 20 Hz by computing the moving average with a 100 ms window and a 50 ms overlap between windows. Although this step was necessary for our research-grade EMG system, which records unprocessed EMG signals, many EMG systems that are commercially available have the option to pre-compute the linear envelope~\cite{SparkFun2023-vs,Brains2017-sc}.
Our downsampling step makes our results more comparable to other systems with lower sampling rates than our research-grade EMG system, highlighting the potential for translation to lower-cost systems.
The IMU and EMG data were then read-in to \alg{} and \rev{CNN+LSTM} algorithms. 
For the \rev{CNN+LSTM} algorithm, the same number of input points must be provided for a given gesture. Therefore, we also resampled each IMU and EMG channel to 64 time points per channel and stacked the signals into a single vector input of 5,632 points: 64 time points per channel $\times$ (16 EMG channels + 72 IMU channels).

\subsubsection{Custom Gesture Segmentation.}

Biosignal-agnostic gesture segmentation is outside the scope of \alg{} and is an important area of future work~\cite{Khanna2024-gk}. 
We implemented a custom gesture segmentation algorithm that automatically identified the beginning and end of a gesture using the raw EMG data (see Appendix~\ref{sec:appdx:segmentation} for pseudocode).
For each gesture, the experimenter prompted the participant to perform the gesture once data collection started; the experimenter stopped data collection after the gesture ended. This procedure resulted in signal padding at either end of the gesture. To remove this padding, we developed a pipeline that automatically identified where areas of activity (i.e., the gesture) started and ended. 
For each gesture, we identified a subset of EMG channels that had a high amplitude, as well as EMG channels that had high variance.
We chose to investigate the EMG channels rather than the IMU channels because prior work demonstrated that individuals with motor disabilities may flex their muscles without visibly moving~\cite{Yamagami2023-xj}.
This meant that IMU data that relies on the sensor physically moving in space to record signals may not pick up the participants' subtle movements.
With this subset of channels, we identified multiple candidate gesture start and end times through a combination of thresholding the signal based on the signal range~\cite{Staude2001-vw} and differentiating the signal to identify large changes in slope. We removed candidate cutoffs that contributed to high variance within and between muscles as well as cutoffs that occurred in the wrong area of the signal (i.e., start times that occurred towards the end of the signal and end times that occurred towards the beginning of the signal). 
From the remaining candidate cutoffs, the first start cutoff and the last end cutoff were selected to ensure that only the gesture data was used for subsequent data analysis. 
The pseudocode for the custom gesture segmentation algorithm can be found in Appendix~\ref{sec:appdx:segmentation}.

\subsection{Algorithm Evaluation and Statistical Analysis}

We adapted methods from~\citet{Vatavu2012-da} and \citet{Wobbrock2007-ul} to compute user-dependent and user-independent recognition error rates. 
User-dependent error rates are computed individually for each participant. 
User-independent error rates are computed over multiple pooled participants.
\rev{The computed recognition error rates were used to compare \alg{} to \rev{SVM} and \rev{CNN+LSTM}.}

\subsubsection{Evaluation Conditions.}

First, we used the \textit{personalized} and \textit{variations} conditions to compute user-dependent recognition error rates. In these conditions, participants came up with \rev{user-defined} gestures that were customized to their abilities.
Similar to prior evaluations~\cite{Vatavu2012-da,Wobbrock2007-ul}, for each gesture type captured in the \textit{personalized} condition, $T$ samples were randomly chosen as templates, and 1 sample was chosen randomly as the candidate gesture for recognition. The same template and candidate gestures were used for each of the three algorithms (\alg{}, \rev{SVM}, \rev{CNN+LSTM}). Additionally, 1 sample from each of the three \textit{variations} conditions (time, speed, size) was chosen randomly as a candidate gesture for recognition to test the effect of gesture articulation variability on recognition error rates. This process was repeated 100 times for each value of $T$, and the average accuracy was computed as the recognition error rate for each participant.  
We varied $T$ from 1 template to 9 templates. 

Second, we used the \textit{standardized} condition to compute user-independent recognition error rates. As we had a limited dataset and we were interested in investigating algorithm recognition error rate differences between participants with and without motor disabilities, we performed a leave-one-out analysis, where all but one participant's data was used to train the models. For each of the 5 gesture types, $T$ samples were randomly selected for each of the participants used for training.
Then, one sample 
was chosen randomly as a candidate gesture from the left-out ``test participant'' for recognition.
This process was repeated 100 times for each participant. The average accuracy was computed as the recognition error rate for each participant.
We varied $T$ among three levels: small ($T=1$), medium ($T=3$), and large ($T=7$) number of templates. 

\subsubsection{Statistical Analysis}

The primary reported outcome for all conditions was recognition error rate. 
The user-dependent data was generated from a mixed factorial design with the following factors and levels: 
\begin{itemize}
    \item \textit{Algorithm} (within-subjects): \alg{}, \rev{SVM}, \rev{CNN+LSTM};
    \item \textit{Templates} (within-subjects): integers in the range [1-9];
    \item \textit{Disability} (between-subjects): with, without.
\end{itemize}
When we examined the residuals from a $3\times 9 \times 2$ mixed factorial repeated measures ANOVA, we found them non-normal according to the Shapiro-Wilk test~\cite{Shapiro1965-gs}.
After fitting various distributions to our residuals, we found that a Gamma regression model was the best fit according to the Kolmogorov-Smirnov test~\cite{An1933-nb, Smirnov1939-sv}.
Therefore, we analyzed the user-dependent data with a mixed-effects Gamma regression model~\cite{Breslow1993-ji}.
Our random factor was our participant's ID to account for correlated responses from participants due to having within-subjects factors. 
Due to the presence of zeros in our dependent variable, recognition error rate, a constant shift of one (+1) was applied to the response~\cite{Osborne2002-xd}, which does not affect our analysis but enables Gamma models to be fitted with their canonical ``inverse'' link function.

The gesture articulation variability data was similarly generated from a mixed factorial design with the same conditions as the user-dependent study, above, and with one additional factor:
\begin{itemize}
    \item \textit{Gesture Articulation} (within-subjects): time, speed, size
\end{itemize}
Similar to the user-dependent data, we found that our residuals were not normally distributed but followed a Gamma distribution. Therefore, we also analyzed the gesture articulation variability data with a mixed-effects Gamma regression model.

Lastly, the user-independent data was similarly generated from a mixed factorial design with the same conditions as the user-dependent study, above, but with only three levels of templates tested instead of nine:
\begin{itemize}
    \item \textit{Templates} (within-subjects): 1, 3, 7 templates\footnote{\blue{All templates [1-9] were included for user dependent and gesture articulation variability and only three levels [1, 3, 7] were included for the user-independent study due to the order of magnitude increase in the training data in the user-independent study making it challenging to run the evaluations in a reasonable time frame.}}
\end{itemize}
We again analyzed the user-independent data with a mixed-effects Gamma regression model.

Unless specified, reported values are mean (standard deviation) across participants and templates. For all three gesture conditions (user-dependent, gesture articulation variability, user-independent), we performed an analysis of variance using a mixed-effects Gamma regression model, which is a type of generalized linear mixed model (GLMM)~\cite{Breslow1993-ji}.
Holm's sequential Bonferroni procedure~\cite{Holm1979-lh} was used to correct for multiple comparisons when performing \textit{post hoc} tests on significant main and interaction effects.
\rev{Cohen's $d$ was used to compute effect sizes for planned, statistically significant comparisons.}

To ensure that our algorithm runs at interactive speeds (i.e., below 500 ms~\cite{Anderson2011-xh, Ritter2015-vn}), we classified 100 gestures using \alg{} for one participant for the \textit{personalized} condition. We chose to test the algorithm with three and nine templates, as a larger numbers of templates means slower run times because each template must be compared against the candidate gesture. We found that the mean and standard deviation run-time for three and nine templates was 235 ms ($SD=25$) and  485 ms ($SD=38$), respectively, reasonable for interactive use.

%
The complete dataset and analysis code is publicly available at~\cite{Yamagami2024-qg}.

\section{\alg{} Evaluation: Results}
We present the results from our evaluation of \alg{}, including the results from its user-dependent and user-independent evaluations. 
In addition, we also used the models developed from the user-dependent data to test the robustness of the three algorithms to substantial variability in gesture articulation.



\subsection{User-Dependent Result: \alg{} Outperforms \rev{SVM} and \rev{CNN+LSTM}}

\begin{figure}[t]
  \centering
  \includegraphics[width=\linewidth]{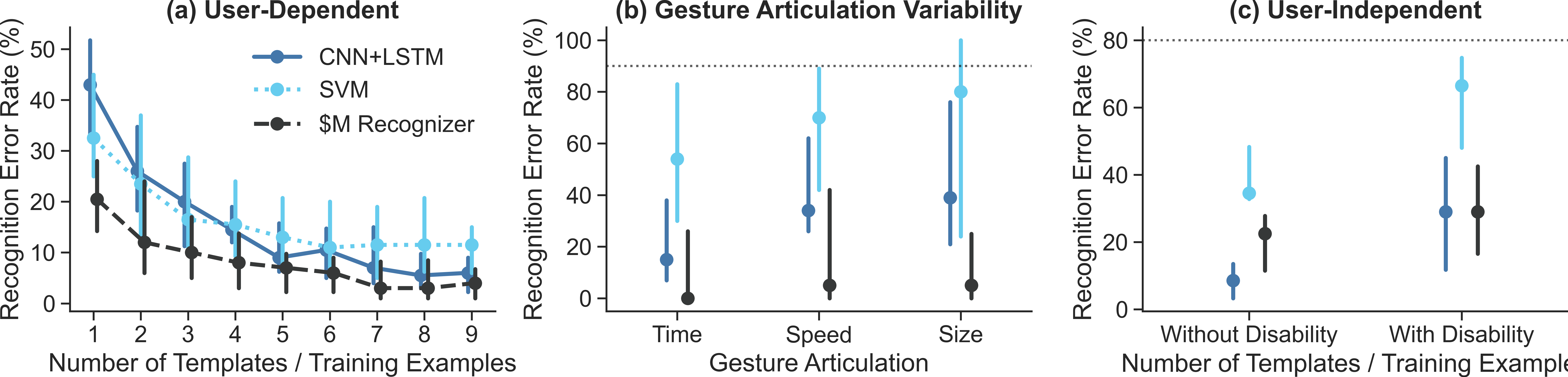}
  \caption{Recognition error rate comparison between \alg{} (black), \rev{SVM} (light blue), and \rev{CNN+LSTM} (dark blue). All Y-axes are recognition error rate, where lower values indicate greater accuracy. Medians are plotted with 25/75  percentile error bars. (\textbf{a}) The effect of \textit{Templates} on user-dependent recognition error rate. 
  As there are 10 possible gesture classes \revision{(i.e., function)}, the pure chance error rate is 90\% (not shown).
  (\textbf{b}) The effect of gesture articulation variability (by time, speed, size) on user-dependent models. 
  As there are 10 possible gesture classes, the pure chance error rate is 90\% (horizontal dotted gray line). Only the results from using the analysis with 7 templates are shown. (\textbf{c}) The effect of \textit{Disability} on user-independent recognition error rate.
  As there are 5 possible gesture classes, the chance error is 80\% (horizontal dotted gray line). Only the results from using the analysis with 7 templates are shown. 
  }\label{fig:main}
  \Description{
  Three figures row from left to right: (a) user-dependent, (b) gesture articulation variability, (c) user-independent. The y axis is recognition error rate as a percentage for all three graphs, the x axis varies. Each graph has three hues representing three different algorithms, \rev{CNN+LSTM} in dark blue, \rev{SVM} in light blue, and B dollar recognizer in grey. (a) user-dependent description: The x axis represents number of templates or training samples ranging from 1 to 9. B dollar recognizer has a median around 20 percent recognition error rate with 1 template and decreases to around 5 percent recognition error rate with 9 templates. \rev{SVM} and \rev{CNN+LSTM} perform similarly, with around 40 percent recognition error rate with 1 template and dropping to around 10 percent recognition error rate with 9 templates. The rate of improving of recognition error rate appears to be steeper for \rev{CNN+LSTM} and \rev{SVM} compared to B dollar recognizer. (b) gesture articulation variability description: The x axis represents the three different gesture articulation variability types that participants were asked to do -- time, speed, and size. For time, the median of B dollar, \rev{SVM} and \rev{CNN+LSTM} appear to be around 0, 60, and 35 percent, respectively. For speed, the median of B dollar, \rev{SVM}, and \rev{CNN+LSTM} appear to be around 5, 60, and 65 percent, respectively. For size, the median of B dollar, \rev{SVM}, and \rev{CNN+LSTM} appear to be around 5, 85, and 55 percent, respectively. (c) user-independent description: The x axis has two labels: without disability and with disability. For without disability, the median of B dollar, \rev{SVM}, and \rev{CNN+LSTM} appear to be around 30, 45, and 5 percent, respectively. For with disability, the median of B dollar, \rev{SVM}, and \rev{CNN+LSTM} appear to be around 35, 70, and 35 percent, respectively. 
  }
\end{figure}

The \alg{} recognizer outperformed the \rev{SVM} and \rev{CNN+LSTM} algorithms, with average recognition error rates of $9\%$ ($SD=9$), $18\%$ ($SD=13$), and $17\%$ ($SD=14$), respectively, averaged across all conditions
(\textit{Algorithm} main effect: $\chi^2_{2,N=918}=170.7, p<.0001$; \figref{fig:main}a).
\textit{Post hoc} pairwise comparisons, corrected with Holm's sequential Bonferroni procedure~\cite{Holm1979-lh}, indicated that all three algorithms had a significantly different recognition error rates ($p<.0001$);
\rev{\alg{} had lower recognition error rates than both \rev{SVM} ($p<.0001, d=1.29$) and \rev{CNN+LSTM} ($p<.0001, d=.93$).}
\textit{Templates} also had a significant main effect on recognition error rate for all tested algorithms ($\chi^2_{1,N=918}=882.4, p<.0001$)
While all three evaluated algorithms improve in recognition error rate as the number of training samples increase, the \alg{} recognizer performs well even with a few templates. 
The recognition error rate for \alg{}, \rev{SVM}, and \rev{CNN+LSTM} ranged from $22\%$ ($SD=11$), $35\%$ ($SD=16$), and $42\%$ ($SD=13$)
with one training sample, respectively, to 
$5\%$ ($SD=5$), $12\%$ ($SD=8$), and $7\%$ ($SD=6$)
for nine training examples, respectively.
\alg{}'s recognition error rate improves more modestly compared to \rev{SVM} and \rev{CNN+LSTM}, so there was an interaction effect between \textit{Algorithm} and \textit{Templates} ($\chi^2_{1,N=918}=69.1, p<.0001$).
There were no main or interaction effects with \textit{Disability}, suggesting that all three algorithms performed similarly across participants with and without disabilities 
($\chi^2_{1,N=918}=0.5, p=.48$).

\subsection{Gesture Articulation Variability Result: \alg{} Performs Robustly}

\alg{} performed robustly even when gesture articulation was varied across time, speed, and size\footnote{Recall that for the “time” variation, we asked participants
to make the \textit{rotate} gesture in the same way they did in the \textit{personalized} condition to test the effect of time between gestures. For the “speed” variation, we asked participants to perform the gesture as fast as possible. For the “size” variation, we asked participants to perform the gesture as large as possible.}
compared to \rev{SVM} and \rev{CNN+LSTM}, 
\rev{with average recognition error rates of 
24\% ($SD=32$), 
62\% ($SD=32$),
and 44\% ($SD=29$), respectively,
averaged across all conditions}
(\textit{Algorithm} main effect: $\chi^2_{2,N=2673}=192.7, p<.0001$; \figref{fig:main}b). 
\textit{Post hoc} pairwise comparisons \rev{indicated that all three algorithms had a significantly different recognition error rates ($p<.0001$);
\alg{} had a lower error rate than both \rev{SVM} ($p<.0001, d=1.29$) and \rev{CNN+LSTM} ($p<.0001, d=.70$).
}
%
There was also a significant main effect of \textit{Gesture Articulation} on recognition error rate 
($\chi^2_{2,N=2673}=15.7, p=.0004$).
The gestures articulated at a different time had a lower recognition error rate \rev{(35\%, $SD=32$)} than gestures that were articulated at a different speed \rev{(48\%, $SD=33$; $p<.0001, d=.35$)} or at a different size \rev{(47\%, $SD=36$; $p<.0001, d=.30$)}.

\rev{There was no interaction effect between \textit{Algorithm} and \textit{Gesture Articulation} ($\chi^2_{4,N=2673}=2.5, p=.65$).}
The average recognition error rate for \alg{}, \rev{SVM}, and \rev{CNN+LSTM} algorithms when the \textit{rotate} gesture was performed at a different time was 
$18\%$ ($SD=26$),  $55\%$ ($SD=32$), and $33\%$ ($SD=27$),
respectively. 
As fatigue~\cite{De_Luca1997-jw} and the length of time that a wearable sensor is placed on the user's body~\cite{Yamagami2018-cg} can affect the biosignal amplitude, this result suggests that \alg{} performs robustly despite these time-based biosignal fluctuations.
Similarly, \alg{}'s performance extrapolated well when \textit{rotate} was performed at a different speed and size that was not included in the training dataset 
(speed: $28\%$ ($SD=34$), size: $25\%$ ($SD=33$)) compared to the \rev{SVM} (speed: $64\%$ ($SD=30$), size: $67\%$ ($SD=33$)) and \rev{CNN+LSTM} (speed: $50\%$ ($SD=25$), size: $50\%$ ($SD=31$)). 
As the number of templates increased, the recognition error rate improved for all three algorithms 
($\chi^2_{1,N=2673}=55.1, p<.0001$).

There was a marginal effect of \textit{Disability} on the recognition error rate 
($\chi^2_{1, N=2673}=3.1, p=.08$), 
with the recognition error rate for participants without disabilities at 33\% ($SD=33$)
and participants with disablities at 
47\% ($SD=34$).
There was also an interaction between \textit{Algorithm} and \textit{Disability} ($\chi^2_{2,N=2673}=29.8, p<.0001$). 
\alg{} and \rev{CNN+LSTM} both had \rev{lower} recognition error rates for participants without disabilities than participants with disabilities \rev{(\alg{}: 23\% ($SD=11$) and 34\% ($SD=23$), respectively,
$p=.005, d=.78$; 
\rev{CNN+LSTM}: 12\% ($SD=9$) and 30\% ($SD=18$), respectively, $p=.02, d=.64$)}.
\rev{SVM} had no significant difference in error recognition rates \rev{between participants without and with disabilities (46\% ($SD=10$) and 64\% ($SD=15$), respectively, $p=1.0$)}.

\subsection{User-Independent Result: Both \alg{} and \rev{CNN+LSTM} Perform Well for Participants with Disabilities}

\rev{CNN+LSTM} outperformed both \alg{} and the \rev{SVM} algorithm, \rev{with average recognition error rates of $22\%$ ($SD=17$), $29\%$ ($SD=19$), and $56\%$ ($SD=16$), respectively, averaged across all conditions} 
(\textit{Algorithm} main effect: $\chi^2_{2,N=162}=449.0, p<.0001$; \figref{fig:main}c).
\textit{Post hoc} pairwise comparisons show all three algorithm performances were significantly different ($p<.0001$); 
\rev{\rev{CNN+LSTM} had lower recognition error rates than both \alg{} ($p<.0001, d=.70$) and \rev{SVM} ($p<.0001, d=2.52$)}. 
More templates resulted in an improved recognition error rate (\textit{Templates} main effect: $\chi^2_{2,N=162}=10.4, p=.005$). 
There was a marginal main effect with \textit{Disability} ($\chi^2_{1,N=162}=3.2, p=.07$).

There was a significant interaction between \textit{Algorithm} and \textit{Disability} 
($\chi^2_{2,N=162}=9.2, p=.01$). 
There was a marginal effect of the \rev{CNN+LSTM} algorithm performing better for participants without disabilities than participants with disabilities 
\rev{($12\%$ ($SD=9$) and $30\%$ $SD=18$, respectively, $p=.08, d=1.52$)}.
Although \rev{CNN+LSTM} outperformed \alg{} for participants without disabilities 
\rev{($12\%$ ($SD=9$) and $23\%$ ($SD=11$), respectively, $p<.0001, d=1.06$)},
there was no difference in recognition error rate between the two algorithms for participants with disabilities 
\rev{($30\%$ ($SD=18$) and $34\%$ ($SD=23$), respectively, $p=.2$)}.
There was also no significant difference in recognition error rate between participants with and without disabilities for \alg{} and \rev{SVM} ($p>.05$). 
This result shows that \alg{} performs robustly across both user groups, whereas \rev{CNN+LSTM} performs significantly better than \alg{} only for participants without disabilities.

\section{\rev{Analysis of \alg{} Components}}\label{sec:analysis}

\rev{
We performed a thorough analysis of \alg{} components, inputs, and parameters by ablating different steps of \alg{} and assessing the effects of sequential prompting and session, data input type, number of resampled points, and number of principal components on \alg{}'s recognition error rate.}

\subsection{Method}
All analyses of ML components were performed with seven templates/training samples.
\rev{Planned comparisons were evaluated using appropriate $t$-tests.
Cohen's $d$ was used to compute effect sizes for statistically significant comparisons.}

\subsubsection{Ablation.}
We performed an ablation study to understand the importance of each step of \alg{}.
We compared a baseline condition of no ablation (None) against ablating Step 2 (Normalize), Step 3 (PCA), and both Steps 2 and 3 in \alg{}.
We did not test the effects of Step 1 (resample) because the \$-family recognizers (including \alg{}) requires a consistent number of input samples for recognition. 
Paired $t$-tests between \alg{} and each ablation recognition error rates were used to evaluate significant differences.

\subsubsection{Gesture Sequence and Session.}
We tested the effect of sequential prompting and session on \alg{}'s recognition.
Three participants without motor disabilities (two men and one woman between the ages of 22 - 31) participated in a multi-session study.
They performed 10 standardized gestures 10 times for a total of 100 gestures per session.\footnote{In addition to the five gestures from the \textit{Standardized} condition, we also tested five more gestures currently integrated into the Apple Watch AssistiveTouch~\cite{Apple_Support_undated-gs}: (6) \textit{single clench}, (7) \textit{double pinch}, (8) \textit{single pinch}, (9) \textit{double clench}, (10) \textit{shake and release}.
Data collection failed for a single gesture for one participant's session so only 99 gestures were collected.}
Instead of prompting participants to perform the five gestures for a given function sequentially (i.e., performing \textit{air tap} gesture 10 times one after the other), we prompted them to perform them non-sequentially in a random order.\footnote{For example, participants were prompted to perform \textit{air tap}, \textit{palm pinch}, \textit{palm pinch}, \textit{air tap}, \textit{palm pinch}...rather than \textit{air tap}, \textit{air tap}, \textit{air tap}...}
Two participants participated in two sessions and one participant participated in three sessions.\footnote{\blue{
The intent was for all three participants to participate in three sessions but had participant dropout between session 2 and 3 with two participants. 
We chose to include all three sessions from the one participant who completed all three sessions to maximize the number of permutations that were available for analysis.
}}
All sessions were performed on separate days. 
To investigate the effect of performing movements sequentially, we computed user-dependent recognition error rates for each of the seven sessions across the three participants.\footnote{(two sessions $\times$ two participants + three sessions $\times$ one participant = seven sessions).}
To investigate the effect of session (including donning and doffing of the wearable sensors), for each of the \rev{three} participants, we trained \alg{} on one session and tested the trained model on the other for ten permutations.\footnote{($_{2 \text{ sessions}}P_{2} \times$ two participants + $_{3 \text{ sessions}}P_{2}\times$ one participants = ten permutations).}
\rev{
Independent $t$-tests for between user-dependent recognition error rates from \alg{}'s main evaluation (where gestures were performed in-sequence) and this study (where gestures were performed non-sequentially) were used to evaluate significant differences for sequential prompting.
Paired $t$-tests between recognition error rates across-session and within-session were used to evaluate significant differences for session.
}

\subsubsection{Input Type and Parameter Evaluation.}
\rev{We tested the effect of input data type and \alg{}'s two parameters (number of resampled points; number of principal components) on recognition error rate. 
For input data type, we tested three conditions: 1) IMU+EMG; 2) IMU only; 3) EMG only.
To ensure that the number of principal components are comparable between the 72 IMU channels and 16 EMG channels, we scaled the number of principal components to be approximately the same ratio of principal components kept for the combined 88 IMU and EMG channels from the original evaluation. 
Since we kept 50 components for the combined IMU+EMG condition, we kept 40 components for the IMU only condition and 10 components for the EMG only condition. 
For the number of resampled points, we tested three conditions: [32, 64, 128] resampled points.
For the PCA step (Step 3) in \alg{} to run properly, the number of principal components must be less than the number of biosignal channels and number of resampled time points. 
Therefore, we tested all conditions with 32 principal components to ensure that the number of principal components remained consistent throughout the evaluation.
For the number of principal components, we tested four conditions: [10, 30, 50, 64] principal components.
We tested all conditions with 64 resampled points to ensure that the number of resampled points remained consistent.
Paired $t$-tests compared the final chosen parameter (e.g., IMU+EMG) in \alg{}'s main evaluation against the other conditions (e.g., IMU only and EMG only).
}

\subsection{Results}

\subsubsection{Ablation Study: All Steps in \alg{} are Important for Recognition.}\label{sec:appdx:ablation}
\begin{figure}[t]
  \centering
  \includegraphics[width=.7\linewidth]{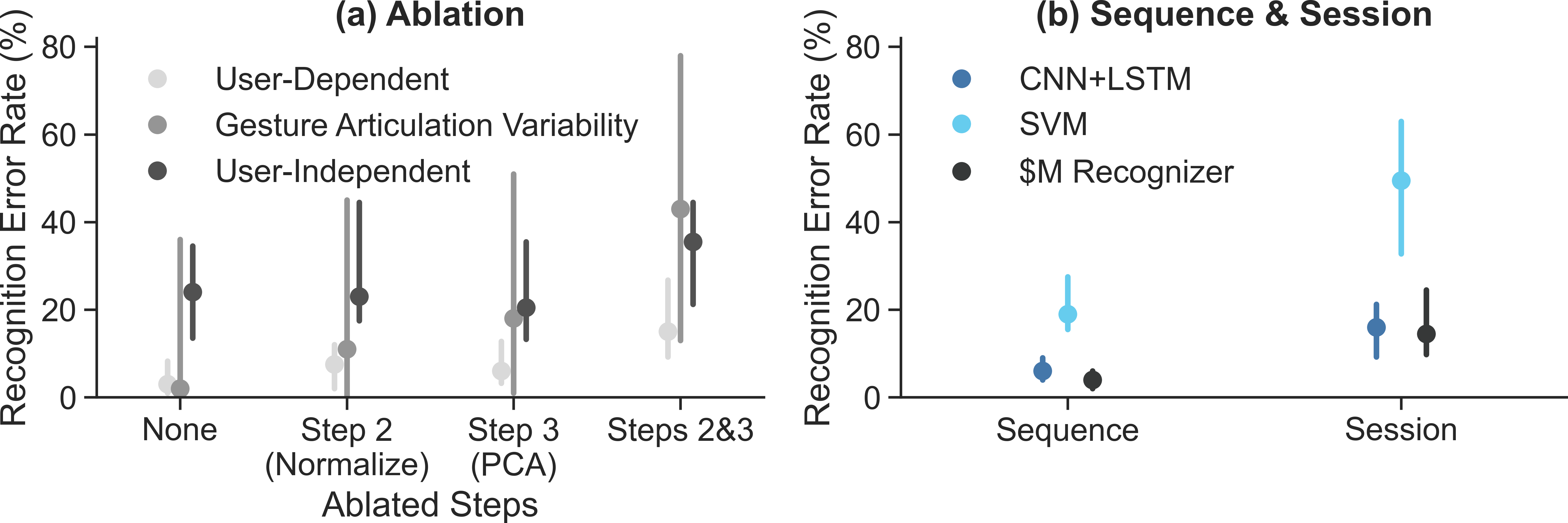}
  \caption{
  \revision{(a) The effects of ablating different steps of \alg{} on recognition error rate for user-dependent (light grey), gesture articulation variability (grey), and user-independent (dark grey).
  (b) The effect of gesture sequence and session on recognition error rate for \alg{} (black), \rev{SVM} (light blue), and \rev{CNN+LSTM} (dark blue). 
  Only the results from using the analysis with 7 templates are shown. 
  The Y-axes are recognition error rate, where lower values indicate greater accuracy. 
  As there are 10 possible gesture classes, the pure chance error is 90\%. 
  Medians are plotted with 25/75  percentile error bars.}
  }\label{fig:sequence}
  \Description{(a) Ablation results. When no steps are ablated (i.e., the results of the original \alg{}) user-dependent, gesture articulation variability, and user-independent recognition error rates range from 0\% to 38\%. When Step 2 (Normalize) is ablated, the upper-range of recognition error increases to 42\%. When Step 3 (PCA) is ablated, the upper-range of the error rate increases to 50\%. Lastly, when both steps are ablated, the error rates range from 10\% - 80\%.
  (b) Effect of gesture sequence and session on recognition error rate. For sequence, the error rates for \rev{CNN+LSTM} and \alg{} ranges from 2\%-10\%; for \rev{SVM}, the error rates range between 12\%-30\%. For session, \alg{} and \rev{CNN+LSTM} range from 10\%-25\%; for \rev{SVM}, the error ranges from 25\%-60\%. }
\end{figure}

Removing both Steps 2 (normalize) and 3 (PCA) increased the recognition error rate for user-dependent, gesture articulation, and user-independent by 13\% ($p<.0001, \rev{d=1.34}$), 25\% ($p<.0001, \rev{d=.65}$), and 8\% ($p=.03, \rev{d=.56}$), respectively (\figref{fig:sequence}a).
Removing solely Step 2 (normalize) increased the recognition error rate for user-dependent by 4\% ($p=.0002, \rev{d=.71}$).
Although the recognition error rate for gesture articulation and user independent increased by 4\% and 3\%, respectively, the results were not statistically significant ($p>.05$).
Removing solely Step 3 (PCA) increased the recognition error rate for user-dependent and gesture articulation by 5\% ($p<.0001, \rev{d=1.05}$) and 10\% ($p<.0001, \rev{d=.52}$), respectively. 
Although the recognition error rate for user-independent decreased by 3\% (i.e., the recognition accuracy improved), the results were not statistically significant ($p=0.3$). 
\blue{In conclusion, both Steps 2 (normalize) and 3 (PCA) perform important roles in improving recognition for user-dependent and gesture articulation.}

\subsubsection{Gesture Sequence and Session: \alg{} and \rev{CNN+LSTM} Perform Robustly.}

\rev{
Both \alg{} and \rev{CNN+LSTM} had comparable recognition error rates when gestures were performed non-sequentially and in-sequence (\alg{}: 5\% ($SD=5$) and 5\% ($SD=5$), respectively, $p=.88$; 
\rev{CNN+LSTM}: 8\% ($SD=6$) and 9\% ($SD=7$), respectively, 
$p=0.57$;
Fig.~\ref{fig:sequence}b).
\rev{SVM} had significantly worse recognition error rates when gestures were performed non-sequentially compared to when gestures were performed in-sequence (23\% ($SD=14$) and 13\% ($SD=9$), respectively; $p=.02, d=1.02$).
\alg{} and \rev{CNN+LSTM} had comparable recognition error rates when the algorithms were trained on one session and tested on another (17\% ($SD=11$) and 16\% ($SD=10$), respectively, $p=.77$).
However, all algorithms had lower recognition error rates across-session compared to within-session (\alg{}: 17\% ($SD=11$) and 5\% ($SD=5$), respectively, $p<.0001, d=1.63$; 
\rev{CNN+LSTM}: 16\% ($SD=10$) and 9\% ($SD=7$), respectively, 
$p=.02, d=.88$;
\rev{SVM}: 46\% ($SD=22$) and 13\% ($SD=9$), respectively, 
$p<.0001, d=2.49$).
}
\blue{In conclusion, both \alg{} and CNN+LSTM performed robustly when the gestures were performed out of sequence and when the movements were recognized across different sessions.}


\subsubsection{\rev{Input Type and Parameter Evaluation: IMU+EMG and Number of Principal Components Influence \alg{}'s Performance.}}

\begin{figure*}[t]
  \centering
  \includegraphics[width=.8\linewidth]{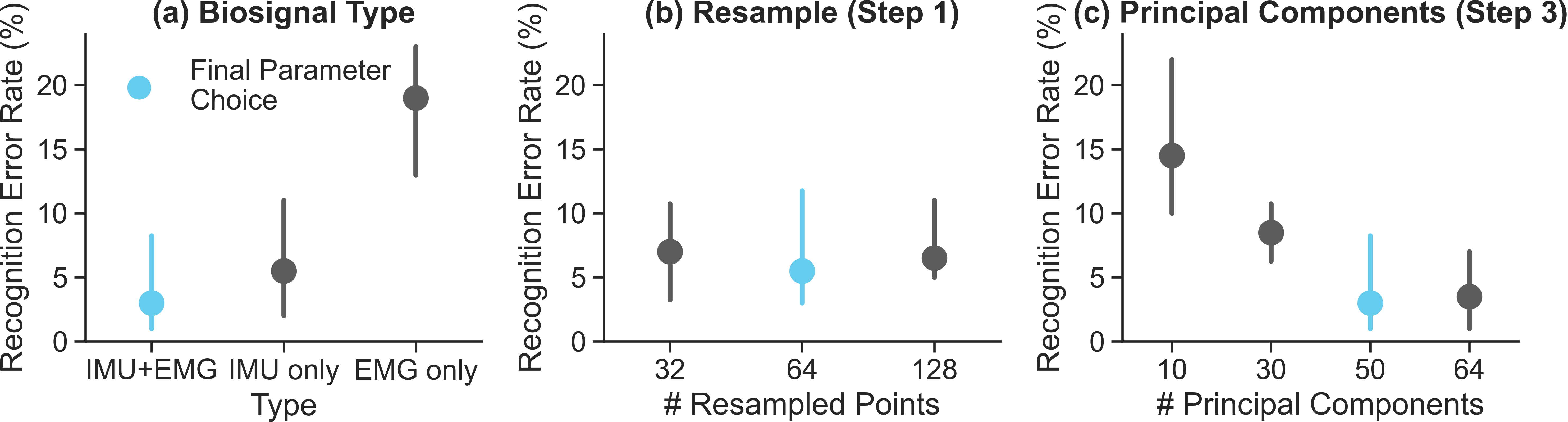}
  \caption{
  The effects of (a) data input type, (b) number of resampled points, and (c) number of principal components on recognition error rate.
  The light blue represents the parameter choice made for the \alg{} evaluation and comparison against \rev{SVM} and \rev{CNN+LSTM}.
  For all figures, only the results from using the analysis with 7 templates are shown. 
  The Y-axes are recognition error rate, where lower values indicate greater accuracy. 
  As there are 10 possible gesture classes, the pure chance error is 90\% (not shown). 
  Medians are plotted with 25/75  percentile error bars.
  }\label{fig:MLanalysis}
  \Description{(a) Biosignal type results. IMU+EMG recognition error rate ranges from 0-10\% (median 3\%), highlighted in light blue. IMU only ranges from 3-12\% (median 5\%). EMG only ranges from 12-23\% (median 20\%). 
  (b) Number of resampled points. All tested number of resampled points (32, 64, and 128) recognition error rate ranges from 3-12\% (median 6-7\%). The median recognition error rate of 64 resampled points highlighted in light blue appears slightly lower than others by 2\%. 
  (c) Number of principal components. 50 (highlighted in light blue) and 60 components recognition error rate ranges from 1-8\% (median 3\%, 50 components appear 1-2\% lower than 64 components). 
  10 principal components recognition error rate ranges from 10-23\% (median 15\%). 
  30 principal components recognition error rate ranges from 6-10\% (median 8\%). 
  }
\end{figure*}

\rev{Combined IMU+EMG outperformed both IMU only ($p=.0006, d=.65$) and EMG only ($p<.0001, d=1.51$) input types (Fig.~\ref{fig:MLanalysis}a).
Baseline IMU+EMG had a recognition error rate of 5\% ($SD=5$), whereas IMU only and EMG only had recognition error rates of 8\% ($SD=8$) and 20\% ($SD=12$), respectively.
}
\rev{The number of resampled points did not significantly affect recognition error rate ($p>.05$; Fig.~\ref{fig:MLanalysis}b).
Baseline 64 resampled points had a recognition error rate of 8\% ($SD=7$), whereas 32 and 128 resampled points had recognition error rates of 8\% ($SD=6$) and 9\% ($SD=6$), respectively.}
\rev{More retained principal components resulted in better performance (Fig.~\ref{fig:MLanalysis}c), with 50 principal components having a lower recognition error rate than 10 components ($p<.0001, d=1.12$) and 30 components ($p<.0001, d=1.27$).
Baseline 50 principal components had a recognition error rate of 5\% ($SD=5$), whereas 10, 30, and 64 principal components had recognition error rates of 18\% ($SD=13$), 9\% ($SD=6$), and 5\% ($SD=6$), respectively.}
\blue{In conclusion, combined IMU and EMG data and a larger number of principal components improved gesture recognition, but a larger number of resampled points did not.}

\section{Code Usability Study}

We investigated the ease-of-implementation of \alg{} through a code usability study with six participants.
\rev{
To investigate the challenges of implementing the model architectures and hyperparameter tuning, we provided participants with detailed coding instructions and template code that already included data loading (See Supplementary Material).
}

\subsection{Method}

\subsubsection{Participants.}
We asked six students between the ages of 18 - 25 with some experience in coding and Python and minimal experience in machine learning (one semester course or less) to implement \alg{} and the \rev{CNN+LSTM} algorithms using Python3.
Three women and three men with 2-6 years of coding experience participated.
All were STEM majors, two were graduate students, and four were freshmen and sophomore undergraduate students.
\rev{
Participants had a median Python/coding experience of 2.75 (range: 2-3.5), where 1 corresponded to one semester of coding experience, 3 corresponded to having completed some coding projects, where students had previously used packages like scikit but not deep learning libraries, and 5 corresponded to expert, where students had familiarities with all major Python libraries.
Three out of six participants had previously used deep learning libraries (e.g., Pytorch) in the past.
Participants had a median level of expertise in machine learning of 1.6 (range: 1-2), where 1 corresponded to novice, 3 corresponded to having previously taken a machine learning class, and 5 corresponded to expert or having taken multiple machine learning classes.
The students who rated themselves a 2 mentioned having previously taken scientific computing classes that touched on similar topics or having watched educational videos on machine learning in the past. 
}

\subsubsection{Task.}
Each participant was assigned to complete \alg{} and \rev{CNN+LSTM} conditions in a random order.
\rev{
\rev{SVM} was not implemented for this study because we wanted to focus on ease-of-implementation of the code. 
The primary challenge of \rev{SVM} is having expertise in what features to extract for the specific input biosignal of interest, rather than expertise in coding, which falls outside the scope of our study.
}
Participants were provided with an instruction sheet and template code, and asked to complete the code based on the provided instructions on a Google Colaboratory Notebook 
\rev{(see Supplementary Material for instruction sheet and template code).
Participants also received mid-air gesture data from our primary \alg{} evaluation from one individual.
The data included five gesture classes repeated three times, for a total of 15 gesture files.
The files were already split into test/train splits, with two repetitions per gesture class for training (10 gesture files total) and one repetition per gesture class for testing (5 gesture files total).  
}

\rev{
The instruction sheet provided an overview of upper-body gesture recognition, data overview including sensor type, gestures that participants completed, data formatting, and pre-defined test/train splits.
We did not provide the pseudocode in Appendix~\ref{sec:appdx:M} for \alg{}.
Instead, we provided an overview of the model architectures of \alg{} and \rev{CNN+LSTM}, hyperparameters to tune, and links to relevant Python functions (e.g., \texttt{torch.nn.Conv1d()}; \texttt{sklearn.decomposition.PCA()}).
Participants were asked to continue hyperparameter tuning and debugging until average testing accuracy was above 60\%.
Participants were asked to directly edit the coding template provided on Google Colaboratory so we could observe the coding history.
The coding template already included reading in the data and primarily focused on implementing each model's architecture and computing the classification accuracy. 
Unit tests were included for each coding condition to ensure that participants were implementing each portion of the model correctly. 
}

\rev{Participants were instructed to use a time tracking application to keep track of how many minutes they spent on each condition.}
Each participant recorded the number of minutes it took for them to implement the code and filled out the NASA TLX~\cite{Hart1988-tz} scores that assessed work load\footnote{NASA TLX ranges from [0-100], with higher scores indicating a higher work load} 
after the completion of each condition. 
We conducted a semi-structured interview after both \alg{} and \rev{CNN+LSTM} coding conditions were completed
\rev{
(See Supplementary Material for interview questions).
The interview covered questions on the challenges the participants experienced implementing each algorithm, which algorithm they would prefer to use in a gesture recognizer prototype, and whether and how they used AI to help with the task. 
We also asked basic demographic questions and Likert-Scale questions on their prior coding and machine learning experience. 
}

\subsubsection{Analysis.}
\rev{The median and inter-quartile range of implementation time and NASA TLX scores for each condition were reported for each participant.}
The semi-structured interviews were transcribed and coded by the first author \rev{using reflexive thematic analysis~\cite{Braun2019-ks, Braun2006-kf}.
Representative quotes were extracted for reporting.
}

\subsection{Results}
The median implementation time for \alg{} and \rev{CNN+LSTM} were 117 (25th and 75th inter-quartile range IQR: 84, 168) and 229 (IQR: 199, 266) minutes, respectively, with 
\alg{} taking 53\% less time to implement than \rev{CNN+LSTM} (IQR: 26\%, 63\%). 
Median NASA TLX for \alg{} and \rev{CNN+LSTM} were 48 (IQR: 37, 53) and 55 (IQR: 52, 59), respectively, 
with \alg{} taking 23\% less workload to implement than \rev{CNN+LSTM} (IQR: 9, 31). 
Participants found \alg{} \textit{``less frustrating''} and \textit{``easier''} to implement, especially since they had to \textit{``look more into the documentation''} for the \rev{CNN+LSTM} algorithm. 

All participants agreed that \alg{} was easier to implement than \rev{CNN+LSTM} because 
\alg{} \textit{``seemed more of a simple solution''} with \textit{``less function calls''}.
In contrast, they felt that \rev{CNN+LSTM} was \textit{``hard to debug...I was confused as to why I was having super low accuracy...[and didn't] know exactly what to change [to get higher accuracy]''}.  
Five out of six participants preferred to implement \alg{} over \rev{CNN+LSTM} if they were to prototype a \rev{mid-air} gesture recognizer because
\textit{``it felt more intuitive''} and \textit{``I only needed to tune one hyperparameter''}.
In contrast, \rev{CNN+LSTM} felt \textit{``very guess-and-check to me''} because \textit{``there's a lot of hyperparameters you need to go through...kernel, hidden size, dropouts, batch size...so it's more complex''}.
One participant preferred \rev{CNN+LSTM} because \textit{``I've learned the concepts for deep learning before [through online videos] so I understood it a little bit better''}. 

Five out of the six participants used generative AI tools like Gemini and ChatGPT to complete the coding tasks. 
Three participants used these tools to:
\textit{``explain the concept''} and 
\textit{``learned a lot about the deep learning and the architecture of the model from AI. It's just like a teacher explaining the concepts to me''.}
One mentioned that 
\textit{``I did try to use AI in a documentation lookup context...but it wasn't actually very useful, and then I ended up going back to the documentation''} that was already provided from SciPy.
One participant used it to \textit{``write the boring for loop stuff''} and \textit{``debug...to adjust the accuracy...[and] tune the model''.}
One participant only used the autocomplete feature in Google Colab. 

\section{Discussion}

The \alg{} recognizer performs well for \rev{user-defined mid-air} gestures \rev{performed by participants with and without motor disabilities}, with over 90\% accuracy and is simple to implement for college-age students with some experience in coding.
\blue{Each step of \alg{} is critical for accurate recognition, and the two parameters (number of resampled points and principal components) can be easily tuned.}
In comparison, other algorithms typically deployed for \rev{mid-air} biosignal gesture recognition, traditional machine learning using expert feature extraction (\rev{SVM}) and deep learning (\rev{CNN+LSTM}), had 82\% and 
83\% 
accuracy, respectively. 
Our results highlight \alg{}'s success in accurately recognizing \rev{mid-air} gestures despite a small number of templates and variability in gesture articulation, especially for participants with motor disabilities.
\blue{Below, we expand on \alg{}'s recognition rate in comparison to state-of-the-art gesture recognizers, 
highlight its usefulness for small mid-air gesture datasets with large heterogeneity, and discuss the benefits of \alg{} in the age of AI.
}

\subsection{\alg{}'s Recognition Rate is Comparable to State-of-the-Art}

\alg{}'s accuracy is comparable to other work that recognizes \rev{mid-air} biosignal gestures using user-dependent models.
For example, EMG gesture recognizers from~\citet{Ctrl-Labs_at_Reality_Labs2024-zb} had a recognition accuracy between 49\% and 73\% across 100 participants without disabilities and nine gesture classes (with 11\% chance accuracy), and \citet{Krasoulis2020-ix} had a recognition accuracy between 70\% and 95\% across 12 participants without disabilities and 2 trans-radial amputees and five gesture classes (with 20\% chance accuracy). 
Comparing our results against prior work that employed few-shot learning to personalize gestures for participants without disabilities, our work performs comparably.
For example, \citet{Xu2022-dw} found a 55\%, 83\%, and 87\% gesture recognition accuracy with one, three, and five templates across 12 new gesture classes (with 8.3\% chance accuracy). \citet{Shahi2024-bb} found 95\% gesture recognition accuracy with four templates across four new gesture classes (with 25\% chance accuracy). 
In comparison, \alg{} had a comparable gesture recognition accuracy of 
91\% and 93\%
with four and five templates across 10 gesture classes (with 10\% chance accuracy), without the need for a large dataset or state-of-the-art machine learning methods.

Similarly, \alg{} is robust to variations in gesture articulation variability, whether that be due to changes in the biosignal due to time, speed, or size, with an average accuracy of 
76\% compared to 38\% and 56\% 
for the \rev{SVM} and \rev{CNN+LSTM} algorithms, respectively.
An interesting finding was that both \rev{SVM} and \rev{CNN+LSTM} algorithms performed poorly when the same gesture was performed at a different time. 
This may be because participants became fatigued throughout the study, and by the time they repeated the gesture, they performed the gesture differently (e.g., slower, smaller, or with less vigor).
Another reason may be due to changes in the EMG signal itself. Prior work has shown that EMG impedance decreases over several hours after the sensor is placed~\cite{Yamagami2018-cg}, and EMG signals also change with fatigue~\cite{De_Luca1997-jw}. Therefore, algorithms must be resilient to time-based changes in biosignal variability.
In our work, we ensured that time-based fluctuations in biosignals do not affect gesture recognition by normalizing the biosignals so that the total standard deviation across all channels for a given biosignal is equal to one \revision{(i.e., different biosignals such as EMG and IMU are normalized separately)}~\cite{Hye2023-xa}. 

\subsection{\rev{\alg{} Works Well With a Small Number of Training Samples and Heterogeneous Movements}}
For our user-dependent evaluation, the largest dataset the algorithms had access to was 9 templates $\times$ 10 gestures = 90 \rev{training samples}. 
For our user-independent evaluation, the largest dataset the algorithms had access to was 7 templates $\times$ 5 gestures $\times$ 17 participants = 595 \rev{training samples}.
The order of magnitude increase in dataset size appeared to significantly improve the gesture recognition accuracy for \rev{CNN+LSTM} for participants without disabilities.
However, we did not see a similar improvement in performance for participants with motor disabilities. Both \alg{} and \rev{CNN+LSTM} performed comparably, with 
66\% and 70\%
recognition accuracy, respectively. 
This result suggests that for users with larger gesture articulation heterogeneity (such as for users with motor disabilities), \alg{} will continue to perform comparably to other recognizers, as it is more robust to gesture articulation variability. 


Qualitatively, we observed differences in how the standardized gestures were articulated between participants with and without disabilities.
First, all participants without disabilities performed their gestures mid-air at chest height, while participants with motor disabilities tended to perform their gestures at waist level, with their hands resting on or barely above their armchair. Second, all participants without disabilities performed the standardized gestures exactly as demonstrated, whereas participants with motor disabilities could choose whichever hand they preferred to do one-handed gestures. 
This meant that participants with disabilities who performed the standardized gesture with the opposite hand than what was demonstrated had fewer training examples. Third, participants with disabilities tended to perform the standardized gestures more slowly and smaller than participants without disabilities. 

\subsection{Benefits of \alg{} in the Age of AI and Practical Applications}

Our code usability study highlights the need for simple algorithms in the age of generative AI. 
No instructions were provided on the use of AI tools for the coding implementation and five out of six participants used AI tools to complete the implementations.
Despite the availability of AI tools, \alg{} still took significantly less time to implement than \rev{CNN+LSTM} for all participants.
Most participants did not find the generative AI outputs particularly useful, with some defaulting to online documentation rather than AI output for coding help. 
Even with AI tools, there is still a need for simple algorithms that prototypers inexperienced with machine learning can use for gesture recognition.

Another potential opportunity for generative AI tools is the ability to quickly convert pseudocode or algorithms implemented in one coding language into another coding language. 
Because \alg{} is a simple algorithm with few lines of code and can be implemented without custom libraries or packages, generative AI tools could take in the pseudocode provided in Appendix~\ref{sec:appdx:M} and generate code in the appropriate coding language. 
For example, Gemini Pro accurately converted \alg{} into Python3 code when prompted with the pseudocode and ``Can you convert this pseudocode into Python?''\footnote{See supplementary materials for generated code and unit testing.}
Because \alg{} only requires fine-tuning of two hyperparameters (number of resampled time points and number of principal components to keep), anyone unfamiliar with gesture recognition can ask a generative AI tool to generate \alg{} code for their coding language of choice, fine-tune two hyperparameters, and continue HCI prototyping. 

\rev{
Why might an application developer want to avoid simply prompting generative AI tools to ``create an accessible gesture recognizer''?
A consideration is the potential for generative AI to reinforce stereotypes such as those surrounding ability~\cite{Kosa2026-av, Glazko2024-gh, Glazko2023-ha}.
For example, \citet{Kosa2026-av} found that when co-making accessible technologies with AI and individuals with visual impairments, AI oversimplified the challenges experienced by individuals with visual impairments, requiring the individual to provide further refinement and AI prompting for AI to fully understand the problem space.
Individuals who are non-experts in accessibility would likely not be able to provide the type of prompt refinement needed to create a fully accessible mid-air gesture recognizer. 
This highlights the need for a simple mid-air gesture recognizer for user-defined gestures, especially for accessibility applications.
}


Next, we briefly discuss potential applications and implications of \alg{}, especially as it pertains to individuals with motor disabilities.
\alg{} is a gesture recognizer specifically designed to support \rev{user-defined mid-air} gesture input for people with and without disabilities. 
Our diverse evaluation dataset highlights the potential for \alg{} to support accessibility for individuals with many different disabilities.
We highlight example use cases in Section~\ref{sec:B:usecase}, including enabling accessible gesture input for current (e.g., smartphone, laptop) and futuristic (e.g., virtual and augmented reality) applications.
For example, current virtual and augmented reality gesture inputs require fine hand and finger motor control, and are not accessible for individuals with motor disabilities~\cite{Mott2020-io,Yamagami2023-xj,Yamagami2022-jl}.
Using \alg{}, students or developers looking to make accessible virtual or augmented reality applications could integrate personalized gesture recognition using kinematic data collected from an extended reality headset or smartphone camera. 
Instead of having a ``Select'' function be mapped to an inaccessible finger tap, ``Select'' could be remapped to a head nod.
Further work is needed to understand what biosignal data quality is needed for accurate recognition with \alg{}. 

\subsection{Limitations and Future Work}

One limitation of \alg{} is classifying gestures across participants without disabilities; \rev{CNN+LSTM} outperformed \alg{} by 
11\%
in the user-independent evaluation. Similarly, comparing \alg{}'s user-independent recognition accuracy with other work that generalized models across participants, \alg{}'s overall gesture recognition accuracy of 
71\%
across five gesture classes (with 25\% chance accuracy) was slightly lower than other work. For example, \citet{Xu2022-dw} had a 96\% accuracy across four gesture classes and over 500 participants. ~\citet{Ctrl-Labs_at_Reality_Labs2024-zb} had a 93\% accuracy across seven gesture classes and over 6,500 participants.
These results highlight the strength of deep learning with much larger datasets~\cite{Ctrl-Labs_at_Reality_Labs2024-zb} 
\rev{and highlights the strength of \alg{} as primarily a mid-air recognizer for user-defined gestures.}

Another limitation with \alg{} is that recognition speed decreases as more templates are added, as each template will have to be compared against the candidate gesture. We do not anticipate this to be an issue for \rev{user-defined mid-air} gesture interfaces as users will save a small number of templates, but it could be a challenge for user-independent applications, where there may be many templates saved from different users.
Future work should consider additions to \alg{} that speed up the recognition, similar to how \$Q~\cite{Vatavu2018-lb} significantly improved computing resources and time compared to \$1~\cite{Wobbrock2007-ul} and \$P~\cite{Vatavu2012-da}.
Next, compared to prior \$-family recognizers, \alg{} achieved a lower gesture recognition accuracy overall. The original \$1 unistroke recognizer had over 99\% accuracy, and the \$P multistroke recognizer had over 95\% accuracy.
In comparison, multiple studies have demonstrated that \rev{mid-air} biosignal gesture recognition ranges from 50\%~\cite{Ctrl-Labs_at_Reality_Labs2024-zb} to 95\%~\cite{Krasoulis2020-ix} for user-dependent algorithms. These recognition rates are despite years of sustained work on developing better expert features~\cite{Abbaspour2020-ei} and deep learning architectures~\cite{Ctrl-Labs_at_Reality_Labs2024-zb, Triwiyanto2023-ub}. In our evaluation, \alg{} had a recognition accuracy of 95\% with only nine templates, on the upper-end of this scale.
This gap in accuracy highlights the challenge of recognizing \rev{mid-air} biosignal gestures compared to 2-D uni- or multistroke gestures. 
To close this gap, new innovations must be made in \rev{mid-air} biosignal gesture recognition for HCI applications.

\revision{
Lastly, a limitation with our evaluation method is that our current sensor array (16 combined IMU and EMG sensors distributed across the upper-body) will not translate well in real-world settings due to the time taken to place the sensors and the current hardware cost.
Despite these limitations, we chose this specific setup to envision what accessible futures might look like if we had access to a distributed wearable IMU and EMG system. 
This setup enabled us to collect data and recognize movements from participants with significant motor disabilities who could not produce the changes in pose (e.g., move hand from left to right) that is needed for a camera system or smartwatch system. 
Conversely, EMG data can be captured when a person attempts to move their arm even if they are unsuccessful, and can sense subtle movements like squeezing a fist that would not be possible with other biosignals like IMUs or kinematic data from cameras.
Although our work's contribution is \alg{}, rather than an entire wearable system that is ready to be deployed into the real world, it is worth considering how such a system might be designed to be deployed in the future.
For example, we could envision a smart textile garment with electronics embedded in them.
Future work could look at new advances in optimizing sensor location for gesture recognition (e.g., if the user only uses their right hand, they do not need sensors on their left hand) or how such a garment could be personalized and fabricated automatically.
}


\section{Towards a New Paradigm for \rev{Mid-Air} Gesture Recognition in HCI}

As highlighted in~\citet{Eddy2023-iz}, biosignal interfaces like electromyography, accelerometers, and depth camera interfaces are rapidly being applied for HCI applications both for users with~\cite{Yamagami2023-xj,Mitchell2022-tm,Franz2023-mn} and without~\cite{Ctrl-Labs_at_Reality_Labs2024-zb, Xu2022-dw,Shahi2024-bb} disabilities.
As many of these biosignals have been previously studied in-depth in areas like assistive technology~\cite{Li2021-mr,Abbaspour2020-ei,Iqbal2018-vz,Atzori2016-ka,Triwiyanto2023-ub} and rehabilitation~\cite{Ho2011-rv,Su2023-sy,Dehzangi2018-ic}, HCI biosignals research have tended to replicate the methods used in these fields~\cite{Eddy2023-iz}.
For example, uninterrupted gesture recognition using a few hundred-millisecond windows is a common method in assistive technology for gesture classification~\cite{Li2021-mr,Abbaspour2020-ei,Iqbal2018-vz,Atzori2016-ka,Triwiyanto2023-ub}. 
This is because in prosthetics research, for example, gestures must be made continuously recognized to determine what position the prosthetic hand should be in at any given moment (e.g., open hand, close hand, flex hand, extend hand)~\cite{Abbaspour2020-ei}. 
However, uninterrupted gesture recognition may not be desirable in HCI applications because technology interactions are often intermittent.
Additionally, recognizing gestures over sliding time windows makes it more challenging to identify gestures that are performed at different speeds, such as when users with motor disabilities perform gestures slower than users without disabilities. 
To account for these different needs in HCI compared to body-machine interface research, we chose to recognize the entire gesture trajectory rather than continuously recognizing sliding time windows gestures. 

Recognizing the entire gesture trajectory instead of sliding time windows poses new research questions for biosignal interfaces that were not considered in this work. 
First, what biosignal features should be used to identify the beginning and end of a gesture and ensure that naturalistic motions such as walking or cooking do not trigger a gesture classification? 
With a 2-D stroke interface, the beginning and end of a gesture are defined as when the person's pen or finger comes into physical contact with the screen.
With mid-air gestures, there are no clear ways to determine when a person begins and ends a gesture.
This highlights the need for new algorithms that can detect when people intend to interact with their technology~\cite{Schwarz2014-pi,Khanna2024-gk}. 

Similarly, extracting expert features from biosignals data (e.g., mean absolute value, average signal power) is a powerful way to ensure that gesture classifiers are robust to natural fluctuations in biosignals~\cite{Li2021-mr,Abbaspour2020-ei}.
However, expert features that work well for assistive technology applications may not work well for HCI applications, especially if the time window over when the features are computed is over the entire gesture window rather than every few hundred milliseconds.
Future work should consider what features should be extracted for biosignal interfaces in HCI, and our dataset provides a starting point for developing IMU and EMG features that are ability-inclusive.
Our work builds on and emphasizes the call to action originally put forth by~\citet{Eddy2023-iz} that new advances in sensors and algorithms are needed to fully integrate biosignals for HCI applications.

\section{Conclusion}

The \alg{} recognizer is a simple and effective algorithm that recognizes \rev{user-defined mid-air} gestures performed by participants with and without motor disabilities, with a recognition accuracy of over 95\% with 9 templates.
It performs well with only a few training samples (78\%
accuracy with 1 template) and recognizes gestures that are articulated in different ways or are articulated by different people.
The accuracy is comparable to other work that classifies biosignal gestures and outperformed algorithms that require domain expertise in biosignals (traditional machine learning using expert feature extraction) or algorithm development (deep learning).
All participants with minimal machine learning experience found \alg{} easier to implement than deep learning gesture recognition algorithms.
Similarly to how the \$-family recognizers transformed 2-D stroke gesture recognition, we hope that \alg{} makes it easier to integrate \rev{mid-air} biosignal gesture recognition into next-generation user interface prototypes. 

\begin{acks}
Huge thanks to our participants for taking their time to participate in our study. 
This work was funded by an unrestricted gift from Meta, 
the National Institute on Disability, Independent Living, and Rehabilitation Research (NIDILRR) ARRT Training grant number 90ARCP0005, 
the National Science Foundation Graduate Research Fellowship Program (NSF GRFP) grant DGE-2140004, 
and the University of Washington Center for Research and Education on Accessible Technology and Experiences (CREATE).
NIDILRR is a Center within the Administration for Community Living (ACL), Department of Health and Human Services (HHS). 
The contents of this paper do not necessarily represent the policy of NIDILRR, ACL, HHS, and you should not assume endorsement by the Federal Government.

\end{acks}

\bibliographystyle{ACM-Reference-Format}
\bibliography{paperpile,ClaireBib}


\begin{thebibliography}{108}


\ifx \showCODEN    \undefined \def \showCODEN     #1{\unskip}     \fi
\ifx \showDOI      \undefined \def \showDOI       #1{#1}\fi
\ifx \showISBNx    \undefined \def \showISBNx     #1{\unskip}     \fi
\ifx \showISBNxiii \undefined \def \showISBNxiii  #1{\unskip}     \fi
\ifx \showISSN     \undefined \def \showISSN      #1{\unskip}     \fi
\ifx \showLCCN     \undefined \def \showLCCN      #1{\unskip}     \fi
\ifx \shownote     \undefined \def \shownote      #1{#1}          \fi
\ifx \showarticletitle \undefined \def \showarticletitle #1{#1}   \fi
\ifx \showURL      \undefined \def \showURL       {\relax}        \fi
\providecommand\bibfield[2]{#2}
\providecommand\bibinfo[2]{#2}
\providecommand\natexlab[1]{#1}
\providecommand\showeprint[2][]{arXiv:#2}

\bibitem[Goo(2016)]%
        {Goodfellow2016-xz}
 \bibinfo{year}{2016}\natexlab{}.
\newblock \showarticletitle{Chapter 14: Autoencoders}.
\newblock In \bibinfo{booktitle}{\emph{Deep Learning}}, \bibfield{editor}{\bibinfo{person}{Ian Goodfellow}, \bibinfo{person}{Yoshua Bengio}, {and} \bibinfo{person}{Aaron Courville}} (Eds.). \bibinfo{publisher}{MIT Press}, \bibinfo{pages}{499--523}.
\newblock


\bibitem[Abbaspour et~al\mbox{.}(2020)]%
        {Abbaspour2020-ei}
\bibfield{author}{\bibinfo{person}{Sara Abbaspour}, \bibinfo{person}{Maria Lindén}, \bibinfo{person}{Hamid Gholamhosseini}, \bibinfo{person}{Autumn Naber}, {and} \bibinfo{person}{Max Ortiz-Catalan}.} \bibinfo{year}{2020}\natexlab{}.
\newblock \showarticletitle{Evaluation of surface {EMG}-based recognition algorithms for decoding hand movements}.
\newblock \bibinfo{journal}{\emph{Med. Biol. Eng. Comput.}} \bibinfo{volume}{58}, \bibinfo{number}{1} (\bibinfo{date}{Jan.} \bibinfo{year}{2020}), \bibinfo{pages}{83--100}.
\newblock


\bibitem[Albertini et~al\mbox{.}(2017)]%
        {Albertini2017-fz}
\bibfield{author}{\bibinfo{person}{Niccolò Albertini}, \bibinfo{person}{Andrea Brogni}, \bibinfo{person}{Riccardo Olivito}, \bibinfo{person}{Emanuele Taccola}, \bibinfo{person}{Baptiste Caramiaux}, {and} \bibinfo{person}{Marco Gillies}.} \bibinfo{year}{2017}\natexlab{}.
\newblock \showarticletitle{Designing natural gesture interaction for archaeological data in immersive environments}.
\newblock \bibinfo{journal}{\emph{Virtual Archaeol. Rev.}} \bibinfo{volume}{8}, \bibinfo{number}{16} (\bibinfo{date}{May} \bibinfo{year}{2017}), \bibinfo{pages}{12}.
\newblock


\bibitem[An(1933)]%
        {An1933-nb}
\bibfield{author}{\bibinfo{person}{K An}.} \bibinfo{year}{1933}\natexlab{}.
\newblock \showarticletitle{Sulla determinazione empirica di una legge didistribuzione}.
\newblock \bibinfo{journal}{\emph{Giorn Dell'inst Ital Degli Att}}  \bibinfo{volume}{4} (\bibinfo{year}{1933}), \bibinfo{pages}{89--91}.
\newblock


\bibitem[Anderson et~al\mbox{.}(2011)]%
        {Anderson2011-xh}
\bibfield{author}{\bibinfo{person}{Glen Anderson}, \bibinfo{person}{Rina Doherty}, {and} \bibinfo{person}{Subhashini Ganapathy}.} \bibinfo{year}{2011}\natexlab{}.
\newblock \showarticletitle{User perception of touch screen latency}.
\newblock In \bibinfo{booktitle}{\emph{Lecture Notes in Computer Science}}. \bibinfo{publisher}{Springer Berlin Heidelberg}, \bibinfo{address}{Berlin, Heidelberg}, \bibinfo{pages}{195--202}.
\newblock


\bibitem[Anthony and Wobbrock(2010)]%
        {Anthony2010-en}
\bibfield{author}{\bibinfo{person}{Lisa Anthony} {and} \bibinfo{person}{J Wobbrock}.} \bibinfo{year}{2010}\natexlab{}.
\newblock \showarticletitle{A lightweight multistroke recognizer for user interface prototypes}.
\newblock \bibinfo{journal}{\emph{Proc. Graph. Interface}} (\bibinfo{date}{May} \bibinfo{year}{2010}), \bibinfo{pages}{245--252}.
\newblock


\bibitem[Anthony and Wobbrock(2012)]%
        {Anthony2012-tk}
\bibfield{author}{\bibinfo{person}{Lisa Anthony} {and} \bibinfo{person}{J Wobbrock}.} \bibinfo{year}{2012}\natexlab{}.
\newblock \showarticletitle{\${N}-protractor: a fast and accurate multistroke recognizer}.
\newblock \bibinfo{journal}{\emph{Proc. Graph. Interface}} (\bibinfo{date}{May} \bibinfo{year}{2012}), \bibinfo{pages}{117--120}.
\newblock


\bibitem[{Apple Support}(2024)]%
        {Apple_Support_undated-gs}
\bibfield{author}{\bibinfo{person}{{Apple Support}}.} \bibinfo{year}{2024}\natexlab{}.
\newblock \bibinfo{title}{Use {AssistiveTouch} on your Apple Watch}.
\newblock \bibinfo{howpublished}{\url{https://support.apple.com/en-us/111111}}.
\newblock
\newblock
\shownote{Accessed: 2024-11-1}.


\bibitem[Ashbrook and Starner(2010)]%
        {Ashbrook2010-fi}
\bibfield{author}{\bibinfo{person}{Daniel Ashbrook} {and} \bibinfo{person}{Thad Starner}.} \bibinfo{year}{2010}\natexlab{}.
\newblock \showarticletitle{{MAGIC}: a motion gesture design tool}. In \bibinfo{booktitle}{\emph{Proceedings of the SIGCHI Conference on Human Factors in Computing Systems}}. \bibinfo{publisher}{ACM}, \bibinfo{address}{New York, NY, USA}.
\newblock


\bibitem[Atzori et~al\mbox{.}(2016)]%
        {Atzori2016-ka}
\bibfield{author}{\bibinfo{person}{Manfredo Atzori}, \bibinfo{person}{Matteo Cognolato}, {and} \bibinfo{person}{Henning Müller}.} \bibinfo{year}{2016}\natexlab{}.
\newblock \showarticletitle{Deep Learning with Convolutional Neural Networks Applied to Electromyography Data: A Resource for the Classification of Movements for Prosthetic Hands}.
\newblock \bibinfo{journal}{\emph{Front. Neurorobot.}} \bibinfo{volume}{10}, \bibinfo{number}{SEP} (\bibinfo{date}{Sept.} \bibinfo{year}{2016}), \bibinfo{pages}{1--10}.
\newblock


\bibitem[Berenzweig et~al\mbox{.}(2024)]%
        {Berenzweig2024-df}
\bibfield{author}{\bibinfo{person}{Adam Berenzweig}, \bibinfo{person}{Sean Bittner}, \bibinfo{person}{Anuoluwapo Bolarinwa}, \bibinfo{person}{Alan Du}, \bibinfo{person}{Alexandre Gramfort}, \bibinfo{person}{Michael Mandel}, \bibinfo{person}{Jeffrey Seely}, {and} \bibinfo{person}{Viswanath Sivakumar}.} \bibinfo{year}{2024}\natexlab{}.
\newblock \showarticletitle{{Emg2qwerty}: A large dataset with baselines for touch typing using surface electromyography}. In \bibinfo{booktitle}{\emph{Advances in Neural Information Processing Systems 37}}, \bibfield{editor}{\bibinfo{person}{A~Globerson}, \bibinfo{person}{L~Mackey}, \bibinfo{person}{D~Belgrave}, \bibinfo{person}{A~Fan}, \bibinfo{person}{U~Paquet}, \bibinfo{person}{J~Tomczak}, {and} \bibinfo{person}{C~Zhang}} (Eds.), Vol.~\bibinfo{volume}{37}. \bibinfo{publisher}{Neural Information Processing Systems Foundation, Inc. (NeurIPS)}, \bibinfo{address}{San Diego, California, USA}, \bibinfo{pages}{91373--91389}.
\newblock


\bibitem[Berman et~al\mbox{.}(2023)]%
        {Berman2023-qu}
\bibfield{author}{\bibinfo{person}{Joseph Berman}, \bibinfo{person}{Robert Hinson}, \bibinfo{person}{I-Chieh Lee}, {and} \bibinfo{person}{He Huang}.} \bibinfo{year}{2023}\natexlab{}.
\newblock \showarticletitle{Harnessing machine learning and physiological knowledge for a novel {EMG}-based neural-machine interface}.
\newblock \bibinfo{journal}{\emph{IEEE Trans. Biomed. Eng.}} \bibinfo{volume}{70}, \bibinfo{number}{4} (\bibinfo{date}{April} \bibinfo{year}{2023}), \bibinfo{pages}{1125--1136}.
\newblock


\bibitem[Bhasin et~al\mbox{.}(2024)]%
        {Bhasin2024-jg}
\bibfield{author}{\bibinfo{person}{Rohin Bhasin}, \bibinfo{person}{Anuoluwapo Bolarinwa}, \bibinfo{person}{Yujun Cai}, \bibinfo{person}{Nathan Danielson}, \bibinfo{person}{Shangchen Han}, \bibinfo{person}{Jesse Marshall}, \bibinfo{person}{Josh Merel}, \bibinfo{person}{Eftychios Pnevmatikakis}, \bibinfo{person}{Sasha Salter}, \bibinfo{person}{Collin Schlager}, \bibinfo{person}{Adrian Spurr}, \bibinfo{person}{Peter Walkington}, \bibinfo{person}{Robert Wang}, {and} \bibinfo{person}{Richard Warren}.} \bibinfo{year}{2024}\natexlab{}.
\newblock \showarticletitle{{emg2pose}: A Large and Diverse Benchmark for Surface Electromyographic Hand Pose Estimation}. In \bibinfo{booktitle}{\emph{Advances in Neural Information Processing Systems 37}}, \bibfield{editor}{\bibinfo{person}{A~Globerson}, \bibinfo{person}{L~Mackey}, \bibinfo{person}{D~Belgrave}, \bibinfo{person}{A~Fan}, \bibinfo{person}{U~Paquet}, \bibinfo{person}{J~Tomczak}, {and} \bibinfo{person}{C~Zhang}} (Eds.), Vol.~\bibinfo{volume}{37}. \bibinfo{publisher}{Neural Information Processing Systems Foundation, Inc. (NeurIPS)}, \bibinfo{address}{San Diego, California, USA}, \bibinfo{pages}{55703--55728}.
\newblock


\bibitem[Brains(2017)]%
        {Brains2017-sc}
\bibfield{author}{\bibinfo{person}{Backyard Brains}.} \bibinfo{year}{2017}\natexlab{}.
\newblock \bibinfo{title}{Muscle {SpikerShield} Bundle}.
\newblock \bibinfo{howpublished}{\url{https://backyardbrains.com/products/muscleSpikershieldBundle}}.
\newblock
\newblock
\shownote{Accessed: 2024-11-1}.


\bibitem[Braun and Clarke(2006)]%
        {Braun2006-kf}
\bibfield{author}{\bibinfo{person}{Virginia Braun} {and} \bibinfo{person}{Victoria Clarke}.} \bibinfo{year}{2006}\natexlab{}.
\newblock \showarticletitle{Using thematic analysis in psychology}.
\newblock \bibinfo{journal}{\emph{Qual. Res. Psychol.}} \bibinfo{volume}{3}, \bibinfo{number}{2} (\bibinfo{date}{Jan.} \bibinfo{year}{2006}), \bibinfo{pages}{77--101}.
\newblock


\bibitem[Braun and Clarke(2019)]%
        {Braun2019-ks}
\bibfield{author}{\bibinfo{person}{Virginia Braun} {and} \bibinfo{person}{Victoria Clarke}.} \bibinfo{year}{2019}\natexlab{}.
\newblock \bibinfo{title}{Reflecting on reflexive thematic analysis}.
\newblock , \bibinfo{numpages}{589--597}~pages.
\newblock


\bibitem[Breslow and Clayton(1993)]%
        {Breslow1993-ji}
\bibfield{author}{\bibinfo{person}{N~E Breslow} {and} \bibinfo{person}{D~G Clayton}.} \bibinfo{year}{1993}\natexlab{}.
\newblock \showarticletitle{Approximate inference in generalized linear mixed models}.
\newblock \bibinfo{journal}{\emph{J. Am. Stat. Assoc.}} \bibinfo{volume}{88}, \bibinfo{number}{421} (\bibinfo{date}{March} \bibinfo{year}{1993}), \bibinfo{pages}{9--25}.
\newblock


\bibitem[Carrington et~al\mbox{.}(2016)]%
        {Carrington2016-by}
\bibfield{author}{\bibinfo{person}{Patrick Carrington}, \bibinfo{person}{Jian-Ming Chang}, \bibinfo{person}{Kevin Chang}, \bibinfo{person}{Catherine Hornback}, \bibinfo{person}{Amy Hurst}, {and} \bibinfo{person}{Shaun~K Kane}.} \bibinfo{year}{2016}\natexlab{}.
\newblock \showarticletitle{The Gest-Rest Family: Exploring input possibilities for wheelchair armrests}.
\newblock \bibinfo{journal}{\emph{ACM Trans. Access. Comput.}} \bibinfo{volume}{8}, \bibinfo{number}{3} (\bibinfo{date}{May} \bibinfo{year}{2016}), \bibinfo{pages}{1--24}.
\newblock


\bibitem[Carrington et~al\mbox{.}(2014)]%
        {Carrington2014-wz}
\bibfield{author}{\bibinfo{person}{Patrick Carrington}, \bibinfo{person}{Amy Hurst}, {and} \bibinfo{person}{Shaun~K Kane}.} \bibinfo{year}{2014}\natexlab{}.
\newblock \showarticletitle{Wearables and chairables: inclusive design of mobile input and output techniques for power wheelchair users}. In \bibinfo{booktitle}{\emph{Proceedings of the SIGCHI Conference on Human Factors in Computing Systems}}. \bibinfo{publisher}{ACM}, \bibinfo{address}{New York, NY, USA}, \bibinfo{pages}{3103--3112}.
\newblock


\bibitem[Christ et~al\mbox{.}(2018)]%
        {Christ2018-tf}
\bibfield{author}{\bibinfo{person}{Maximilian Christ}, \bibinfo{person}{Nils Braun}, \bibinfo{person}{Julius Neuffer}, {and} \bibinfo{person}{Andreas~W Kempa-Liehr}.} \bibinfo{year}{2018}\natexlab{}.
\newblock \showarticletitle{Time Series {FeatuRe} Extraction on basis of Scalable Hypothesis tests (tsfresh – A Python package)}.
\newblock \bibinfo{journal}{\emph{Neurocomputing}}  \bibinfo{volume}{307} (\bibinfo{date}{Sept.} \bibinfo{year}{2018}), \bibinfo{pages}{72--77}.
\newblock


\bibitem[Cormen et~al\mbox{.}(2022)]%
        {Cormen2022-vv}
\bibfield{author}{\bibinfo{person}{T~H Cormen}, \bibinfo{person}{C~E Leiserson}, \bibinfo{person}{R~L Rivest}, {and} \bibinfo{person}{C Stein}.} \bibinfo{year}{2022}\natexlab{}.
\newblock \bibinfo{booktitle}{\emph{Introduction to Algorithms, fourth edition}}.
\newblock \bibinfo{publisher}{MIT Press}.
\newblock


\bibitem[De~Luca(1997)]%
        {De_Luca1997-jw}
\bibfield{author}{\bibinfo{person}{Carlo~J De~Luca}.} \bibinfo{year}{1997}\natexlab{}.
\newblock \showarticletitle{The Use of Surface Electromyography in Biomechanics}.
\newblock \bibinfo{journal}{\emph{J. Appl. Biomech.}} \bibinfo{volume}{13}, \bibinfo{number}{2} (\bibinfo{date}{May} \bibinfo{year}{1997}), \bibinfo{pages}{135--163}.
\newblock


\bibitem[De~Luca et~al\mbox{.}(2010)]%
        {De_Luca2010-ul}
\bibfield{author}{\bibinfo{person}{Carlo~J De~Luca}, \bibinfo{person}{L~Donald Gilmore}, \bibinfo{person}{Mikhail Kuznetsov}, {and} \bibinfo{person}{Serge~H Roy}.} \bibinfo{year}{2010}\natexlab{}.
\newblock \showarticletitle{Filtering the surface {EMG} signal: Movement artifact and baseline noise contamination}.
\newblock \bibinfo{journal}{\emph{J. Biomech.}} \bibinfo{volume}{43}, \bibinfo{number}{8} (\bibinfo{date}{May} \bibinfo{year}{2010}), \bibinfo{pages}{1573--1579}.
\newblock


\bibitem[Dehzangi and Sahu(2018)]%
        {Dehzangi2018-ic}
\bibfield{author}{\bibinfo{person}{Omid Dehzangi} {and} \bibinfo{person}{Vaishali Sahu}.} \bibinfo{year}{2018}\natexlab{}.
\newblock \showarticletitle{{IMU}-based robust human activity recognition using feature analysis, extraction, and reduction}. In \bibinfo{booktitle}{\emph{2018 24th International Conference on Pattern Recognition (ICPR)}}. \bibinfo{publisher}{IEEE}, \bibinfo{pages}{1402--1407}.
\newblock


\bibitem[Deldari et~al\mbox{.}(2022)]%
        {Deldari2022-nm}
\bibfield{author}{\bibinfo{person}{Shohreh Deldari}, \bibinfo{person}{Hao Xue}, \bibinfo{person}{Aaqib Saeed}, \bibinfo{person}{Daniel~V Smith}, {and} \bibinfo{person}{Flora~D Salim}.} \bibinfo{year}{2022}\natexlab{}.
\newblock \showarticletitle{{COCOA}: Cross {mOdality} {COntrastive} {leArning} for sensor data}.
\newblock \bibinfo{journal}{\emph{Proc. ACM Interact. Mob. Wearable Ubiquitous Technol.}} \bibinfo{volume}{6}, \bibinfo{number}{3} (\bibinfo{date}{Sept.} \bibinfo{year}{2022}), \bibinfo{pages}{1--28}.
\newblock


\bibitem[Developers(2024)]%
        {Developers_undated-bb}
\bibfield{author}{\bibinfo{person}{Scikit-Learn Developers}.} \bibinfo{year}{2024}\natexlab{}.
\newblock \bibinfo{title}{{PCA}}.
\newblock \bibinfo{howpublished}{\url{https://scikit-learn.org/stable/modules/generated/sklearn.decomposition.PCA.html}}.
\newblock
\newblock
\shownote{Accessed: 2024-11-1}.


\bibitem[Eddy et~al\mbox{.}(2025)]%
        {Eddy2025-ir}
\bibfield{author}{\bibinfo{person}{Ethan Eddy}, \bibinfo{person}{Evan Campbell}, \bibinfo{person}{Christian Morrell}, \bibinfo{person}{Heather Williams}, \bibinfo{person}{Scott Bateman}, {and} \bibinfo{person}{Erik Scheme}.} \bibinfo{year}{2025}\natexlab{}.
\newblock \showarticletitle{Raising the standard: an open source benchmarking platform and data repository to accelerate myoelectric control research}.
\newblock \bibinfo{journal}{\emph{Mach. Learn.: Health}} \bibinfo{volume}{1}, \bibinfo{number}{1} (\bibinfo{date}{Sept.} \bibinfo{year}{2025}), \bibinfo{pages}{010601}.
\newblock


\bibitem[Eddy et~al\mbox{.}(2023)]%
        {Eddy2023-iz}
\bibfield{author}{\bibinfo{person}{Ethan Eddy}, \bibinfo{person}{Erik~J Scheme}, {and} \bibinfo{person}{Scott Bateman}.} \bibinfo{year}{2023}\natexlab{}.
\newblock \showarticletitle{A framework and call to action for the future development of {EMG}-based input in {HCI}}. In \bibinfo{booktitle}{\emph{Proceedings of the 2023 CHI Conference on Human Factors in Computing Systems}} \emph{(\bibinfo{series}{CHI '23}, Vol.~\bibinfo{volume}{72})}. \bibinfo{publisher}{ACM}, \bibinfo{address}{New York, NY, USA}, \bibinfo{pages}{1--23}.
\newblock


\bibitem[Fan et~al\mbox{.}(2018)]%
        {Fan2018-ng}
\bibfield{author}{\bibinfo{person}{Junjun Fan}, \bibinfo{person}{Xiangmin Fan}, \bibinfo{person}{Feng Tian}, \bibinfo{person}{Yang Li}, \bibinfo{person}{Zitao Liu}, \bibinfo{person}{Wei Sun}, {and} \bibinfo{person}{Hongan Wang}.} \bibinfo{year}{2018}\natexlab{}.
\newblock \showarticletitle{What is that in your hand?: Recognizing grasped objects via forearm electromyography sensing}.
\newblock \bibinfo{journal}{\emph{Proc. ACM Interact. Mob. Wearable Ubiquitous Technol.}} \bibinfo{volume}{2}, \bibinfo{number}{4} (\bibinfo{date}{Dec.} \bibinfo{year}{2018}), \bibinfo{pages}{1--24}.
\newblock


\bibitem[Fan et~al\mbox{.}(2020)]%
        {Fan2020-bu}
\bibfield{author}{\bibinfo{person}{Mingming Fan}, \bibinfo{person}{Zhen Li}, {and} \bibinfo{person}{Franklin~Mingzhe Li}.} \bibinfo{year}{2020}\natexlab{}.
\newblock \showarticletitle{Eyelid gestures on mobile devices for people with motor impairments}. In \bibinfo{booktitle}{\emph{Proceedings of the 22nd International ACM SIGACCESS Conference on Computers and Accessibility}} \emph{(\bibinfo{series}{ASSETS '20}, \bibinfo{number}{Article 15})}. \bibinfo{publisher}{ACM}, \bibinfo{address}{New York, NY, USA}, \bibinfo{pages}{1--8}.
\newblock


\bibitem[Franz et~al\mbox{.}(2023)]%
        {Franz2023-mn}
\bibfield{author}{\bibinfo{person}{Rachel~L Franz}, \bibinfo{person}{Jinghan Yu}, {and} \bibinfo{person}{Jacob~O Wobbrock}.} \bibinfo{year}{2023}\natexlab{}.
\newblock \showarticletitle{Comparing locomotion techniques in virtual reality for people with upper-body motor impairments}. In \bibinfo{booktitle}{\emph{The 25th International ACM SIGACCESS Conference on Computers and Accessibility}} \emph{(\bibinfo{series}{ASSETS '23}, \bibinfo{number}{Article 39})}. \bibinfo{publisher}{ACM}, \bibinfo{address}{New York, NY, USA}, \bibinfo{pages}{1--15}.
\newblock


\bibitem[Gashi et~al\mbox{.}(2020)]%
        {gashi_2020_detection}
\bibfield{author}{\bibinfo{person}{Shkurta Gashi}, \bibinfo{person}{Elena Di~Lascio}, \bibinfo{person}{Bianca Stancu}, \bibinfo{person}{Vedant~Das Swain}, \bibinfo{person}{Varun Mishra}, \bibinfo{person}{Martin Gjoreski}, {and} \bibinfo{person}{Silvia Santini}.} \bibinfo{year}{2020}\natexlab{}.
\newblock \showarticletitle{Detection of Artifacts in Ambulatory Electrodermal Activity Data}.
\newblock \bibinfo{journal}{\emph{Proc. ACM Interact. Mob. Wearable Ubiquitous Technol.}} \bibinfo{volume}{4}, \bibinfo{number}{2}, Article \bibinfo{articleno}{44} (\bibinfo{date}{June} \bibinfo{year}{2020}), \bibinfo{numpages}{31}~pages.
\newblock


\bibitem[Glazko et~al\mbox{.}(2024)]%
        {Glazko2024-gh}
\bibfield{author}{\bibinfo{person}{Kate Glazko}, \bibinfo{person}{Yusuf Mohammed}, \bibinfo{person}{Ben Kosa}, \bibinfo{person}{Venkatesh Potluri}, {and} \bibinfo{person}{Jennifer Mankoff}.} \bibinfo{year}{2024}\natexlab{}.
\newblock \showarticletitle{Identifying and improving disability bias in {GPT}-based resume screening}. In \bibinfo{booktitle}{\emph{The 2024 ACM Conference on Fairness Accountability and Transparency}}. \bibinfo{publisher}{ACM}, \bibinfo{address}{New York, NY, USA}, \bibinfo{pages}{687--700}.
\newblock


\bibitem[Glazko et~al\mbox{.}(2023)]%
        {Glazko2023-ha}
\bibfield{author}{\bibinfo{person}{Kate~S Glazko}, \bibinfo{person}{Momona Yamagami}, \bibinfo{person}{Aashaka Desai}, \bibinfo{person}{Kelly~Avery Mack}, \bibinfo{person}{Venkatesh Potluri}, \bibinfo{person}{Xuhai Xu}, {and} \bibinfo{person}{Jennifer Mankoff}.} \bibinfo{year}{2023}\natexlab{}.
\newblock \showarticletitle{An autoethnographic case study of generative artificial intelligence's utility for accessibility}. In \bibinfo{booktitle}{\emph{The 25th International ACM SIGACCESS Conference on Computers and Accessibility}}. \bibinfo{publisher}{ACM}, \bibinfo{address}{New York, NY, USA}, \bibinfo{pages}{1--8}.
\newblock


\bibitem[Greenacre et~al\mbox{.}(2022)]%
        {Greenacre2022-co}
\bibfield{author}{\bibinfo{person}{Michael Greenacre}, \bibinfo{person}{Patrick J~F Groenen}, \bibinfo{person}{Trevor Hastie}, \bibinfo{person}{Alfonso~Iodice D'Enza}, \bibinfo{person}{Angelos Markos}, {and} \bibinfo{person}{Elena Tuzhilina}.} \bibinfo{year}{2022}\natexlab{}.
\newblock \showarticletitle{Principal component analysis}.
\newblock \bibinfo{journal}{\emph{Nat. Rev. Methods Primers}} \bibinfo{volume}{2}, \bibinfo{number}{1} (\bibinfo{date}{Dec.} \bibinfo{year}{2022}), \bibinfo{pages}{1--21}.
\newblock


\bibitem[Guan and Pl\"{o}tz(2017)]%
        {guan_2017_ensembles}
\bibfield{author}{\bibinfo{person}{Yu Guan} {and} \bibinfo{person}{Thomas Pl\"{o}tz}.} \bibinfo{year}{2017}\natexlab{}.
\newblock \showarticletitle{Ensembles of Deep LSTM Learners for Activity Recognition using Wearables}.
\newblock \bibinfo{journal}{\emph{Proc. ACM Interact. Mob. Wearable Ubiquitous Technol.}} \bibinfo{volume}{1}, \bibinfo{number}{2}, Article \bibinfo{articleno}{11} (\bibinfo{date}{June} \bibinfo{year}{2017}), \bibinfo{numpages}{28}~pages.
\newblock


\bibitem[Gummesson et~al\mbox{.}(2006)]%
        {Gummesson2006-xx}
\bibfield{author}{\bibinfo{person}{Christina Gummesson}, \bibinfo{person}{Michael~M Ward}, {and} \bibinfo{person}{Isam Atroshi}.} \bibinfo{year}{2006}\natexlab{}.
\newblock \showarticletitle{The shortened disabilities of the arm, shoulder and hand questionnaire ({QuickDASH}): validity and reliability based on responses within the full-length {DASH}}.
\newblock \bibinfo{journal}{\emph{BMC Musculoskelet. Disord.}} \bibinfo{volume}{7}, \bibinfo{number}{1} (\bibinfo{date}{May} \bibinfo{year}{2006}), \bibinfo{pages}{44}.
\newblock


\bibitem[Hao et~al\mbox{.}(2017)]%
        {hao_2017_MindfulWatch}
\bibfield{author}{\bibinfo{person}{Tian Hao}, \bibinfo{person}{Chongguang Bi}, \bibinfo{person}{Guoliang Xing}, \bibinfo{person}{Roxane Chan}, {and} \bibinfo{person}{Linlin Tu}.} \bibinfo{year}{2017}\natexlab{}.
\newblock \showarticletitle{MindfulWatch: A Smartwatch-Based System For Real-Time Respiration Monitoring During Meditation}.
\newblock \bibinfo{journal}{\emph{Proc. ACM Interact. Mob. Wearable Ubiquitous Technol.}} \bibinfo{volume}{1}, \bibinfo{number}{3}, Article \bibinfo{articleno}{57} (\bibinfo{date}{Sept.} \bibinfo{year}{2017}), \bibinfo{numpages}{19}~pages.
\newblock


\bibitem[Hart(1988)]%
        {Hart1988-tz}
\bibfield{author}{\bibinfo{person}{S~G Hart}.} \bibinfo{year}{1988}\natexlab{}.
\newblock \showarticletitle{Development of {NASA}-{TLX} (Task Load Index): Results of empirical and theoretical research}.
\newblock \bibinfo{journal}{\emph{Human mental workload/Elsevier}} (\bibinfo{year}{1988}).
\newblock


\bibitem[Ho et~al\mbox{.}(2011)]%
        {Ho2011-rv}
\bibfield{author}{\bibinfo{person}{N~S~K Ho}, \bibinfo{person}{K~Y Tong}, \bibinfo{person}{X~L Hu}, \bibinfo{person}{K~L Fung}, \bibinfo{person}{X~J Wei}, \bibinfo{person}{W Rong}, {and} \bibinfo{person}{E~A Susanto}.} \bibinfo{year}{2011}\natexlab{}.
\newblock \showarticletitle{An {EMG}-driven exoskeleton hand robotic training device on chronic stroke subjects: task training system for stroke rehabilitation}.
\newblock \bibinfo{journal}{\emph{IEEE Int. Conf. Rehabil. Robot.}}  \bibinfo{volume}{2011} (\bibinfo{year}{2011}), \bibinfo{pages}{5975340}.
\newblock


\bibitem[Holm(1979)]%
        {Holm1979-lh}
\bibfield{author}{\bibinfo{person}{S Holm}.} \bibinfo{year}{1979}\natexlab{}.
\newblock \showarticletitle{A Simple Sequentially Rejective Multiple Test Procedure}.
\newblock \bibinfo{journal}{\emph{Scandinavian Journal of Statistics}} \bibinfo{volume}{6}, \bibinfo{number}{2} (\bibinfo{year}{1979}), \bibinfo{pages}{65--70}.
\newblock


\bibitem[Huang et~al\mbox{.}(2015)]%
        {Huang2015-uh}
\bibfield{author}{\bibinfo{person}{Donny Huang}, \bibinfo{person}{Xiaoyi Zhang}, \bibinfo{person}{T~Scott Saponas}, \bibinfo{person}{James Fogarty}, {and} \bibinfo{person}{Shyamnath Gollakota}.} \bibinfo{year}{2015}\natexlab{}.
\newblock \showarticletitle{Leveraging dual-observable input for fine-grained thumb interaction using forearm {EMG}}. In \bibinfo{booktitle}{\emph{Proceedings of the 28th Annual ACM Symposium on User Interface Software \& Technology}}. \bibinfo{publisher}{ACM}, \bibinfo{address}{New York, NY, USA}, \bibinfo{pages}{523--528}.
\newblock


\bibitem[Hye et~al\mbox{.}(2023)]%
        {Hye2023-xa}
\bibfield{author}{\bibinfo{person}{Nafe~Muhtasim Hye}, \bibinfo{person}{Umma Hany}, \bibinfo{person}{Sumit Chakravarty}, \bibinfo{person}{Lutfa Akter}, {and} \bibinfo{person}{Imtiaz Ahmed}.} \bibinfo{year}{2023}\natexlab{}.
\newblock \showarticletitle{Artificial intelligence for {sEMG}-based muscular movement recognition for hand prosthesis}.
\newblock \bibinfo{journal}{\emph{IEEE Access}}  \bibinfo{volume}{11} (\bibinfo{year}{2023}), \bibinfo{pages}{38850--38863}.
\newblock


\bibitem[Iqbal et~al\mbox{.}(2018)]%
        {Iqbal2018-vz}
\bibfield{author}{\bibinfo{person}{Nisheena~V Iqbal}, \bibinfo{person}{Kamalraj Subramaniam}, {and} \bibinfo{person}{Shaniba~Asmi P.}} \bibinfo{year}{2018}\natexlab{}.
\newblock \showarticletitle{A review on upper-limb myoelectric prosthetic control}.
\newblock \bibinfo{journal}{\emph{IETE J. Res.}} \bibinfo{volume}{64}, \bibinfo{number}{6} (\bibinfo{date}{Nov.} \bibinfo{year}{2018}), \bibinfo{pages}{740--752}.
\newblock


\bibitem[Kaifosh et~al\mbox{.}(2025)]%
        {Ctrl-Labs_at_Reality_Labs2024-zb}
\bibfield{author}{\bibinfo{person}{Patrick Kaifosh}, \bibinfo{person}{Thomas~R Reardon}, {and} \bibinfo{person}{{CTRL-labs at Reality Labs}}.} \bibinfo{year}{2025}\natexlab{}.
\newblock \showarticletitle{A generic non-invasive neuromotor interface for human-computer interaction}.
\newblock \bibinfo{journal}{\emph{Nature}} \bibinfo{volume}{645}, \bibinfo{number}{8081} (\bibinfo{date}{Sept.} \bibinfo{year}{2025}), \bibinfo{pages}{702--711}.
\newblock


\bibitem[Karnam et~al\mbox{.}(2022)]%
        {Karnam2022-wf}
\bibfield{author}{\bibinfo{person}{Naveen~Kumar Karnam}, \bibinfo{person}{Shiv~Ram Dubey}, \bibinfo{person}{Anish~Chand Turlapaty}, {and} \bibinfo{person}{Balakrishna Gokaraju}.} \bibinfo{year}{2022}\natexlab{}.
\newblock \showarticletitle{{EMGHandNet}: A hybrid {CNN} and Bi-{LSTM} architecture for hand activity classification using surface {EMG} signals}.
\newblock \bibinfo{journal}{\emph{Biocybern. Biomed. Eng.}} \bibinfo{volume}{42}, \bibinfo{number}{1} (\bibinfo{date}{Jan.} \bibinfo{year}{2022}), \bibinfo{pages}{325--340}.
\newblock


\bibitem[Khanna et~al\mbox{.}(2024)]%
        {Khanna2024-gk}
\bibfield{author}{\bibinfo{person}{Prerna Khanna}, \bibinfo{person}{I~V Ramakrishnan}, \bibinfo{person}{Shubham Jain}, \bibinfo{person}{Xiaojun Bi}, {and} \bibinfo{person}{Aruna Balasubramanian}.} \bibinfo{year}{2024}\natexlab{}.
\newblock \showarticletitle{Hand gesture recognition for blind users by tracking {3D} gesture trajectory}.
\newblock \bibinfo{journal}{\emph{Proc. SIGCHI Conf. Hum. Factor. Comput. Syst.}}  \bibinfo{volume}{2024} (\bibinfo{date}{May} \bibinfo{year}{2024}).
\newblock


\bibitem[Kimura et~al\mbox{.}(2022)]%
        {kimura_2022_SilentSpeller}
\bibfield{author}{\bibinfo{person}{Naoki Kimura}, \bibinfo{person}{Tan Gemicioglu}, \bibinfo{person}{Jonathan Womack}, \bibinfo{person}{Richard Li}, \bibinfo{person}{Yuhui Zhao}, \bibinfo{person}{Abdelkareem Bedri}, \bibinfo{person}{Zixiong Su}, \bibinfo{person}{Alex Olwal}, \bibinfo{person}{Jun Rekimoto}, {and} \bibinfo{person}{Thad Starner}.} \bibinfo{year}{2022}\natexlab{}.
\newblock \showarticletitle{SilentSpeller: Towards mobile, hands-free, silent speech text entry using electropalatography}. In \bibinfo{booktitle}{\emph{Proceedings of the 2022 CHI Conference on Human Factors in Computing Systems}} (New Orleans, LA, USA) \emph{(\bibinfo{series}{CHI '22})}. \bibinfo{publisher}{Association for Computing Machinery}, \bibinfo{address}{New York, NY, USA}, Article \bibinfo{articleno}{288}, \bibinfo{numpages}{19}~pages.
\newblock
\showISBNx{9781450391573}


\bibitem[Kingma and Ba(2014)]%
        {Kingma2014-wg}
\bibfield{author}{\bibinfo{person}{Diederik~P Kingma} {and} \bibinfo{person}{Jimmy Ba}.} \bibinfo{year}{2014}\natexlab{}.
\newblock \showarticletitle{Adam: A method for stochastic optimization}.
\newblock \bibinfo{journal}{\emph{arXiv [cs.LG]}} (\bibinfo{date}{Dec.} \bibinfo{year}{2014}).
\newblock


\bibitem[Kohlsdorf and Starner(2013)]%
        {Kohlsdorf2013-qz}
\bibfield{author}{\bibinfo{person}{Daniel Kohlsdorf} {and} \bibinfo{person}{Thad Starner}.} \bibinfo{year}{2013}\natexlab{}.
\newblock \showarticletitle{{MAGIC} summoning: towards automatic suggesting and testing of gestures with low probability of false positives during use}.
\newblock \bibinfo{journal}{\emph{J. Mach. Learn. Res.}}  \bibinfo{volume}{14} (\bibinfo{year}{2013}), \bibinfo{pages}{209--242}.
\newblock


\bibitem[Kosa et~al\mbox{.}(2026)]%
        {Kosa2026-av}
\bibfield{author}{\bibinfo{person}{Ben Kosa}, \bibinfo{person}{Hsuanling Lee}, \bibinfo{person}{Jasmine Li}, \bibinfo{person}{Sanbrita Mondal}, \bibinfo{person}{Yuhang Zhao}, {and} \bibinfo{person}{Liang He}.} \bibinfo{year}{2026}\natexlab{}.
\newblock \showarticletitle{Not seeing the whole picture: Challenges and opportunities in using {AI} for co-making physical, {DIY}-{AT} for people with visual impairments}. In \bibinfo{booktitle}{\emph{Proceedings of the 2026 CHI Conference on Human Factors in Computing Systems}}. \bibinfo{publisher}{ACM}, \bibinfo{address}{New York, NY, USA}, \bibinfo{pages}{1--20}.
\newblock


\bibitem[Krasoulis et~al\mbox{.}(2017)]%
        {Krasoulis2017-jb}
\bibfield{author}{\bibinfo{person}{Agamemnon Krasoulis}, \bibinfo{person}{Iris Kyranou}, \bibinfo{person}{Mustapha~Suphi Erden}, \bibinfo{person}{Kianoush Nazarpour}, {and} \bibinfo{person}{Sethu Vijayakumar}.} \bibinfo{year}{2017}\natexlab{}.
\newblock \showarticletitle{Improved prosthetic hand control with concurrent use of myoelectric and inertial measurements}.
\newblock \bibinfo{journal}{\emph{J. Neuroeng. Rehabil.}} \bibinfo{volume}{14}, \bibinfo{number}{1} (\bibinfo{date}{July} \bibinfo{year}{2017}), \bibinfo{pages}{71}.
\newblock


\bibitem[Krasoulis et~al\mbox{.}(2020)]%
        {Krasoulis2020-ix}
\bibfield{author}{\bibinfo{person}{Agamemnon Krasoulis}, \bibinfo{person}{Sethu Vijayakumar}, {and} \bibinfo{person}{Kianoush Nazarpour}.} \bibinfo{year}{2020}\natexlab{}.
\newblock \showarticletitle{Multi-grip classification-based prosthesis control with two {EMG}-{IMU} sensors}.
\newblock \bibinfo{journal}{\emph{IEEE Trans. Neural Syst. Rehabil. Eng.}} \bibinfo{volume}{28}, \bibinfo{number}{2} (\bibinfo{date}{Feb.} \bibinfo{year}{2020}), \bibinfo{pages}{508--518}.
\newblock


\bibitem[Lee et~al\mbox{.}(2019)]%
        {Lee2019-jq}
\bibfield{author}{\bibinfo{person}{Juyoung Lee}, \bibinfo{person}{Shaurye Aggarwal}, \bibinfo{person}{Jason Wu}, \bibinfo{person}{Thad Starner}, {and} \bibinfo{person}{Woontack Woo}.} \bibinfo{year}{2019}\natexlab{}.
\newblock \showarticletitle{{SelfSync}: exploring self-synchronous body-based hotword gestures for initiating interaction}. In \bibinfo{booktitle}{\emph{Proceedings of the 23rd International Symposium on Wearable Computers}}. \bibinfo{publisher}{ACM}, \bibinfo{address}{New York, NY, USA}.
\newblock


\bibitem[Li et~al\mbox{.}(2022)]%
        {Li2022-he}
\bibfield{author}{\bibinfo{person}{Franklin~Mingzhe Li}, \bibinfo{person}{Michael~Xieyang Liu}, \bibinfo{person}{Yang Zhang}, {and} \bibinfo{person}{Patrick Carrington}.} \bibinfo{year}{2022}\natexlab{}.
\newblock \showarticletitle{Freedom to choose: Understanding input modality preferences of people with upper-body motor impairments for activities of daily living}. In \bibinfo{booktitle}{\emph{Proceedings of the 24th International ACM SIGACCESS Conference on Computers and Accessibility}}. \bibinfo{publisher}{ACM}, \bibinfo{address}{New York, NY, USA}.
\newblock


\bibitem[Li et~al\mbox{.}(2023)]%
        {Li2023-ca}
\bibfield{author}{\bibinfo{person}{Jiyang Li}, \bibinfo{person}{Lin Huang}, \bibinfo{person}{Siddharth Shah}, \bibinfo{person}{Sean~J Jones}, \bibinfo{person}{Yincheng Jin}, \bibinfo{person}{Dingran Wang}, \bibinfo{person}{Adam Russell}, \bibinfo{person}{Seokmin Choi}, \bibinfo{person}{Yang Gao}, \bibinfo{person}{Junsong Yuan}, {and} \bibinfo{person}{Zhanpeng Jin}.} \bibinfo{year}{2023}\natexlab{}.
\newblock \showarticletitle{{SignRing}: Continuous American sign language recognition using {IMU} rings and virtual {IMU} data}.
\newblock \bibinfo{journal}{\emph{Proc. ACM Interact. Mob. Wearable Ubiquitous Technol.}} \bibinfo{volume}{7}, \bibinfo{number}{3} (\bibinfo{date}{Sept.} \bibinfo{year}{2023}), \bibinfo{pages}{1--29}.
\newblock


\bibitem[Li et~al\mbox{.}(2021)]%
        {Li2021-mr}
\bibfield{author}{\bibinfo{person}{Wei Li}, \bibinfo{person}{Ping Shi}, {and} \bibinfo{person}{Hongliu Yu}.} \bibinfo{year}{2021}\natexlab{}.
\newblock \showarticletitle{Gesture Recognition Using Surface Electromyography and Deep Learning for Prostheses Hand: State-of-the-Art, Challenges, and Future}.
\newblock \bibinfo{journal}{\emph{Front. Neurosci.}}  \bibinfo{volume}{15} (\bibinfo{date}{April} \bibinfo{year}{2021}), \bibinfo{pages}{621885}.
\newblock


\bibitem[Li et~al\mbox{.}(2024)]%
        {li_2024_WheelPoser}
\bibfield{author}{\bibinfo{person}{Yunzhi Li}, \bibinfo{person}{Vimal Mollyn}, \bibinfo{person}{Kuang Yuan}, {and} \bibinfo{person}{Patrick Carrington}.} \bibinfo{year}{2024}\natexlab{}.
\newblock \showarticletitle{WheelPoser: Sparse-IMU Based Body Pose Estimation for Wheelchair Users}. In \bibinfo{booktitle}{\emph{Proceedings of the 26th International ACM SIGACCESS Conference on Computers and Accessibility}} (St. John's, NL, Canada) \emph{(\bibinfo{series}{ASSETS '24})}. \bibinfo{publisher}{Association for Computing Machinery}, \bibinfo{address}{New York, NY, USA}, Article \bibinfo{articleno}{8}, \bibinfo{numpages}{17}~pages.
\newblock
\showISBNx{9798400706776}


\bibitem[Liu et~al\mbox{.}(2017)]%
        {liu_2017_supporting}
\bibfield{author}{\bibinfo{person}{Fannie Liu}, \bibinfo{person}{Laura Dabbish}, {and} \bibinfo{person}{Geoff Kaufman}.} \bibinfo{year}{2017}\natexlab{}.
\newblock \showarticletitle{Supporting Social Interactions with an Expressive Heart Rate Sharing Application}.
\newblock \bibinfo{journal}{\emph{Proc. ACM Interact. Mob. Wearable Ubiquitous Technol.}} \bibinfo{volume}{1}, \bibinfo{number}{3}, Article \bibinfo{articleno}{77} (\bibinfo{date}{Sept.} \bibinfo{year}{2017}), \bibinfo{numpages}{26}~pages.
\newblock


\bibitem[Lugaresi et~al\mbox{.}(2019)]%
        {Lugaresi2019-zi}
\bibfield{author}{\bibinfo{person}{Camillo Lugaresi}, \bibinfo{person}{Jiuqiang Tang}, \bibinfo{person}{Hadon Nash}, \bibinfo{person}{Chris McClanahan}, \bibinfo{person}{Esha Uboweja}, \bibinfo{person}{Michael Hays}, \bibinfo{person}{Fan Zhang}, \bibinfo{person}{Chuo-Ling Chang}, \bibinfo{person}{Ming Yong}, {and} \bibinfo{person}{Juhyun Lee}.} \bibinfo{year}{2019}\natexlab{}.
\newblock \showarticletitle{Mediapipe: A framework for perceiving and processing reality}. In \bibinfo{booktitle}{\emph{Third workshop on computer vision for AR/VR at IEEE computer vision and pattern recognition (CVPR)}}.
\newblock


\bibitem[Malu et~al\mbox{.}(2018)]%
        {Malu2018-uo}
\bibfield{author}{\bibinfo{person}{Meethu Malu}, \bibinfo{person}{Pramod Chundury}, {and} \bibinfo{person}{Leah Findlater}.} \bibinfo{year}{2018}\natexlab{}.
\newblock \showarticletitle{Exploring accessible smartwatch interactions for people with upper body motor impairments}. In \bibinfo{booktitle}{\emph{Proceedings of the 2018 CHI Conference on Human Factors in Computing Systems}} \emph{(\bibinfo{series}{CHI '18}, \bibinfo{number}{Paper 488})}. \bibinfo{publisher}{ACM}, \bibinfo{address}{New York, NY, USA}, \bibinfo{pages}{1--12}.
\newblock


\bibitem[{Meta}(2024)]%
        {Meta_undated-hw}
\bibfield{author}{\bibinfo{person}{{Meta}}.} \bibinfo{year}{2024}\natexlab{}.
\newblock \bibinfo{title}{Getting started with Hand and Body Tracking on Meta Quest headsets}.
\newblock \bibinfo{howpublished}{\url{https://www.meta.com/help/quest/articles/headsets-and-accessories/controllers-and-hand-tracking/hand-tracking/}}.
\newblock
\newblock
\shownote{Accessed: 2024-11-1}.


\bibitem[Mezari and Maglogiannis(2018)]%
        {Mezari2018-tn}
\bibfield{author}{\bibinfo{person}{Antigoni Mezari} {and} \bibinfo{person}{Ilias Maglogiannis}.} \bibinfo{year}{2018}\natexlab{}.
\newblock \showarticletitle{An easily customized gesture recognizer for assisted living using commodity mobile devices}.
\newblock \bibinfo{journal}{\emph{J. Healthc. Eng.}}  \bibinfo{volume}{2018} (\bibinfo{date}{July} \bibinfo{year}{2018}), \bibinfo{pages}{3180652}.
\newblock


\bibitem[{Microsoft}(2021)]%
        {Microsoft2021-zx}
\bibfield{author}{\bibinfo{person}{{Microsoft}}.} \bibinfo{year}{2021}\natexlab{}.
\newblock \bibinfo{title}{{HoloLens} 2 gestures for authoring and navigating in Dynamics 365 Guides}.
\newblock \bibinfo{howpublished}{\url{https://learn.microsoft.com/en-us/dynamics365/mixed-reality/guides/authoring-gestures-hl2}}.
\newblock
\newblock
\shownote{Accessed: 2024-11-1}.


\bibitem[Mitchell et~al\mbox{.}(2022)]%
        {Mitchell2022-tm}
\bibfield{author}{\bibinfo{person}{Claire~L Mitchell}, \bibinfo{person}{Gabriel~J Cler}, \bibinfo{person}{Susan~K Fager}, \bibinfo{person}{Paola Contessa}, \bibinfo{person}{Serge~H Roy}, \bibinfo{person}{Gianluca De~Luca}, \bibinfo{person}{Joshua~C Kline}, {and} \bibinfo{person}{Jennifer~M Vojtech}.} \bibinfo{year}{2022}\natexlab{}.
\newblock \showarticletitle{Ability-based keyboards for augmentative and alternative communication: Understanding how individuals' movement patterns translate to more efficient keyboards: Methods to generate keyboards tailored to user-specific motor abilities}.
\newblock \bibinfo{journal}{\emph{Ext. Abstr. Hum. Factors Computing Syst.}}  \bibinfo{volume}{2022} (\bibinfo{date}{April} \bibinfo{year}{2022}).
\newblock


\bibitem[Mott et~al\mbox{.}(2020)]%
        {Mott2020-io}
\bibfield{author}{\bibinfo{person}{Martez Mott}, \bibinfo{person}{John Tang}, \bibinfo{person}{Shaun Kane}, \bibinfo{person}{Edward Cutrell}, {and} \bibinfo{person}{Meredith Ringel~Morris}.} \bibinfo{year}{2020}\natexlab{}.
\newblock \showarticletitle{“I just went into it assuming that {I} wouldn't be able to have the full experience”: Understanding the Accessibility of Virtual Reality for People with Limited Mobility}. In \bibinfo{booktitle}{\emph{The 22nd International ACM SIGACCESS Conference on Computers and Accessibility}} \emph{(\bibinfo{series}{ASSETS '20}, \bibinfo{number}{Article 43})}. \bibinfo{publisher}{ACM}, \bibinfo{address}{New York, NY, USA}, \bibinfo{pages}{1--13}.
\newblock


\bibitem[Nacenta et~al\mbox{.}(2013)]%
        {Nacenta2013-ao}
\bibfield{author}{\bibinfo{person}{Miguel~A Nacenta}, \bibinfo{person}{Yemliha Kamber}, \bibinfo{person}{Yizhou Qiang}, {and} \bibinfo{person}{Per~Ola Kristensson}.} \bibinfo{year}{2013}\natexlab{}.
\newblock \showarticletitle{Memorability of pre-designed and user-defined gesture sets}. In \bibinfo{booktitle}{\emph{Proceedings of the SIGCHI Conference on Human Factors in Computing Systems}} \emph{(\bibinfo{series}{CHI '13})}. \bibinfo{publisher}{ACM}, \bibinfo{address}{New York, NY, USA}, \bibinfo{pages}{1099--1108}.
\newblock


\bibitem[Osborne(2002)]%
        {Osborne2002-xd}
\bibfield{author}{\bibinfo{person}{Jason Osborne}.} \bibinfo{year}{2002}\natexlab{}.
\newblock \showarticletitle{Notes on the use of data transformations}.
\newblock \bibinfo{journal}{\emph{Pract. Assess. Res. Eval.}} \bibinfo{volume}{8}, \bibinfo{number}{1} (\bibinfo{year}{2002}).
\newblock


\bibitem[Pasqual and Wobbrock(2014)]%
        {Pasqual2014-xt}
\bibfield{author}{\bibinfo{person}{Phillip~T Pasqual} {and} \bibinfo{person}{Jacob~O Wobbrock}.} \bibinfo{year}{2014}\natexlab{}.
\newblock \showarticletitle{Mouse pointing endpoint prediction using kinematic template matching}. In \bibinfo{booktitle}{\emph{Proceedings of the SIGCHI Conference on Human Factors in Computing Systems}}. \bibinfo{publisher}{ACM}, \bibinfo{address}{New York, NY, USA}, \bibinfo{pages}{743--752}.
\newblock


\bibitem[Peters et~al\mbox{.}(2018)]%
        {Peters2018-sv}
\bibfield{author}{\bibinfo{person}{Keshia~M Peters}, \bibinfo{person}{Valerie~E Kelly}, \bibinfo{person}{Tasha Chang}, \bibinfo{person}{Madeline~C Weismann}, \bibinfo{person}{Sarah Westcott~McCoy}, {and} \bibinfo{person}{Katherine~M Steele}.} \bibinfo{year}{2018}\natexlab{}.
\newblock \showarticletitle{Muscle recruitment and coordination during upper-extremity functional tests}.
\newblock \bibinfo{journal}{\emph{J. Electromyogr. Kinesiol.}}  \bibinfo{volume}{38} (\bibinfo{date}{Feb.} \bibinfo{year}{2018}), \bibinfo{pages}{143--150}.
\newblock


\bibitem[Rahimian et~al\mbox{.}(2021)]%
        {Rahimian2021-sq}
\bibfield{author}{\bibinfo{person}{Elahe Rahimian}, \bibinfo{person}{Soheil Zabihi}, \bibinfo{person}{Amir Asif}, \bibinfo{person}{Dario Farina}, \bibinfo{person}{Seyed~Farokh Atashzar}, {and} \bibinfo{person}{Arash Mohammadi}.} \bibinfo{year}{2021}\natexlab{}.
\newblock \showarticletitle{{FS}-{HGR}: Few-shot learning for Hand Gesture Recognition via electromyography}.
\newblock \bibinfo{journal}{\emph{IEEE Trans. Neural Syst. Rehabil. Eng.}}  \bibinfo{volume}{29} (\bibinfo{date}{June} \bibinfo{year}{2021}), \bibinfo{pages}{1004--1015}.
\newblock


\bibitem[Ritter et~al\mbox{.}(2015)]%
        {Ritter2015-vn}
\bibfield{author}{\bibinfo{person}{Walter Ritter}, \bibinfo{person}{Guido Kempter}, {and} \bibinfo{person}{Tobias Werner}.} \bibinfo{year}{2015}\natexlab{}.
\newblock \showarticletitle{User-acceptance of latency in touch interactions}.
\newblock In \bibinfo{booktitle}{\emph{Universal Access in Human-Computer Interaction. Access to Interaction}}. \bibinfo{publisher}{Springer International Publishing}, \bibinfo{address}{Cham}, \bibinfo{pages}{139--147}.
\newblock


\bibitem[Rizzoglio et~al\mbox{.}(2021)]%
        {Rizzoglio2021-gj}
\bibfield{author}{\bibinfo{person}{Fabio Rizzoglio}, \bibinfo{person}{Maura Casadio}, \bibinfo{person}{Dalia De~Santis}, {and} \bibinfo{person}{Ferdinando~A Mussa-Ivaldi}.} \bibinfo{year}{2021}\natexlab{}.
\newblock \showarticletitle{Building an adaptive interface via unsupervised tracking of latent manifolds}.
\newblock \bibinfo{journal}{\emph{Neural Netw.}}  \bibinfo{volume}{137} (\bibinfo{date}{May} \bibinfo{year}{2021}), \bibinfo{pages}{174--187}.
\newblock


\bibitem[Rubine(1991)]%
        {Rubine1991-yo}
\bibfield{author}{\bibinfo{person}{Dean Rubine}.} \bibinfo{year}{1991}\natexlab{}.
\newblock \showarticletitle{Specifying gestures by example}.
\newblock \bibinfo{journal}{\emph{Comput. Graph. (ACM)}} \bibinfo{volume}{25}, \bibinfo{number}{4} (\bibinfo{date}{July} \bibinfo{year}{1991}), \bibinfo{pages}{329--337}.
\newblock


\bibitem[Saponas et~al\mbox{.}(2008)]%
        {Saponas2008-dt}
\bibfield{author}{\bibinfo{person}{T~Scott Saponas}, \bibinfo{person}{Desney~S Tan}, \bibinfo{person}{Dan Morris}, {and} \bibinfo{person}{Ravin Balakrishnan}.} \bibinfo{year}{2008}\natexlab{}.
\newblock \showarticletitle{Demonstrating the feasibility of using forearm electromyography for muscle-computer interfaces}. In \bibinfo{booktitle}{\emph{Proceedings of the SIGCHI Conference on Human Factors in Computing Systems}}, Vol.~\bibinfo{volume}{00}. \bibinfo{publisher}{ACM}, \bibinfo{address}{New York, NY, USA}, \bibinfo{pages}{515}.
\newblock


\bibitem[Schwarz et~al\mbox{.}(2014)]%
        {Schwarz2014-pi}
\bibfield{author}{\bibinfo{person}{Julia Schwarz}, \bibinfo{person}{Charles~Claudius Marais}, \bibinfo{person}{Tommer Leyvand}, \bibinfo{person}{Scott~E Hudson}, {and} \bibinfo{person}{Jennifer Mankoff}.} \bibinfo{year}{2014}\natexlab{}.
\newblock \showarticletitle{Combining body pose, gaze, and gesture to determine intention to interact in vision-based interfaces}. In \bibinfo{booktitle}{\emph{Proceedings of the SIGCHI Conference on Human Factors in Computing Systems}} \emph{(\bibinfo{series}{CHI '14})}. \bibinfo{publisher}{ACM}, \bibinfo{address}{New York, NY, USA}, \bibinfo{pages}{3443--3452}.
\newblock


\bibitem[Shahi et~al\mbox{.}(2024)]%
        {Shahi2024-bb}
\bibfield{author}{\bibinfo{person}{Soroush Shahi}, \bibinfo{person}{Vimal Mollyn}, \bibinfo{person}{Cori Tymoszek~Park}, \bibinfo{person}{Runchang Kang}, \bibinfo{person}{Asaf Liberman}, \bibinfo{person}{Oron Levy}, \bibinfo{person}{Jun Gong}, \bibinfo{person}{Abdelkareem Bedri}, {and} \bibinfo{person}{Gierad Laput}.} \bibinfo{year}{2024}\natexlab{}.
\newblock \showarticletitle{Vision-based hand gesture customization from a single demonstration}. In \bibinfo{booktitle}{\emph{Proceedings of the 37th Annual ACM Symposium on User Interface Software and Technology}}, Vol.~\bibinfo{volume}{33}. \bibinfo{publisher}{ACM}, \bibinfo{address}{New York, NY, USA}, \bibinfo{pages}{1--14}.
\newblock


\bibitem[Shapiro and Wilk(1965)]%
        {Shapiro1965-gs}
\bibfield{author}{\bibinfo{person}{S~S Shapiro} {and} \bibinfo{person}{M~B Wilk}.} \bibinfo{year}{1965}\natexlab{}.
\newblock \showarticletitle{An analysis of variance test for normality (complete samples)}.
\newblock \bibinfo{journal}{\emph{Biometrika}} \bibinfo{volume}{52}, \bibinfo{number}{3-4} (\bibinfo{date}{Dec.} \bibinfo{year}{1965}), \bibinfo{pages}{591--611}.
\newblock


\bibitem[Sharma et~al\mbox{.}(2023)]%
        {Sharma2023-gf}
\bibfield{author}{\bibinfo{person}{Adwait Sharma}, \bibinfo{person}{Christina Salchow-Hömmen}, \bibinfo{person}{Vimal~Suresh Mollyn}, \bibinfo{person}{Aditya~Shekhar Nittala}, \bibinfo{person}{Michael~A Hedderich}, \bibinfo{person}{Marion Koelle}, \bibinfo{person}{Thomas Seel}, {and} \bibinfo{person}{Jürgen Steimle}.} \bibinfo{year}{2023}\natexlab{}.
\newblock \showarticletitle{{SparseIMU}: Computational design of sparse {IMU} layouts for sensing fine-grained finger microgestures}.
\newblock \bibinfo{journal}{\emph{ACM Trans. Comput. Hum. Interact.}} \bibinfo{volume}{30}, \bibinfo{number}{3} (\bibinfo{date}{June} \bibinfo{year}{2023}), \bibinfo{pages}{1--40}.
\newblock


\bibitem[Smirnov(1939)]%
        {Smirnov1939-sv}
\bibfield{author}{\bibinfo{person}{N Smirnov}.} \bibinfo{year}{1939}\natexlab{}.
\newblock \showarticletitle{Sur les écarts de la courbe de distribution empirique: Recueil Mathématique (Matematiceskii Sbornik), v}.
\newblock \bibinfo{journal}{\emph{New Solut.}}  \bibinfo{volume}{6} (\bibinfo{year}{1939}), \bibinfo{pages}{3--26}.
\newblock


\bibitem[{SparkFun}(2023)]%
        {SparkFun2023-vs}
\bibfield{author}{\bibinfo{person}{{SparkFun}}.} \bibinfo{year}{2023}\natexlab{}.
\newblock \bibinfo{title}{{MyoWare} 2.0 Muscle Sensor}.
\newblock \bibinfo{howpublished}{\url{https://www.sparkfun.com/products/21265}}.
\newblock
\newblock
\shownote{Accessed: 2024-11-1}.


\bibitem[Staude et~al\mbox{.}(2001)]%
        {Staude2001-vw}
\bibfield{author}{\bibinfo{person}{Gerhard Staude}, \bibinfo{person}{Claus Flachenecker}, \bibinfo{person}{Martin Daumer}, {and} \bibinfo{person}{Werner Wolf}.} \bibinfo{year}{2001}\natexlab{}.
\newblock \showarticletitle{Onset Detection in Surface Electromyographic Signals: A Systematic Comparison of Methods}.
\newblock \bibinfo{journal}{\emph{EURASIP J. Adv. Signal Process.}} \bibinfo{volume}{2001}, \bibinfo{number}{2} (\bibinfo{date}{June} \bibinfo{year}{2001}), \bibinfo{pages}{1--15}.
\newblock


\bibitem[Su et~al\mbox{.}(2023)]%
        {Su2023-sy}
\bibfield{author}{\bibinfo{person}{Dongnan Su}, \bibinfo{person}{Zhigang Hu}, \bibinfo{person}{Jipeng Wu}, \bibinfo{person}{Peng Shang}, {and} \bibinfo{person}{Zhaohui Luo}.} \bibinfo{year}{2023}\natexlab{}.
\newblock \showarticletitle{Review of adaptive control for stroke lower limb exoskeleton rehabilitation robot based on motion intention recognition}.
\newblock \bibinfo{journal}{\emph{Front. Neurorobot.}}  \bibinfo{volume}{17} (\bibinfo{date}{July} \bibinfo{year}{2023}), \bibinfo{pages}{1186175}.
\newblock


\bibitem[Tong et~al\mbox{.}(2021)]%
        {Tong2021-pr}
\bibfield{author}{\bibinfo{person}{Catherine Tong}, \bibinfo{person}{Jinchen Ge}, {and} \bibinfo{person}{Nicholas~D Lane}.} \bibinfo{year}{2021}\natexlab{}.
\newblock \showarticletitle{Zero-shot learning for {IMU}-based activity recognition using video embeddings}.
\newblock \bibinfo{journal}{\emph{Proc. ACM Interact. Mob. Wearable Ubiquitous Technol.}} \bibinfo{volume}{5}, \bibinfo{number}{4} (\bibinfo{date}{Dec.} \bibinfo{year}{2021}), \bibinfo{pages}{1--23}.
\newblock


\bibitem[Triwiyanto et~al\mbox{.}(2023)]%
        {Triwiyanto2023-ub}
\bibfield{author}{\bibinfo{person}{Triwiyanto Triwiyanto}, \bibinfo{person}{Wahyu Caesarendra}, \bibinfo{person}{Abdussalam~Ali Ahmed}, {and} \bibinfo{person}{{Abdullayev}}.} \bibinfo{year}{2023}\natexlab{}.
\newblock \showarticletitle{How deep learning and neural networks can improve prosthetics and exoskeletons: A review of state-of-the-art methods and challenges}.
\newblock \bibinfo{journal}{\emph{Journal of Electronics, Electromedical Engineering, and Medical Informatics}} \bibinfo{volume}{5}, \bibinfo{number}{4} (\bibinfo{date}{Oct.} \bibinfo{year}{2023}), \bibinfo{pages}{277--289}.
\newblock


\bibitem[Vatavu(2017)]%
        {Vatavu2017-lx}
\bibfield{author}{\bibinfo{person}{Radu-Daniel Vatavu}.} \bibinfo{year}{2017}\natexlab{}.
\newblock \showarticletitle{Improving gesture recognition accuracy on touch screens for users with low vision}. In \bibinfo{booktitle}{\emph{Proceedings of the 2017 CHI Conference on Human Factors in Computing Systems}} \emph{(\bibinfo{series}{CHI '17}, Vol.~\bibinfo{volume}{2017-May})}. \bibinfo{publisher}{ACM}, \bibinfo{address}{New York, NY, USA}, \bibinfo{pages}{4667--4679}.
\newblock


\bibitem[Vatavu et~al\mbox{.}(2012)]%
        {Vatavu2012-da}
\bibfield{author}{\bibinfo{person}{Radu-Daniel Vatavu}, \bibinfo{person}{Lisa Anthony}, {and} \bibinfo{person}{Jacob~O Wobbrock}.} \bibinfo{year}{2012}\natexlab{}.
\newblock \showarticletitle{Gestures as point clouds: a \${P} recognizer for user interface prototypes}. In \bibinfo{booktitle}{\emph{Proceedings of the 14th ACM international conference on Multimodal interaction}} \emph{(\bibinfo{series}{ICMI '12})}. \bibinfo{publisher}{ACM}, \bibinfo{address}{New York, NY, USA}, \bibinfo{pages}{273--280}.
\newblock


\bibitem[Vatavu et~al\mbox{.}(2018)]%
        {Vatavu2018-lb}
\bibfield{author}{\bibinfo{person}{Radu-Daniel Vatavu}, \bibinfo{person}{Lisa Anthony}, {and} \bibinfo{person}{Jacob~O Wobbrock}.} \bibinfo{year}{2018}\natexlab{}.
\newblock \showarticletitle{\${Q}: a super-quick, articulation-invariant stroke-gesture recognizer for low-resource devices}. In \bibinfo{booktitle}{\emph{Proceedings of the 20th International Conference on Human-Computer Interaction with Mobile Devices and Services}} \emph{(\bibinfo{series}{MobileHCI '18}, \bibinfo{number}{Article 23})}. \bibinfo{publisher}{ACM}, \bibinfo{address}{New York, NY, USA}, \bibinfo{pages}{1--12}.
\newblock


\bibitem[Vatavu and Ungurean(2022)]%
        {Vatavu2022-qm}
\bibfield{author}{\bibinfo{person}{Radu-Daniel Vatavu} {and} \bibinfo{person}{Ovidiu-Ciprian Ungurean}.} \bibinfo{year}{2022}\natexlab{}.
\newblock \showarticletitle{Understanding gesture input articulation with upper-body wearables for users with upper-body motor impairments}. In \bibinfo{booktitle}{\emph{CHI Conference on Human Factors in Computing Systems}}. \bibinfo{publisher}{ACM}, \bibinfo{address}{New York, NY, USA}.
\newblock


\bibitem[Wang et~al\mbox{.}(2021)]%
        {Wang2020-ua}
\bibfield{author}{\bibinfo{person}{Yaqing Wang}, \bibinfo{person}{Quanming Yao}, \bibinfo{person}{James~T Kwok}, {and} \bibinfo{person}{Lionel~M Ni}.} \bibinfo{year}{2021}\natexlab{}.
\newblock \showarticletitle{Generalizing from a few examples: A survey on few-shot learning}.
\newblock \bibinfo{journal}{\emph{ACM Comput. Surv.}} \bibinfo{volume}{53}, \bibinfo{number}{3} (\bibinfo{date}{May} \bibinfo{year}{2021}), \bibinfo{pages}{1--34}.
\newblock


\bibitem[Wobbrock et~al\mbox{.}(2018)]%
        {Wobbrock2018-cz}
\bibfield{author}{\bibinfo{person}{Jacob~O Wobbrock}, \bibinfo{person}{Krzysztof~Z Gajos}, \bibinfo{person}{Shaun~K Kane}, {and} \bibinfo{person}{Gregg~C Vanderheiden}.} \bibinfo{year}{2018}\natexlab{}.
\newblock \showarticletitle{Ability-based design}.
\newblock \bibinfo{journal}{\emph{Commun. ACM}} \bibinfo{volume}{61}, \bibinfo{number}{6} (\bibinfo{date}{May} \bibinfo{year}{2018}), \bibinfo{pages}{62--71}.
\newblock


\bibitem[Wobbrock et~al\mbox{.}(2011)]%
        {Wobbrock2011-wl}
\bibfield{author}{\bibinfo{person}{Jacob~O Wobbrock}, \bibinfo{person}{Shaun~K Kane}, \bibinfo{person}{Krzysztof~Z Gajos}, \bibinfo{person}{Susumu Harada}, {and} \bibinfo{person}{Jon Froehlich}.} \bibinfo{year}{2011}\natexlab{}.
\newblock \showarticletitle{Ability-based design: Concept, principles and examples}.
\newblock \bibinfo{journal}{\emph{ACM Trans. Access. Comput.}} \bibinfo{volume}{3}, \bibinfo{number}{3} (\bibinfo{date}{April} \bibinfo{year}{2011}), \bibinfo{pages}{1--27}.
\newblock


\bibitem[Wobbrock and Kientz(2016)]%
        {Wobbrock2016-to}
\bibfield{author}{\bibinfo{person}{Jacob~O Wobbrock} {and} \bibinfo{person}{Julie~A Kientz}.} \bibinfo{year}{2016}\natexlab{}.
\newblock \showarticletitle{Research contributions in human-computer interaction}.
\newblock \bibinfo{journal}{\emph{Interactions}} \bibinfo{volume}{23}, \bibinfo{number}{3} (\bibinfo{date}{April} \bibinfo{year}{2016}), \bibinfo{pages}{38--44}.
\newblock


\bibitem[Wobbrock et~al\mbox{.}(2009)]%
        {Wobbrock2009-md}
\bibfield{author}{\bibinfo{person}{Jacob~O Wobbrock}, \bibinfo{person}{Meredith~Ringel Morris}, {and} \bibinfo{person}{Andrew~D Wilson}.} \bibinfo{year}{2009}\natexlab{}.
\newblock \showarticletitle{User-defined gestures for surface computing}. In \bibinfo{booktitle}{\emph{Proceedings of the SIGCHI Conference on Human Factors in Computing Systems}}. \bibinfo{publisher}{ACM}, \bibinfo{address}{New York, NY, USA}, \bibinfo{pages}{1083--1092}.
\newblock


\bibitem[Wobbrock et~al\mbox{.}(2007)]%
        {Wobbrock2007-ul}
\bibfield{author}{\bibinfo{person}{Jacob~O Wobbrock}, \bibinfo{person}{Andrew~D Wilson}, {and} \bibinfo{person}{Yang Li}.} \bibinfo{year}{2007}\natexlab{}.
\newblock \showarticletitle{Gestures without libraries, toolkits or training: a \$1 recognizer for user interface prototypes}. In \bibinfo{booktitle}{\emph{Proceedings of the 20th annual ACM symposium on User interface software and technology}}. \bibinfo{publisher}{ACM}, \bibinfo{address}{New York, NY, USA}, \bibinfo{pages}{159–168}.
\newblock


\bibitem[Wu et~al\mbox{.}(2019)]%
        {Wu2019-fd}
\bibfield{author}{\bibinfo{person}{Huiyue Wu}, \bibinfo{person}{Shaoke Zhang}, \bibinfo{person}{Jiayi Liu}, \bibinfo{person}{Jiali Qiu}, {and} \bibinfo{person}{Xiaolong~(luke) Zhang}.} \bibinfo{year}{2019}\natexlab{}.
\newblock \showarticletitle{The gesture disagreement problem in free-hand gesture interaction}.
\newblock \bibinfo{journal}{\emph{Int. J. Hum. Comput. Interact.}} \bibinfo{volume}{35}, \bibinfo{number}{12} (\bibinfo{date}{July} \bibinfo{year}{2019}), \bibinfo{pages}{1102--1114}.
\newblock


\bibitem[Xie et~al\mbox{.}(2018)]%
        {Xie2018-gz}
\bibfield{author}{\bibinfo{person}{Baao Xie}, \bibinfo{person}{Baihua Li}, {and} \bibinfo{person}{Andy Harland}.} \bibinfo{year}{2018}\natexlab{}.
\newblock \showarticletitle{Movement and gesture recognition using deep learning and wearable-sensor technology}. In \bibinfo{booktitle}{\emph{Proceedings of the 2018 International Conference on Artificial Intelligence and Pattern Recognition}} \emph{(\bibinfo{series}{AIPR '18})}. \bibinfo{publisher}{ACM}, \bibinfo{address}{New York, NY, USA}, \bibinfo{pages}{26--31}.
\newblock


\bibitem[Xu et~al\mbox{.}(2022)]%
        {Xu2022-dw}
\bibfield{author}{\bibinfo{person}{Xuhai Xu}, \bibinfo{person}{Jun Gong}, \bibinfo{person}{Carolina Brum}, \bibinfo{person}{Lilian Liang}, \bibinfo{person}{Bongsoo Suh}, \bibinfo{person}{Shivam~Kumar Gupta}, \bibinfo{person}{Yash Agarwal}, \bibinfo{person}{Laurence Lindsey}, \bibinfo{person}{Runchang Kang}, \bibinfo{person}{Behrooz Shahsavari}, \bibinfo{person}{Tu Nguyen}, \bibinfo{person}{Heriberto Nieto}, \bibinfo{person}{Scott~E Hudson}, \bibinfo{person}{Charlie Maalouf}, \bibinfo{person}{Jax~Seyed Mousavi}, {and} \bibinfo{person}{Gierad Laput}.} \bibinfo{year}{2022}\natexlab{}.
\newblock \showarticletitle{Enabling hand gesture customization on wrist-worn devices}. In \bibinfo{booktitle}{\emph{Proceedings of the SIGCHI Conference on Human Factors in Computing Systems}}. \bibinfo{publisher}{ACM}, \bibinfo{address}{New York, NY, USA}.
\newblock


\bibitem[Yamagami(2024)]%
        {Yamagami2024-qg}
\bibfield{author}{\bibinfo{person}{Momona Yamagami}.} \bibinfo{year}{2024}\natexlab{}.
\newblock \bibinfo{title}{{M}-analysis}.
\newblock \bibinfo{howpublished}{\url{https://github.com/YamagamiLab/M-analysis}}.
\newblock
\newblock
\shownote{Accessed: 2026-10-30}.


\bibitem[Yamagami et~al\mbox{.}(2022)]%
        {Yamagami2022-jl}
\bibfield{author}{\bibinfo{person}{Momona Yamagami}, \bibinfo{person}{Sasa Junuzovic}, \bibinfo{person}{Mar Gonzalez-Franco}, \bibinfo{person}{Eyal Ofek}, \bibinfo{person}{Edward Cutrell}, \bibinfo{person}{John~R Porter}, \bibinfo{person}{Andrew~D Wilson}, {and} \bibinfo{person}{Martez~E Mott}.} \bibinfo{year}{2022}\natexlab{}.
\newblock \showarticletitle{Two-in-one: A design space for mapping unimanual input into bimanual interactions in {VR} for users with limited movement}.
\newblock \bibinfo{journal}{\emph{ACM Transactions on Accessible Computing (TACCESS)}} \bibinfo{volume}{15}, \bibinfo{number}{3} (\bibinfo{date}{July} \bibinfo{year}{2022}), \bibinfo{pages}{1--25}.
\newblock


\bibitem[Yamagami et~al\mbox{.}(2018)]%
        {Yamagami2018-cg}
\bibfield{author}{\bibinfo{person}{Momona Yamagami}, \bibinfo{person}{Keshia~M Peters}, \bibinfo{person}{Ivana Milovanovic}, \bibinfo{person}{Irene Kuang}, \bibinfo{person}{Zeyu Yang}, \bibinfo{person}{Nanshu Lu}, {and} \bibinfo{person}{Katherine~M Steele}.} \bibinfo{year}{2018}\natexlab{}.
\newblock \showarticletitle{Assessment of dry epidermal electrodes for long-term electromyography measurements}.
\newblock \bibinfo{journal}{\emph{Sensors}} \bibinfo{volume}{18}, \bibinfo{number}{4} (\bibinfo{date}{April} \bibinfo{year}{2018}), \bibinfo{pages}{1--15}.
\newblock


\bibitem[Yamagami et~al\mbox{.}(2023)]%
        {Yamagami2023-xj}
\bibfield{author}{\bibinfo{person}{Momona Yamagami}, \bibinfo{person}{Alexandra~A Portnova-Fahreeva}, \bibinfo{person}{Junhan Kong}, \bibinfo{person}{Jacob~O Wobbrock}, {and} \bibinfo{person}{Jennifer Mankof}.} \bibinfo{year}{2023}\natexlab{}.
\newblock \showarticletitle{How do people with limited movement personalize upper-body gestures? Considerations for the design of personalized and accessible gesture interfaces}. In \bibinfo{booktitle}{\emph{Proceedings of the 25th International ACM SIGACCESS Conference on Computers and Accessibility (ASSETS '23).}} \emph{(\bibinfo{series}{ASSETS '23}, Vol.~\bibinfo{volume}{2023})}. \bibinfo{publisher}{Association for Computing Machinery}, \bibinfo{address}{New York, NY, USA}, \bibinfo{pages}{1--15}.
\newblock


\bibitem[Yeo et~al\mbox{.}(2020)]%
        {Yeo2020-zp}
\bibfield{author}{\bibinfo{person}{Hui-Shyong Yeo}, \bibinfo{person}{Wenxin Feng}, {and} \bibinfo{person}{Michael~Xuelin Huang}.} \bibinfo{year}{2020}\natexlab{}.
\newblock \showarticletitle{{WATouCH}: Enabling direct input on non-touchscreen using smartwatch's photoplethysmogram and {IMU} sensor fusion}. In \bibinfo{booktitle}{\emph{Proceedings of the 2020 CHI Conference on Human Factors in Computing Systems}}. \bibinfo{publisher}{ACM}, \bibinfo{address}{New York, NY, USA}.
\newblock


\bibitem[Zhang et~al\mbox{.}(2022)]%
        {Zhang2022-nw}
\bibfield{author}{\bibinfo{person}{Qian Zhang}, \bibinfo{person}{Jiazhen Jing}, \bibinfo{person}{Dong Wang}, {and} \bibinfo{person}{Run Zhao}.} \bibinfo{year}{2022}\natexlab{}.
\newblock \showarticletitle{{WearSign}: Pushing the limit of sign language translation using inertial and {EMG} wearables}.
\newblock \bibinfo{journal}{\emph{Proc. ACM Interact. Mob. Wearable Ubiquitous Technol.}} \bibinfo{volume}{6}, \bibinfo{number}{1} (\bibinfo{date}{March} \bibinfo{year}{2022}), \bibinfo{pages}{1--27}.
\newblock


\bibitem[Zhang et~al\mbox{.}(2021)]%
        {Zhang2021-vr}
\bibfield{author}{\bibinfo{person}{Yingwei Zhang}, \bibinfo{person}{Yiqiang Chen}, \bibinfo{person}{Hanchao Yu}, \bibinfo{person}{Xiaodong Yang}, \bibinfo{person}{Ruizhe Sun}, {and} \bibinfo{person}{Bixiao Zeng}.} \bibinfo{year}{2021}\natexlab{}.
\newblock \showarticletitle{A feature adaptive learning method for high-density {sEMG}-based gesture recognition}.
\newblock \bibinfo{journal}{\emph{Proc. ACM Interact. Mob. Wearable Ubiquitous Technol.}} \bibinfo{volume}{5}, \bibinfo{number}{1} (\bibinfo{date}{March} \bibinfo{year}{2021}), \bibinfo{pages}{1--26}.
\newblock


\bibitem[Zhao et~al\mbox{.}(2022)]%
        {Zhao2022-gc}
\bibfield{author}{\bibinfo{person}{Xuan Zhao}, \bibinfo{person}{Mingming Fan}, {and} \bibinfo{person}{Teng Han}.} \bibinfo{year}{2022}\natexlab{}.
\newblock \showarticletitle{“I don’t want people to look at me differently”: Designing user-defined above-the-neck gestures for people with upper body motor impairments}. In \bibinfo{booktitle}{\emph{CHI Conference on Human Factors in Computing Systems}}. \bibinfo{publisher}{ACM}, \bibinfo{address}{New York, NY, USA}.
\newblock


\bibitem[Zhou et~al\mbox{.}(2024)]%
        {Zhou2024-io}
\bibfield{author}{\bibinfo{person}{Yexu Zhou}, \bibinfo{person}{Haibin Zhao}, \bibinfo{person}{Yiran Huang}, \bibinfo{person}{Tobias Röddiger}, \bibinfo{person}{Murat Kurnaz}, \bibinfo{person}{Till Riedel}, {and} \bibinfo{person}{Michael Beigl}.} \bibinfo{year}{2024}\natexlab{}.
\newblock \showarticletitle{{AutoAugHAR}: Automated data augmentation for sensor-based human activity recognition}.
\newblock \bibinfo{journal}{\emph{Proc. ACM Interact. Mob. Wearable Ubiquitous Technol.}} \bibinfo{volume}{8}, \bibinfo{number}{2} (\bibinfo{date}{May} \bibinfo{year}{2024}), \bibinfo{pages}{1--27}.
\newblock


\bibitem[Zhu et~al\mbox{.}(2018)]%
        {zhu_2018_EIS}
\bibfield{author}{\bibinfo{person}{Zijie Zhu}, \bibinfo{person}{Xuewei Wang}, \bibinfo{person}{Aakaash Kapoor}, \bibinfo{person}{Zhichao Zhang}, \bibinfo{person}{Tingrui Pan}, {and} \bibinfo{person}{Zhou Yu}.} \bibinfo{year}{2018}\natexlab{}.
\newblock \showarticletitle{EIS: A Wearable Device for Epidermal American Sign Language Recognition}.
\newblock \bibinfo{journal}{\emph{Proc. ACM Interact. Mob. Wearable Ubiquitous Technol.}} \bibinfo{volume}{2}, \bibinfo{number}{4}, Article \bibinfo{articleno}{202} (\bibinfo{date}{Dec.} \bibinfo{year}{2018}), \bibinfo{numpages}{22}~pages.
\newblock


\end{thebibliography}

\appendix
\newpage

\renewcommand{\thefigure}{A-\arabic{figure}} 
\setcounter{figure}{0}  

\section{\alg{} Recognizer}\label{sec:appdx:M}

A biosignal candidate gesture $G$ or template $T_i$ is composed of multiple $biosignals$ $b_j$.
Each $biosignal$ may have one or more data $channels$ $c$.
Each channel is composed of a one-dimensional time-series $points$ path (Fig.~\ref{fig:teaser}).
The code is formatted according to~\citet{Cormen2022-vv}.

\newcommand{\tnew}{newTime}
\newcommand{\told}{oldTime}

\begin{algorithm}
\raggedright
RESAMPLE($points, n$)
\\
\textbf{input}: $points$ is an array of length $N$\\
\-\hspace{.95cm}$n$ is the desired number of resampled time points\\
\textbf{output}: $newPoints$ is an array of length $n$\\
\begin{algorithmic}[1]
    \State $N\gets$ LENGTH($points$)
    \State $\told,\tnew  \gets \textrm{TIME}(N,n)$
    \State $t_o\gets 0$
    \State \textbf{initialize} $newPoints$ array
    \For{time point $t_n$ in $\tnew$}
        \While{$\told_{t_o}<t_n$}
            \State $t_o\gets t_o+1$
        \EndWhile
        \State $y_1,y_2\gets points_{t_o-1},points_{t_o}$
        \State $x_1,x_2\gets \told_{t_o-1},\told_{t_o}$
        \State $m\gets \frac{y_2-y_1}{x_2-x_1}$ \Comment{calculate slope}
        \State $b\gets y_2 - m\times x_2$\Comment{calculate intercept}
        \State $q\gets m\times t_n+b$ \Comment{calculate new point}
        \State $\textrm{APPEND}(newPoints,q)$
    \EndFor
    \State \textbf{return} $newPoints$
\end{algorithmic}
\raggedright
TIME($N,n$) 
\\
\textbf{input}: $N,n$ are integers corresponding to the array length of $points$ and desired number of resampled time points, \\\-\hspace{.95cm} respectively\\
\textbf{output}: $oldTime$, $newTime$ are arrays of length $N$ and $n$, with the same initial and final time points 0 and $N$\\
\begin{algorithmic}[1]
    \State index $i\gets 0$
    \State \textbf{initialize} $\told$ array
    \While{index $i < N$} \Comment{create array for old time}
        \State APPEND$(\told,i)$
        \State $i\gets i+1$
    \EndWhile
    \State index $j\gets 0$
    \State \textbf{initialize} $\tnew$ array
    \While{index $j < n$} \Comment{create array for new time}
        \State APPEND$(\tnew,\frac{j(N-1)}{n-1})$
        \State $j\gets j+1$
    \EndWhile
    \State \textbf{return} $\told,\tnew$
\end{algorithmic}
\raggedright
LENGTH($points$)
\\
\textbf{input}: $points$ is an array of length $N$\\
\textbf{output}: $N$, the length of the input array\\
\begin{algorithmic}[1]
    \State $N\gets 0$ 
    \For{point $p_i$ in $points$}
        \State $N\gets N+1$
    \EndFor
    \State \textbf{return} $N$
\end{algorithmic}
\caption*{\textbf{Step 1.} For each channel $c$, resample the $points$ path into $n$ evenly spaced points using piecewise linear interpolation.} 
\label{alg:resample}
\end{algorithm}

\newpage
\begin{algorithm}
\raggedright
DEMEAN($points, n$)
\\
\textbf{input}: $points$ are all points in a biosignal channel $c_i$ $n$ is the length of $points$\\
\textbf{output}: $newPoints$ is an array with a mean of 0\\
\begin{algorithmic}[1]
    \State \textbf{initialize} $newPoints$ array
    \State $\bar{p}\gets \textrm{MEAN}(points)$
    \For{point $p_i$ in $points$}
        \State $q \gets p_i - \bar{p}$
        \State APPEND($newPoints,q$)
    \EndFor
    \State \textbf{return} $newPoints$
\end{algorithmic}
\raggedright
NORMALIZE($biosignal$)
\\
\textbf{input}: $biosignal$ is an array of all the channels $c_i$ after the DEMEAN function is applied for a given biosignal $b_j$ \\\-\hspace{.95cm}concatenated into one array\\
\textbf{output}: $newBiosignal$ is an array of all the channels for a given biosignal with a standard deviation of 1\\
\begin{algorithmic}[1]
    \State $\sigma\gets\mathrm{STD}(biosignal)$
    \State \textbf{initialize} $newPoints$ array
    \For{$points$ in each $biosignal$ $channel$}
        \For{point $p_i$ in $points$}
            \State $q \gets \frac{p_i}{\sigma}$
            \State APPEND($newPoint,q$)
        \EndFor
        \State APPEND($newBiosignal,newPoints$)
    \EndFor
    \State \textbf{return} $newBiosignal$
\end{algorithmic}
\raggedright
MEAN($points,n$)
\\
\textbf{input}: $points$ are all points in a biosignal channel $c_i$
\\\-\hspace{0.95cm} $n$ is the length of $points$\\
\textbf{output}: $\bar{p}$ is the average value of $points$\\
\begin{algorithmic}[1]
    \State $\bar{p} \gets 0$
    \For{point $p_i$ in $points$}
        \State $\bar{p} \gets \bar{p} + p_i$
    \EndFor
    \State $\bar{p} \gets \frac{\bar{p}}{n}$
    \State \textbf{return} $\bar{p}$
\end{algorithmic}
\raggedright
STD($biosignal$)
\\
\textbf{input}:  $biosignal$ is an array of all the channels $c_i$ after the DEMEAN function is applied for a given biosignal $b_j$ \\\-\hspace{.95cm}concatenated into one array\\
\textbf{output}: $\sigma$ is the standard deviation of $biosignal$\\
\begin{algorithmic}[1]
    \State $ix \gets \textrm{LENGTH}(biosignal)$
    \State $\bar{p}\gets \textrm{MEAN}(biosignal,ix)$ \Comment{get mean of all signals}
    \State $\sigma\gets 0$
    \For{point $p_i$ in $biosignal$}
        \State $\sigma \gets \sigma + (p_i - \bar{p})^2$
    \EndFor
    \State $\sigma \gets \sqrt{\frac{1}{ix}\sigma}$
    \State \textbf{return} $\sigma$
\end{algorithmic}
\caption*{\textbf{Step 2.} Scale $points$ for channels $c_i$ for a given biosignal $b_j$.
DEMEAN sets each point $p_i$ in $points$ such that the mean of a channel $c_i$ is equal to 0. 
STD takes the standard deviation of all channels for a given biosignal $b_j$ . 
NORMALIZE scales the $points$ such that the standard deviation of all channels $c$ for a given biosignal $b_j$ is equal to 1.} 
\label{alg:normalize}
\end{algorithm}

\newpage
\begin{algorithm}
\raggedright
PCA($D,c,n, nPC$)
\begin{algorithmic}[1]
    \State $covariance \gets \mathrm{COV}(D,c,n) $\Comment{Compute covariance matrix}
    \State $eigenvalues,eigenvectors\gets \mathrm{EIG(covariance)}$ \Comment{Compute eigenvalues and eigenvectors}
    \State $sorted\_eigenvectors\gets$ sort eigenvectors column-wise in order corresponding to the largest eigenvalue to the smallest in $eigenvalues$
    \State $U\gets$ the first $nPC$ columns of $sorted\_eigenvectors$
    \State $points\gets\mathrm{MATMUL}(D,U)$ \Comment{get transformed biosignal}
    \State $points\gets \mathrm{FLATTEN}(points)$ \Comment{flatten transformed biosignal into a one-dimensional array}
    \State \textbf{return} principal components $U$, transformed one-dimensional $points$
\end{algorithmic}
\raggedright
COV($D,c,n$)\\
\textbf{input}: $D$ is a matrix of size $c$ by $n$\\
\textbf{output}: $covariance$ is a matrix of size $c$ by $c$\\
\begin{algorithmic}[1]
    \State \textbf{initialize} $covariance$ matrix
    \For{index $j=1$ to $c$}
        \State \textbf{initialize} $cov$ array
        \For{index $k=1$ to $c$}
            \State $sum \gets 0$
            \For{index $i=1$ to $n$}
                \State $sum\gets sum + D_{j,i} D_{k,i}$
            \EndFor
            \State $sum\gets \frac{sum}{n-1}$
            \State APPEND($cov, sum$)
        \EndFor
        \State APPEND($covariance,cov$)
    \EndFor
    \State \textbf{return} $covariance$ 
\end{algorithmic}
\raggedright
EIG($covariance$)\\
\textbf{input}: $covariance$ is a matrix of size $c$ by $c$\\
\textbf{output}: $eigenvalues$ will be an array of length $c$, $eigenvectors$ will be a matrix of size $c$ by $c$\\
\begin{algorithmic}[1]
    \State use a linear algebra package to compute eigenvalues and eigenvectors of covariance matrix
    \State \textbf{return} $eigenvalues,eigenvectors$ 
\end{algorithmic}
\caption*{\textbf{Step 3.} Compute template principal components and transformed points for $nPC$ principal components. 
The input to the PCA is all points $p$ across all channels $c$ for all biosignals $b$ concatenated into a matrix $D$ with $n$ timepoints and $c$ biosignal channels (where $c$ is the total number of channels for all measured biosignals). 
For these steps, we assume that all biosignal channels $c$ are concatenated vertically such that the resulting data matrix $D$ is comprised of $c$ rows and $n$ columns (i.e.,, $D$ is a matrix of size $c\times n$).
After this step, the $template$ $T_i$ points $p$ are a concatenated one-dimensional signal.
Note that while this step has multiple computations, there are many built-in algorithms (e.g., Python: \texttt{sklearn.decomposition.pca}; Java: \texttt{weka.attributeSelection.PrincipalComponents}; R: \texttt{stats.princomp}; Matlab: \texttt{pca}) that compute the PCs and transformed points.
The built-in algorithms are often optimized such that the runtime is faster than implementing the PCA by hand~\cite{Developers_undated-bb}.} 
\label{alg:pca1}
\end{algorithm}

\newpage
\begin{algorithm}
\raggedright
MATMUL($D,U$)\\
\textbf{input}: $D^T$ is a matrix of size $n$ by $c$\\
\-\hspace{.95cm}$U$ is a matrix of size $c$ by $nPC$\\
\textbf{output}: $newD$ is an array of length $nPC \times n$\\
\begin{algorithmic}[1]
    \State $nPC\gets$ column length of $U$
    \State $n\gets$ column length of $D$
    \State \textbf{initialize} $newD$ matrix with $nPC$ rows and $n$ columns \\ \-\hspace{1.3cm}(i.e., a matrix of size $nPC$ by $n$)
    \For{index $i=1$ to $n$}
        \For{index $j=1$ to $n$}
            \For{index $k=1$ to $nPC$}
                \State $newD_{i,j}\gets D^T_{i,k} \times U_{k,j}$
            \EndFor
        \EndFor
    \EndFor
    \State \textbf{return} $newPoints$
\end{algorithmic}
\raggedright
FLATTEN($D,nPC,n$)\\
\textbf{input}: $D$ is a matrix of size $nPC$ by $n$\\
\textbf{output}: $newPoints$ is an array of length $nPC \times n$\\
\begin{algorithmic}[1]
    \State \textbf{initialize} $newPoints$ array
    \State $nPC,n\gets$ row and column length of $D$
    \For{index $i=1$ to $n$}
        \For{index $j=1$ to $n$}
            \State APPEND($newPoints,D_{i,j}$)
        \EndFor
    \EndFor
    \State \textbf{return} $newPoints$
\end{algorithmic}
\caption*{\textbf{Step 3 cont.}} 
\label{alg:pca2}
\end{algorithm}

\begin{algorithm}
\raggedright
RECOGNIZE(gesture $G, templates$)\\
\textbf{input}: $G$ is a matrix of size $c$ by $n$ $templates$ are the set of templates $T_i$, each of which contains one-dimensional array \\\-\hspace{.95cm}of points $p$ of length $nPC\times n$\\
\textbf{output}: matched template $T'$ $score$ of final matched template\\
\begin{algorithmic}[1]
    \State $b\gets \infty$
    \ForEach{template $T_i$ in $templates$}
        \State $U\gets$ principal components of $T_i$ computed in Step 3
        \State $points\gets \mathrm{MATMUL}(G^T,U)$
        \State $points\gets \mathrm{FLATTEN}(points)$
        \State $d\gets \mathrm{PATH-DISTANCE}(points,T_i)$
        \If{$d<b$}
            \State $b\gets d$
            \State $T' \gets T_i$
        \EndIf
    \EndFor
    \State \textbf{return} $<T'>$ 
\end{algorithmic}
\raggedright
PATH-DISTANCE($A,B$)\\
\textbf{input}: $A,B$ are both one-dimensional arrays\\
\textbf{output}: distance $d$ as a floating point
\begin{algorithmic}[1] 
    \State $d\gets 0$
    \ForEach{$A_i,B_i$ in $A,B$}
        \State $d\gets d + \mathrm{DISTANCE}(A_i,B_i)$
    \EndFor
    \State \textbf{return} $d$
\end{algorithmic}
\raggedright
DISTANCE($A_i, B_i$)\\
\textbf{input}: $A_i,B_i$ are both floating points\\
\textbf{output}: distance $d$ as a floating point
\begin{algorithmic}[1] 
    \State $d\gets$ absolute value of ($A_i-B_i$)
    \State \textbf{return} $d$
\end{algorithmic}
\caption*{\textbf{Step 4.} Match candidate gesture $G$ against each stored template $T_i$. 
$G$ is resampled and normalized.
The principal components $U$ have already been computed for the templates and have already been transformed and flattened into a one-dimensional points array $T_i$. 
The candidate gesture $G$ will be transformed and flattened into a one-dimensional points array during the template matching process.}
\label{alg:match}
\end{algorithm}

\newpage
\section{Data Segmentation Pseudocode}\label{sec:appdx:segmentation}

We implemented a custom gesture segmentation method for our dataset based on the recorded raw EMG channels.  
We solely used the biosignal channels $c$ from EMG data (16 EMG channels per gesture). 

\begin{algorithm}
\raggedright
SEGMENT($G$)\\
\textbf{input}: $G$ is a matrix of size $n$ time points by $c$ channels of the absolute value of the raw EMG data\\
\textbf{output}: $start, stop$, segmentation cutoff indices\\
\begin{algorithmic}[1]
    \State $G_{rms} \gets RMS(G)$ 
    \State $highest\_amplitude\_channels \gets$ get 3 highest amplitude channels across all $G$ for one participant
    \State $highest\_variance\_channels \gets$ get 3 highest variance channels within one $G$
    \State $relevant\_channels \gets$ combine $highest\_amplitude\_channels$ and $highest\_variance\_channels$, remove duplicate \\\-\hspace{2.9cm}channels and channels with variance below $0.1$ of the maximum variance across channels
    \State \textbf{initialize} $allStart, allStop$ arrays
    \ForEach{channel $c$ in $relevant\_channels$}
        \State $Dc \gets$ DIFF(c)  \Comment{take first difference across channel $c$}
        \State $D^2c \gets$ DIFF(Dc)  \Comment{take second difference}
        %
        \State \textbf{Compute different possible $start, stop$ based on heuristics for $c$:}
        \State $start_{Dc} \gets $ LARGEST\_SLOPE($c,max=True$)
        \State $stop_{Dc} \gets $ LARGEST\_SLOPE($c,max=False$)
        \State $startBelow_1,stopBelow_1 \gets$ THRESHOLD($c,value=0.3$) \Comment{Compute $start, stop$ for aggressive threshold}
        \State $startBelow_2,stopBelow_2 \gets$ THRESHOLD($c,value=0.15$) \Comment{Compute $start, stop$ for conservative threshold}
        \State $startAbove_1,stopAbove_1 \gets$ EDGE($c$,THRESHOLD($c,value=0.3$))
        \State \Comment{Check if initial and final values in $c$ surpass threshold}
        \State $startDc_1 \gets$  LARGEST\_SLOPE($Dc,max=True$), restrict $Dc$ to indices between $0$ and $startBelow_1$ 
        \State $stopDc_1 \gets$ LARGEST\_SLOPE($Dc,max=False$), restrict $Dc$ to indices between $stopBelow_1$ and array end
        \State $startDc_2 \gets$  
        LARGEST\_SLOPE($Dc,max=True$), restrict $Dc$ to indices between $0$ and $startBelow_1$
        \State $stopDc_2 \gets$ LARGEST\_SLOPE($Dc,max=False$), restrict $Dc$ to indices between $stopBelow_2$ and array end
        \State $startD^2c_1 \gets$  
        LARGEST\_SLOPE($D^2c,max=True$), restrict $D^2c$ to indices between $0$ and $startBelow_1$ 
        \State $stopD^2c_1 \gets$ LARGEST\_SLOPE($D^2c,max=False$), restrict $D^2c$ to indices between $stopBelow_1$ and array end
        \State APPEND($allStart,[startDc, startAbove_1, startBelow_1, startDc_1, startDc_2, startD^2c_1]$)
        \State APPEND($allStop,[stopDc, stopAbove_1, stopBelow_1, stopDc_1, stopDc_2, stopD^2c_1]$)
        \EndFor
        %
    \State \textbf{Choose most conservative (i.e., the smallest index for $start$ and largest index for $stop$.) $start,stop$ in}
    \State \-\hspace{0.5cm}$relevant\_channels$:
    \State $start \gets$ MIN($allStart$)
    \State $stop \gets$ MAX($allStop$)
    \State \textbf{return} $start, stop$
\end{algorithmic}
\raggedright
RMS($points,window,overlap$)\\
\textbf{input}: $points$ are all points in a biosignal channel $c_i$\\
\-\hspace{.95cm}$window$ is the window size over which to compute the RMS over\\
\-\hspace{.95cm}$overlap$ is the percent of overlap of each time window\\
\textbf{output}: $newPoints$, the root-mean-square of $points$
\begin{algorithmic}[1] 
    \State \textbf{initialize} $newPoints$ array
    \For{$i$ in $\frac{\textrm{LENGTH}(points)-window}{overlap\times window}$}
        \State $winPoints \gets $ all $points$ between $i\times overlap\times window$ to 
        $(i+1)\times window$
    \EndFor
    \For{point $p_j$ in $winPoints$}
        \State $q \gets \sqrt{\frac{1}{n}\sum_j p_j^2}$
        \State APPEND($newPoints,q$)
    \EndFor
    \State \textbf{return} $newPoints$
\end{algorithmic}
\caption*{\textbf{Data Segmentation.} Our custom data segmentation method obtains different possible cutoff values based on heuristics and chooses the most conservative cutoff indices.} 
\label{alg:segmentation}
\end{algorithm}

\newpage

\begin{algorithm}
\raggedright
LARGEST\_SLOPE($points,max=True$)\\
\textbf{input}: $points$ are all points in a biosignal channel $c_i$\\
\textbf{output}: $start, stop$, largest slope index for $c$
\begin{algorithmic}[1] 
    \If{$max=True$}
        \State $start \gets$ ARGMAX($Dc$) \Comment{Get largest positive increase}
        \State \textbf{return} $start$       
    \Else
        \State $stop \gets$ ARGMIN($Dc$) +1 \Comment{Get largest negative increase}
        \State \textbf{return} $stop$      
    \EndIf
\end{algorithmic}
\raggedright
EDGE($c,thresh$)\\
\textbf{input}: $points$ are all points in a biosignal channel $c_i$\\
\-\hspace{.95cm}$thresh$ is the threshold \\
\textbf{output}: $start, stop$, edge cases 
\begin{algorithmic}[1] 
    \State \textbf{initialize} $start \gets$ NaN
    \If{$c_1$ > $thresh$}
        \State $start$ = 1
    \EndIf
    \State $stop \gets$ NaN
    \If{$c_N$ > $thresh$} \Comment{Where $N$==LENGTH(c)}
        \State $stop$ = N
    \EndIf
\end{algorithmic}
\raggedright
THRESHOLD($c, value$)\\
\textbf{input}: $c$, a one-dimensional array\\
\textbf{output}: $start, stop$ for a given threshold $value$
\begin{algorithmic}[1] 
    \State $c_{max} \gets$ MAX($c$) \Comment{get maximum value in $c$}
    \State $c_{min} \gets$ MIN($c$) \Comment{get minimum value in $c$}
    \State $threshold \gets$ $value$($c_{max}$ - $c_{min}$) + $c_{min}$ 
    \State $cBelow \gets$ WHERE($c$ < $threshold$)  \Comment{Get all values in c below $threshold$}
    \State $start \gets$ first index of the output of WHERE(DIFF($cBelow$)>1)   \Comment{Return first instance where difference is \\\-\hspace{9.45cm}below threshold}
    \State $stop \gets$ last index of the output of WHERE(DIFF($cBelow$)>1)  \Comment{Return last instance where difference is\\\-\hspace{9.45cm} below threshold}
    \State \textbf{return} $start, stop$
\end{algorithmic}
\caption*{\textbf{Heuristic $start, stop$ Computations.} Each computation is a different way to heuristically compute $start, stop$ indices from our dataset.
These heuristic methods were chosen based on extensive data exploration.} 
\label{alg:heuristic}
\end{algorithm}

\section{PCA Explanation}\label{sec:appdx:pca}

\revision{Pseudocode of the PCA steps can be found in Appendix~\ref{sec:appdx:M}.
We will compute the principal components of a data matrix $D$ with $c$ rows and $n$ columns, where $c$ represents the total number of channels across all measured biosignal types and $n$ represents the number of resampled time points (\figref{fig:PCA}).}
First, the covariance matrix $cov$ of the data matrix $D$ is computed: $cov=\frac{1}{n-1}DD^T$. 
$cov$ is a matrix of size $c$ by $c$ and $D^T$ represents the transform of matrix $D$. 
Then, the $eigenvalues$ and $eigenvectors$ of $cov$ are computed with a built-in function (e.g., Python: \texttt{numpy.linalg.eig}; Java: \texttt{org.apache.commons.math3.linear.EigenDecomposition};  Matlab: \texttt{eig}; R: \texttt{matlib.eigen}). 
The eigenvalue decomposition of the covariance matrix ensures that the data represented in the latent space contains no redundancies (i.e., all eigenvectors are linearly independent).  
Next, the $eigenvectors$ are sorted column-wise in order corresponding from the largest \textit{eigenvalue} to the smallest $eigenvalue$, such that the first column in the sorted eigenvectors corresponds to the largest eigenvalue and the second column in the sorted eigenvectors corresponds to the second-largest eigenvalue and so on.
The sorting ensures that the first principal component transforms the original data matrix $D$ such that the first principal component contains the data of maximum variance, the second principal component contains the data of second maximum variance, and so on.  
The principal components $U$ are then defined as the first $nPC$ columns of the sorted eigenvectors.
$nPC$, or the number of principal components to keep, is a parameter that must be explored and defined by the prototyper when implementing \alg{}.
Lastly, the transpose of the original data matrix $D^T$ is multiplied with the principal components $U$ (i.e., $D^TU$) and flattened into a one-dimensional vector $points$ of length $nPC\times n$.

\end{document}